\documentclass[preprint]{elsarticle}

\scrollmode
\usepackage{amsmath}
\usepackage{amsfonts}
\usepackage{amssymb}
\usepackage{latexsym}
\usepackage{stmaryrd}
\usepackage{array}
\usepackage{exscale}
\newcommand{\nc}{\newcommand}
\newcommand{\ol}{\overline}
\newcommand{\ul}{\underline}
\newcommand{\es}{\emptyset}
\newcommand{\sm}{\setminus}
\newcommand{\ve}{\varepsilon}
\newcommand{\vp}{\varphi}
\newcommand{\bw}{\bigwedge}
\newcommand{\bv}{\bigvee}
\newcommand{\bc}{\bigcup}
\newcommand{\bca}{\bigcap}
\newcommand{\Lra}{\Leftrightarrow}

\newcommand{\Ra}{\Rightarrow}

\newcommand{\ra}{\rightarrow}

\newcommand{\la}{\leftarrow}
\newcommand{\lra}{\leftrightarrow}

\newcommand{\sse}{\subseteq}

\newcommand{\spe}{\supseteq}
\newcommand{\fa}{\forall}
\newcommand{\ex}{\exists}
\newcommand{\mr}{\mathrm}
\newcommand{\mc}{\mathcal}
\newcommand{\mf}{\mathfrak}

\newcommand{\DMO}{\DeclareMathOperator}
\newcommand{\DST}{\displaystyle}

\newcommand{\ZZ}{\mathbb{Z}}
\newcommand{\NN}{\mathbb{N}}
\newcommand{\NNZ}{\NN_0}

\newcommand{\RR}{\mathbb{R}}
\newcommand{\CC}{\mathbb{C}}

\newcommand{\aru}{\ar @{-}} 
\usepackage{listings}
\newcommand{\inl}[1]{\lstinline$#1$}
\newcommand{\und}{{\:\wedge\:}} 
\newcommand{\oder}{{\:\vee\:}} 
\newcommand{\mb}{{\:|\:}} 
\newcommand{\set}[1]{\{ #1 \}}
\newcommand{\setb}[1]{\big \{ \, #1 \, \big \}}
\DeclareMathOperator{\dom}{dom}

\nc{\simlvi}[1]{\!\sim_{#1}}
\DeclareMathOperator{\symdif}{\vartriangle} 
\nc{\apprel}[3]{{#1}(#2)_{(#3)}} 
\nc{\cmpli}[1]{\complement^1_{#1}} 
\nc{\cmplzi}[1]{\complement^0_{#1}} 
\nc{\cmplzoi}[1]{\complement^*_{#1}} 

\nc{\zf}{\mr{ZF}}
\nc{\zfmf}{\zf^0} 
\nc{\zfc}{\mr{ZFC}}
\nc{\zfcmf}{\zfc^0} 
\nc{\bst}{\mr{BST}} 
\newcommand{\tb}[2]{\set{#1, \dots, #2}} 
\providecommand{\abs}[1]{\lvert #1 \rvert} 
\providecommand{\norm}[1]{\lVert #1 \rVert} 
\providecommand{\inprod}[1]{\left\langle #1 \right\rangle} 
\newcommand{\trans}[1]{#1^{\hspace{0.05em}\mr{t}}} 
\makeatletter
\DeclareRobustCommand{\genericinterval}[2]{%
  \@ifstar{\genericinterval@star{#1}{#2}}{\genericinterval@nostar{#1}{#2}}}
\newcommand{\genericinterval@star}[4]{\mathopen{}\mathclose{\left#1#3,#4\right#2}}
\newcommand{\genericinterval@nostar}[4]{\mathopen{#1}#3,#4\mathclose{#2}}

\makeatother
\nc{\untit}[2]{{#1}^{#2 \downarrow}} 
\nc{\obit}[2]{{#1}^{#2 \uparrow}} 
\nc{\inzEKi}[1]{\mc{I}^{\mr{V}}_{#1}}

\nc{\inzKEi}[1]{\mc{I}^{\mr{E}}_{#1}}

\nc{\adjEi}[1]{\mc{A}^{\mr{V}}_{#1}}

\nc{\BD}[1]{{#1}\text{-}\mr{BD}}

\nc{\konv}[2]{{#1}[{#2}]} 
\nc{\actpres}[1]{\Phi_{#1}} 
\newcommand{\floor}[1]{\lfloor #1 \rfloor}

\nc{\Prim}{\mc{PR}} 

\nc{\sselr}{\sse^{\mapsto}}
\nc{\sserl}{\sse^{\mapsfrom}}
\nc{\spelr}{\spe^{\mapsto}}
\nc{\sperl}{\spe^{\mapsfrom}}
\nc{\ball}[1]{\mr{B}^{#1}} 
\nc{\oball}[1]{\breve{\mr{B}}^{#1}} 
\nc{\pball}[1]{\dot{\mr{B}}^{#1}} 
\nc{\prr}[1]{\dot{\RR}^{#1}} 
\nc{\sph}[1]{\mr{S}^{#1}} 
\nc{\ssim}[1]{s\sigma_{#1}} 
\nc{\koerper}[1]{\norm{#1}}
\nc{\Ccovdim}{\mc{CD}}
\nc{\Cinddim}{\mc{SID}}

\nc{\CInddim}{\mc{LID}}

\DeclareMathOperator{\diffop}{D} 
\DeclareMathOperator*{\diffoplimit}{D} 
\nc{\diffopc}[1]{\sideset{_{#1}}{}\diffoplimit} 
\nc{\diffopp}[1]{\diffop_{#1}} 
\nc{\diffopcp}[2]{\sideset{_{#2}}{_{#1}}\diffoplimit} 
\nc{\meanH}[2]{\mf{M}_{#1,#2}} 
\nc{\emean}[2]{\mf{M}_{\exp_{#1},#2}} 
\DeclareMathOperator{\mor}{Mor}
\DeclareMathOperator{\Hom}{Hom} 
\nc{\autoerw}[1]{\mr{Aut}^{#1}} 
\nc{\komma}[2]{(#1 \downarrow #2)} 
\nc{\Kmat}{\mf{MAT}} 
\nc{\Khmat}{\mf{HMAT}} 
\nc{\homfun}[1]{\mor_{#1}(-_1,-_2)} 
\nc{\homfunae}[1]{\mor_{#1}(-_1)} 
\nc{\homfunbe}[1]{\mor_{#1}(-_2)} 
\nc{\homfunxy}[3]{\mor_{#1}(#2(-_1), #3(-_2))}
\nc{\homfunx}[2]{\mor_{#1}(#2(-_1), -_2)}
\nc{\homfuny}[2]{\mor_{#1}(-_1, #2(-_2))}
\nc{\homfuna}[2]{\mor_{#1}(#2, -)} 
\nc{\homfunb}[2]{\mor_{#1}(-, #2)} 
\nc{\hhomfuna}[2]{\Hom_{#1}(#2, -)} 
\nc{\hhomfunb}[2]{\Hom_{#1}(-, #2)} 
\newcommand{\Va}{\mc{V\hspace{-0.1em}A}}

\newcommand{\Lit}{\mc{LIT}}
\newcommand{\Cl}{\mc{CL}}
\newcommand{\Cls}{\mc{CLS}}

\newcommand{\Pcls}[1]{#1\mbox{--}\Cls}

\newcommand{\Pass}{\mc{P\hspace{-0.32em}ASS}}
\newcommand{\epa}{\pab{}} 
\newcommand{\Tass}{\mc{T\hspace{-0.35em}ASS}}
\newcommand{\Sat}{\mc{SAT}}

\newcommand{\Usat}{\mc{USAT}}

\newcommand{\Musat}{\mc{M\hspace{0.8pt}U}} 
\newcommand{\Musati}[1]{\Musat_{\!#1}} 
\newcommand{\Smusat}{\mc{S}\Musat} 
\newcommand{\Smusati}[1]{\Smusat_{\!#1}}

\nc{\Clsoo}{\Cls^{1,1}} 
\DeclareMathOperator{\lit}{lit}
\DeclareMathOperator{\var}{var}

\newcommand{\Clash}{\mc{HIT}} 

\newcommand{\Uclash}{\mc{U}\Clash} 
\newcommand{\Uclashi}[1]{\Uclash_{\!\!#1}}

\newcommand{\Ho}{\mc{HO}} 

\DeclareMathOperator{\res}{\diamond} 
\DeclareMathOperator{\dpl}{DP} 
\newcommand{\dpi}[1]{\dpl_{\!#1}}
\DMO{\premr}{ax} 
\DMO{\concr}{C} 
\DMO{\allcr}{cl} 

\DeclareMathOperator{\hardness}{hd}
\DMO{\thardness}{thd} 
\DMO{\phardness}{phd} 
\DeclareMathOperator{\wid}{wid} 
\DMO{\whardness}{awid} 
\DMO{\dep}{dep} 
\DMO{\hts}{hs} 
\DMO{\semspace}{css} 
\DMO{\resspace}{crs} 
\DMO{\treespace}{cts} 
\newcommand{\php}{\mathrm{PHP}}
\newcommand{\ephp}{\mathrm{EPHP}} 
\newcommand{\pab}[1]{\langle #1 \rangle}
\newcommand{\pao}[2]{\langle #1 \ra #2 \rangle}

\nc{\bth}[1]{\langle{#1}\rangle} 
\DMO{\rsub}{r_S} 
\DMO{\rk}{r} 
\DMO{\ro}{\rk_1} 
\DMO{\rki}{\rk_{\infty}} 
\DMO{\rpl}{r^{pl}} 
\DMO{\ropl}{\rk_1^{pl}} 
\nc{\rslur}{\xrightarrow{\text{SLUR}}} 
\nc{\rslurs}{\rslur_{\!*}} 
\DMO{\slur}{slur} 
\nc{\Slur}{\mc{SLUR}} 
\nc{\rkslur}[1]{\xrightarrow{\text{SLUR}_{#1}}} 
\nc{\rkslurs}[1]{\rkslur{#1}_{\!*}} 
\nc{\Altsluri}[1]{\Slur(#1)}
\nc{\Altslurstari}[1]{\Slur\text{\textasteriskcentered}(#1)}
\nc{\Canoni}[1]{\mr{CANON}(#1)}
\nc{\rkslurstar}[1]{\xrightarrow{\text{SLUR\textasteriskcentered}#1}} 
\nc{\rkslursstar}[1]{\rkslurstar{#1}_{\!*}} 
\DMO{\slurstar}{\slur\!\text{\textasteriskcentered}}
\nc{\Urefc}{\mc{UC}}
\nc{\Propc}{\mc{PC}}
\nc{\Wrefc}{\mc{WC}} 
\DeclareMathOperator{\vdeg}{vd} 
\DeclareMathOperator{\minvdeg}{\mu\!\vdeg} 
\DMO{\varmvd}{\var_{\minvdeg}} 
\DMO{\nfc}{fc} 
\DMO{\maxnfc}{\nu\!\nfc} 
\DMO{\inonmer}{i_{nM}} 
\nc{\svbf}{\mc{VB}} 
\nc{\svbfs}{\mc{VB}^*} 
\DMO{\potp}{pp} 
\DMO{\potprec}{NM} 
\DMO{\minnonmer}{\mu{}nM} 
\DMO{\varsing}{\var_s} 
\DMO{\varosing}{\var_{1s}} 
\DMO{\varnosing}{\var_{\neg1s}} 
\nc{\Musatns}{\Musat'} 
\nc{\Musatnsi}[1]{\Musati{#1}'}
\nc{\Smusatns}{\Smusat'} 
\nc{\Smusatnsi}[1]{\Smusati{#1}'}
\nc{\Uclashns}{\Uclash'} 
\nc{\Uclashnsi}[1]{\Uclashi{#1}'}
\nc{\tsdp}{\xrightarrow{\text{sDP}}}
\nc{\tsdps}{\tsdp_{\!*}}
\nc{\tosdp}{\xrightarrow{\text{1sDP}}}
\nc{\tosdps}{\tosdp_{\!*}}
\DMO{\sdp}{sDP} 
\DMO{\osdp}{sDP_1} 
\nc{\cflmusat}{\mc{CF}\Musat} 
\nc{\cflmusati}[1]{\mc{CF}\Musati{#1}}
\nc{\cflimusat}{\mc{CFI}\Musat} 
\DMO{\sNF}{sNF} 
\DMO{\eqp}{eqp} 
\DMO{\sgp}{sp} 
\DMO{\singind}{si} 
\DMO{\osingind}{si_1} 
\DMO{\shyp}{svh} 
\DMO{\sdph}{ssh} 
\DMO{\msdph}{mss} 
\DMO{\osdph}{ssh_1} 
\DMO{\mosdph}{mss_1} 
\DMO{\mps}{mps} 
\DMO{\purec}{puc} 
\DMO{\doping}{D}
\DeclareMathOperator{\primec}{prc} 
\nc{\glue}[4]{\mr{glue}((#1,#2), (#3,#4))} 
\DMO{\fvdglue}{\boxplus} 
\nc{\gluea}[3]{#1 \boxplus_{#3} #2} 
\DMO{\frl}{fl} 
\nc{\Con}{\mr{Con}}
\nc{\Log}{\mr{Log}}
\nc{\Lin}{\mr{Lin}}
\nc{\Pol}{\mr{Pol}}
\nc{\ExL}{\mr{ExL}}
\nc{\ExP}{\mr{ExP}}
\nc{\CTime}{\mr{CTime}}
\nc{\CSpace}{\mr{CSpace}}
\nc{\LTime}{\mr{LTime}}
\nc{\LSpace}{\mr{L}}
\nc{\NLSpace}{\mr{NL}}
\nc{\LinTime}{\mr{LinTime}}
\nc{\LinSpace}{\mr{LinSpace}}
\nc{\PTime}{\mr{P}}
\nc{\PSpace}{\mr{PSpace}}
\nc{\Np}{\mr{NP}}
\nc{\Conp}{\text{coNP}}
\nc{\NPSpace}{\mr{NPSpace}}
\nc{\CoNPSpace}{\mr{coNPSpace}}
\nc{\ELTime}{\mr{ELTime}}
\nc{\ELSpace}{\mr{ELSpace}}
\nc{\EPTime}{\mr{EPTime}}
\nc{\EPSpace}{\mr{EPSpace}}
\nc{\NEPTime}{\mr{NEPTime}}
\nc{\polydelta}[1]{\Delta_{#1}^{\mr P}}
\nc{\polypi}[1]{\Pi_{#1}^{\mr P}}
\nc{\polysigma}[1]{\Sigma_{#1}^{\mr P}}
\nc{\Ph}{\mr{PH}}

\nc{\Dp}{D^P}
\nc{\PllC}[2]{{\text{$\mr{PT}$/$\mr{WK}$}(#1, #2)}} 
\nc{\Nc}{\mr{NC}}
\nc{\Nci}[1]{\Nc^{#1}}
\nc{\Ac}{\mr{AC}}
\nc{\Aci}[1]{\Ac^{#1}}
\nc{\pmodpoly}{P / \mathrm{poly}}
\nc{\Wh}[1]{\mr{W}[#1]} 
\nc{\Rl}{\mr{RL}}
\nc{\coRl}{\mr{coRL}}
\nc{\Rp}{\mr{RP}}
\nc{\coRp}{\mr{coRP}}
\nc{\Zpp}{\mr{ZPP}}
\nc{\Bpp}{\mr{BPP}}
\nc{\Pp}{\mr{PP}}
\nc{\Reach}{\mr{STCON}} 
\nc{\Undreach}{\mr{USTCON}} 
\nc{\Pcol}[2]{\mr{COL}(#1,#2)} 
\nc{\Pscol}[2]{\mr{SCOL}(#1,#2)} 
\nc{\Psorcol}[2]{\mr{SORCOL}(#1,#2)} 
\DMO{\slp}{slp}
\nc{\Mss}{\mr{MSS}}
\nc{\Key}{\mr{KEY}}
\nc{\Keyi}[1]{\Key_{\!#1}}
\nc{\Nbmss}{N_{\mr{bm}}} 
\nc{\Nbkey}{N_{\mr{bk}}} 
\nc{\Rnb}{N_{\mr{b}}}
\nc{\Rnk}{N_{\mr{k}}}
\nc{\Rnr}{N_{\mr{r}}}

\nc{\Byte}{\mr{B}[8]}
\nc{\QByte}{\mr{B}[4,8]}
\nc{\KByte}{\mc{B}} 
\nc{\RQByte}{\mc{QB}} 

\nc{\ramz}[3]{\mr{ram}_{#1}^{#2}(#3)} 
\nc{\waez}[2]{\mr{vdw}_{#1}(#2)} 
\nc{\gtz}[2]{\mr{grt}_{#1}(#2)} 
\nc{\pdwaez}[2]{\mr{vdw}_{#1}^{\mr{pd}}(#2)} 
\nc{\absfeh}[1]{\delta_{#1}} 
\nc{\relfeh}[1]{\ve_{#1}} 
\usepackage{theorem} 
\usepackage[driverfallback=hypertex]{hyperref}
\newtheorem{defi}{Definition}[section]
\newtheorem{lem}[defi]{Lemma}
\newtheorem{thm}[defi]{Theorem}
\newtheorem{corol}[defi]{Corollary}

\newtheorem{conj}[defi]{Conjecture}

\newtheorem{examp}[defi]{Example}

\theorembodyfont{\rmfamily}

\theorembodyfont{}
\newenvironment{prf}{\noindent\textbf{Proof:}\;}{\par\noindent\ignorespacesafterend}

\newcommand{\Qed}{\hfill $\square$}
\newcounter{dDef} 

\newcounter{dLem} 

\newcounter{dThm} 

\newcounter{dPro} 

\newcounter{Beispielzaehler}

\nc{\bm}{\boldmath}
\nc{\bmm}[1]{\mbox{\bm$\DST #1$}}
\nc{\mi}[1]{\bmm{\mathrm{(#1):}} \quad}

\usepackage[active]{srcltx}
\usepackage{a4}
\usepackage[all,poly]{xy}
\usepackage{float}
\usepackage{bussproofs}

\newcommand{\Schrift}{report}
\newcommand{\Zusatz}{}
\newcommand{\Liste}{Satisfiability (SAT) \sep XOR \sep parity constraints \sep equivalence reasoning \sep arc consistency \sep hyperarc consistency (GAC) \sep SAT encoding \sep SAT representation \sep unit-propagation completeness \sep unit-refutation completeness \sep forcing representation \sep acyclic incidence graph \sep hardness \sep p-hardness \sep asymmetric width \sep resolution width \sep Tseitin translation \sep fixed-parameter tractability (fpt) \sep monotone circuits \sep monotone span programs \sep lower bounds}

\DMO{\EUrefc}{\exists\,\Urefc}

\DMO{\twidth}{tw}
\DMO{\twidthin}{tw^*}
\DMO{\twidthpr}{tw}

\DMO{\rows}{rs}
\DMO{\cols}{cs}

\DMO{\income}{in}
\DMO{\tstr}{tt}
\DMO{\tstrr}{tt_r}

\begin{document}

\title{A framework for good SAT translations, with applications to CNF representations of XOR constraints}

\author[sw]{Matthew Gwynne}
\ead[url]{http://cs.swan.ac.uk/~csmg/}

\author[sw]{Oliver Kullmann\corref{cor1}}
\ead[url]{http://cs.swan.ac.uk/~csoliver}

\cortext[cor1]{Corresponding author}

\address[sw]{Computer Science Department, Swansea University, UK}

\begin{abstract}
  We present a general framework for ``good CNF-representations'' of boolean constraints, to be used for translating decision problems into SAT problems (i.e., deciding satisfiability for conjunctive normal forms). We apply it to the representation of systems of XOR-constraints (``exclusive-or''), also known as systems of linear equations over the two-element field, or systems of parity constraints, or as systems of equivalences (XOR is the negation of an equivalence).

  The general framework defines the notion of ``representation'', and provides several methods to measure the quality of the representation, by measuring the complexity (``hardness'') needed for making implicit ``knowledge'' of the representation explicit (to a SAT-solving mechanism). We obtain general upper and lower bounds.

  Applied to systems of XOR-constraints, we show a super-polynomial lower bound on ``good'' representations under very general circumstances. A corresponding upper bound shows fixed-parameter tractability in the number of constraints.

  The measurement underlying this upper bound ignores the auxiliary variables needed for shorter representations of XOR-constraints. Improved upper bounds for special cases take them into account, and a rich picture begins to emerge, under the various hardness measurements.
\end{abstract}


\maketitle

\tableofcontents

\section{Introduction}
\label{sec:intro}

SAT solving has developed within the last 15 years a strong applied side, as witnessed by the Handbook of Satisfiability \cite{2008HandbuchSAT}. An important aspect here is the ``encoding'' of the original problem into a SAT problem, as has been emphasised perhaps first in \cite{BailleuzBoufkhad2003CardinalityConstraints}. ``Encoding'' is used in general in a wide sense, using any kind of useful relation between the original problem and its SAT encoding. Recent examples are \cite{TamuraTagaKitagawaBanbara2009OrderEncoding,2011CompactOrderEncoding} for work on general constraint translations. However often the encoding, or ``translation'', involves the usage of various constraints, like for example cardinality constraints (constraints on the number of 0's and 1's for some set of boolean variables) and/or XOR-constraints. When using general translations of such constraints, then the ``encoding'' must neither loose nor add satisfying assignments for the constraints, and we then speak of a \emph{(CNF-)representation}. Examples are given by \cite{BailleuzBoufkhad2003CardinalityConstraints,Sinz2005CardinalityConstraints,Een2006Translating} for work on cardinality constraints, and \cite{JovanovicKreuzer2010AlgAttackSAT,GwynneKullmann2011TranslationsPrelim} for investigations into different translations in cryptography.

In this \Schrift{} we concentrate on boolean constraints (so we do not consider the problem of encoding non-boolean variables into boolean variables), i.e., we represent \emph{boolean functions}. We present a general framework for ``good'' representations $F$ of boolean functions $f(v_1,\dots,v_n)$, considering upper and lower bounds, and apply this general theory to the special case of representing XOR constraints.

The basic quality criterion for a representation $F$ is ``generalised arc consistency'' (GAC; \cite{Gent2002ArcConsistency,BailleuzBoufkhad2003CardinalityConstraints,Een2006Translating}), that is, for every partial assignment $\vp$ to the variables $v_1,\dots,v_n$, all assignments $v_i = \ve$ forced by $\vp$ are determined by unit-clause propagation on the result $\vp * F$ of the application. We call a representation $F$ fulfilling this condition a ``GAC-representation''. In other words, a CNF-representation $F$ of $f$ is GAC iff every valid implication $x_1 \und \dots \und x_p \ra y$ for $f$, where $x_i, y$ are literals over $v_1,\dots,v_n$, is detected by unit-clause propagation on $F$. Indeed many of our results are formulated for more general conditions, but for now we just concentrate on the most prominent condition, GAC.

After setting up the general theory, we study the problem of finding good CNF-representations $F$ of systems of linear equations $S$ over the two-element field, also known as systems of XOR-constraints $x_1 \oplus \dots \oplus x_k = \ve$, $\ve \in \set{0,1}$, or systems of parity-constraints, or systems of equivalences (using $a \lra b = \neg (a \oplus b)$ instead). The number of equations in $S$ is $m$, the number of variables is $n$. These representations are used as parts of SAT problems $F^* \supset F$, such that $F$ has ``good'' properties for SAT solving in the context of $F^*$; here $F^*$ may for example represent the problem of finding the key for a cryptographic cipher. The task of ``good'' representations of $S$ by conjunctive normal forms $F$ (clause-sets, to be precise), for the purpose of SAT solving, shows up in many applications, for example cryptanalysing the Data Encryption Standard and the MD5 hashing algorithm in \cite{BardCourtois2007AlgebraicDES}, translating Pseudo-Boolean constraints to SAT in \cite{Een2006Translating}, and in roughly $1$ in $6$ benchmarks from SAT 2005 to 2011 according to \cite{LaitinenJunttilaNiemelae2012Parity}.

In more detail, recall that the two-element field $\ZZ_2$ (also written as $\mathrm{GF}(2)$) has elements $0,1$, where addition is XOR, which we write as $\oplus$, while multiplication is AND, written $\cdot$. A linear system $S$ of equations over $\ZZ_2$, in matrix form $A \cdot x = b$, where $A$ is an $m \times n$ matrix over $\set{0,1}$, with $b \in \set{0,1}^m$, yields a boolean function $f_S$, which assigns $1$ to a total assignments of the $n$ variables of $S$ iff that assignment is a solution of $S$. This boolean function is to be represented.

  We show there is no GAC-representation of polynomial size for arbitrary $S$. We use the lower bound on monotone circuits for monotone span programs from Babai et al 1999 \cite{BabaiGalWigderson1999MonoteSpanPrograms}, and we show a close relation between monotone circuits and GAC-representations, based on the work in Bessiere et al 2009 \cite{BKNW2009CircuitComplexity}. We then turn to constructing good representations. We analyse the basic translation $F = X_1(S)$, which translates each constraint on its own, by splitting up $x_1 \oplus \dots \oplus x_k = \ve$ into sums $x_1 \oplus x_2 = y_2, y_2 \oplus x_3 = y_3, \dots, y_{k-1} \oplus x_k = \ve$, introducing auxiliary variables $y_i$. We show that $X_1(S^*)$, where $S^*$ is obtained from $S$ by considering all derived equations, is a GAC-representation of $S$. The derived equations are obtained by adding up the equations of all sub-systems $S' \sse S$. There are $2^m$ such $S'$, and computing a GAC-representation is fixed-parameter tractable (fpt) in the parameter $m$, improving Laitinen et al 2013 \cite{LaitinenJunttilaNiemelae2013Parity}, which showed fpt in $n$.

  To obtain stronger representations, instead of mere GAC we consider the class $\Propc$ of propagation-complete clause-sets, as introduced in Bordeaux et al 2012 \cite{BordeauxMarquesSilva2012KnowledgeCompilation}. The stronger criterion is $F \in \Propc$, which requires for \emph{all} partial assignments, possibly involving also the auxiliary (new) variables in $F$, that forced assignments can be determined by unit-clause propagation. Equivalently, every valid implication $x_1 \und \dots \und x_p \ra y$ for $F$ (not just for $f$), where $x_i, y$ are (arbitrary) literals, is detected by unit-clause propagation on $F$.

  Using ``propagation hardness'' $\phardness(F) \in \NNZ$ as mentioned in Gwynne et al 2012 \cite{GwynneKullmann2012SlurJ}, we have $F \in \Propc \Lra \phardness(F) \le 1$. By \cite[Proposition 5]{BordeauxMarquesSilva2012KnowledgeCompilation} we know that $X_1$ applied to a single equation ($m=1$) yields a translation in $\Propc$, i.e., $\phardness(X_1(S)) \le 1$, and we strengthen and generalise this result. Then we study $m=2$. Now $S^*$ (as above) has two equations more, and $X_1(S^*)$ is a GAC-representation, but the ``distance'' to $\Propc$ is arbitrarily high, i.e., $\phardness(X_1(S^*))$ is unbounded (using results from Beyersdorff et al 2014 \cite{BeyersdorffGwynneKullmann2013PHP}). We show two possibilities to remedy this (for $m=2$). On the one hand, if instead of unit-clause propagation we allow (arbitrary) resolution with clauses of length at most $3$ (i.e., $3$-resolution), and only require refutation of inconsistencies after (arbitrary, partial) instantiations, then even just $X_1(S)$ suffices. On the other hand, with a more intelligent translation, which avoids duplication of equivalent auxiliary variables $y_i$, we obtain a (short) representation in $\Propc$. We conjecture that also the general case can be handled this way, that is, computing a representation $F \in \Propc$ of $S$ is fpt in $m$.

\subsection{The idea for the lower bound}
\label{sec:introlowb}

An overview on the proof, that there is no polynomial-size GAC-representation of arbitrary linear systems $S$ (Theorem \ref{thm:xorclsrel}), is as follows. We apply the lower bound on monotone circuit sizes for monotone span programs (MSPs) from \cite{BabaiGalWigderson1999MonoteSpanPrograms}, by translating MSPs into linear systems. An MSP computes a boolean function $f(x_1,\dots,x_n) \in \set{0,1}$ (with $x_i \in \set{0,1}$), by using auxiliary boolean variables $y_1,\dots,y_m$, and for each $i \in \tb 1n$ a linear system $A_i \cdot y = b_i$, where $A_i$ is an $m_i \times m$ matrix over $\ZZ_2$. For the computation of $f(x_1,\dots,x_n)$, a value $x_i = 0$ means the system $A_i \cdot y = b_i$ is active, while otherwise it's inactive; the value of $f$ is $0$ if all the active systems together are unsatisfiable, and $1$ otherwise. Obviously $f$ is monotonically increasing. The task is now to put that machinery into a single system $S$ of (XOR) equations. The main idea is to ``relax'' each equation of every $A_i \cdot y = b_i$ by a dedicated new boolean variable added to the equation, making that equation trivially satisfiable, independently of everything else; all these auxiliary variables together are called $z_1, \dots, z_N$, where $N = \sum_{i=i}^n m_i$ is the number of equations in $S$.

If all the relaxation variables used for a system $A_i \cdot y = b_i$ are set to $0$, then they disappear and the system is active, while if they are not set, then this system is trivially satisfiable, and thus is deactivated. Now consider a GAC-representation $F$ of $S$. Note that the $x_i$ are not part of $F$, but the (primary) variables of $F$ are $y_1,\dots,y_m$ together with $z_1,\dots,z_N$, where the latter represent in a sense the $x_1,\dots,x_n$, plus possibly auxiliary variables. From $F$ we can compute $f$ by setting the $z_j$ accordingly (if $x_i = 0$, then all $z_j$ belonging to $A_i \cdot y = b_i$ are set to $0$, if $x_i = 1$, then these variables stay unassigned), running unit-clause propagation $\ro$ on the system, and output $0$ iff the empty clause was produced by $\ro$. So we can evaluate MSPs by applying partial instantiation to GAC-representations $F$ of the above linear system and running $\ro$. The second pillar of the lower-bound proof is a general polynomial-time translation of GAC-representations of (arbitrary) boolean functions into monotone circuits computing a monotonisation of the boolean function (Theorem \ref{thm:acmono}; strengthening \cite{BKNW2009CircuitComplexity} for the boolean case), where this monotonisation precisely enables partial instantiation, and thus for our case enables computation of $f$. So from $F$ we obtain a monotone circuit $\mc{C}$ computing $f$, whose size is polynomial in $\ell(F)$, where by \cite{BabaiGalWigderson1999MonoteSpanPrograms} the size of $\mc{C}$ is $N^{\Omega(\log N)}$ for certain MSPs.

As shown in \cite{Kullmann2014Collapse}, this superpolynomial lower bound also holds, if we consider any fixed $k \in \NNZ$, and instead of requiring unit-clause propagation to detect all forced assignments, we only ask that ``asymmetric width-bounded resolution'', i.e., $k$-resolution, is sufficient to derive all contradictions obtained by (partial) instantiation (to the variables in $S$; see Corollary \ref{cor:xorcls}). Here $k$-resolution is the appropriate generalisation of width-bounded resolution for handling long clauses (see \cite{Kl93,Ku99b,Ku00g,BeyersdorffKullmann2014PHP}), where for each resolution step at least one parent clause has length at most $k$ (while the standard ``symmetric width'' requires both parent clauses plus the resolvent to have length at most $k$).

We now turn to a discussion of the advantages of having a representation in $\Propc$ compared to mere GAC, also placing this in a wider framework.

\subsection{Measuring ``good'' representations}
\label{sec:intromeasuring}

We have seen yet two criteria for good representations of boolean functions, namely GAC and the stronger condition of unit-propagation completeness (captured by the class $\Propc$). In Subsection \ref{sec:arcvsprop} we discuss some fundamental aspects of these two criteria, while in Subsection \ref{sec:URC} we consider another criterion, namely unit-\emph{refutation} completeness. These conditions are embedded in Subsection \ref{sec:gauge} into a general framework, where satisfying assignments are additionally taken into in Subsection \ref{sec:introsata}. We conclude this overview on the general approach towards ``good representations'' with a reflection on ``partial versus total'' (assignments) in Subsection \ref{sec:prelimrefl}.

\subsubsection{GAC versus $\Propc$}
\label{sec:arcvsprop}

It has been shown that the practical performance of SAT solvers can depend heavily on the SAT representation used. In order to obtain ``good'' representations, the basic concept is that of a GAC-representation, as we have already explained. The task is to ensure that for all (partial) assignments to the variables of the constraint, if there is a forced assignment for a variable of the constraint (i.e., a variable which must be set to a particular value to avoid inconsistency), then unit-clause propagation ($\ro$) is sufficient to find and set this assignment. In a similar vein, there is the class $\Propc$ of propagation-complete clause-sets, containing all clause-sets for which unit-clause propagation is sufficient to detect all forced assignments; the class $\Propc$ was introduced in \cite{BordeauxMarquesSilva2012KnowledgeCompilation}, while in \cite{BBCGKV2013Propc} it is shown that membership decision is coNP-complete.

When translating a constraint into SAT, typically one does not just use the variables of the constraint, but one adds auxiliary variables to allow for a compact representation. Now when speaking of GAC, one only cares about assignments to the (primary) \emph{constraint variables}. But propagation-completeness deals only with the representing clause-set, thus can not know about the distinction between primary and auxiliary variables, and thus it is a property on the (partial) assignments over \emph{all} variables. So a SAT representation, which is GAC, will in general not fulfil the stronger property of propagation-completeness, due to assignments over both constraint \emph{and} auxiliary variables possibly yielding a forced assignment (now also over auxiliary variables) or even an inconsistency which $\ro$ doesn't detect.\footnote{Note the possibility of a partial assignment (only) to the constraint variables forcing an assignment on an auxiliary variable --- this is not considered by GAC. In the conference version \cite{GwynneKullmann2013GoodRepresentationsIILata} by mistake we included this requirement in the definition of GAC; see Example \ref{exp:relphd} for a discussion.}

In \cite{JarvisaloJunttila2009LimitRestrictedLearning} it is shown that conflict-driven SAT solvers with branching restricted to input variables, more precisely, ignoring the auxiliary variables in Tseitin translations of circuits, can not even polynomially simulate tree-resolution, i.e., an enormous reduction in proof-theoretic strengths can occur.\footnote{More precisely it is shown, that a natural proof system corresponding to clause-learning SAT solvers (with arbitrary restarts), which branches only on input variables, has only exponential refutations of $\ephp_n'$, an (extreme) Extended Resolution extension to the pigeon-hole formulas, where in fact $\hardness(\ephp_n') = 2$ holds, as shown in \cite[Subsection 8.4]{BeyersdorffGwynneKullmann2013PHPER}.} Also experimentally it is demonstrated in \cite{JarvisaloJunttila2009LimitRestrictedLearning}, and more extensively in \cite{JarvisaloNiemala2008StructuralBranchingExperiments}, that input-restricted branching can have a strong detrimental effect on solver times and proof sizes for CDCL solvers. This adds motivation to considering \emph{all} variables (rather than just input variables), when deciding what properties we want for SAT translations. We call this the \emph{absolute (representation) condition}, taking also the auxiliary variables into account, while the \emph{relative condition} only considers the original variables.

Besides avoiding the creation of hard unsatisfiable sub-problems, the absolute condition also enables one to study the ``target classes'', like $\Propc$, on their own, without relation to what is represented. Target classes different from $\Propc$ have been proposed, and are reviewed in the following. The underlying idea of GAC- and pro\-pa\-ga\-tion-com\-ple\-te translations is to compress all of the constraint knowledge into the SAT translation, and then to use $\ro$ to extract this knowledge when appropriate. In Subsection \ref{sec:URC} we present a weaker notion of what ``constraint knowledge'' could mean, while in Subsection \ref{sec:gauge} we present different extraction mechanisms.

\subsubsection{UR versus $\Urefc$}
\label{sec:URC}

In \cite{GwynneKullmann2012SlurSOFSEM,GwynneKullmann2012Slur,GwynneKullmann2012SlurJ} we considered the somewhat more fundamental class $\Urefc \supset \Propc$ of ``unit-refutation complete'' clause-sets, introduced in \cite{Val1994UnitResolutionComplete} as a method for propositional knowledge compilation. Rather than requiring that $\ro$ detects all forced assignments (as for $\Propc$), a clause-set is in $\Urefc$ iff for all partial assignments resulting in an unsatisfiable clause-set, $\ro$ detects this. As shown in \cite{GwynneKullmann2012SlurSOFSEM,GwynneKullmann2012Slur,GwynneKullmann2012SlurJ}, the equation $\Urefc = \Slur$ holds, where $\Slur$, introduced in \cite{SAFS95}, is a fundamental class of clause-sets for which SAT is decidable in polynomial time; in \cite{CepekKuceraVlcek2012SLUR} it was shown that membership decision for $\Slur$ is coNP-complete.

GAC-representations are the relative versions of the absolute (class) $\Propc$, while the relative version of the class $\Urefc$ we call a ``UR-representation'', more precisely: a UR-representation of a boolean function $f$ is a CNF-representation $F$ of $f$ such that for all partial assignments $\vp$ to variables of $f$ we have $\vp * F \in \Usat \Lra \ro(\vp * F) = \set{\bot}$. This concept has various (similar) names in the literature: \cite{BordeauxJanotaMarquesSilvaMarquis2012UC} calls it ``existential unit-refutation completeness'', while \cite{Bailleux2012UnitConVsUnitProp} calls it ``unit contradiction''.

All these considerations can be extended to a general ``measurement'' approach, where we do not just have $F$ in or out for some target classes, but where a ``hardness'' measure tells us how far $F$ is from $\Propc$ resp.\ $\Urefc$ (in some sense), and this general approach is discussed next.

\subsubsection{How to gauge representations?}
\label{sec:gauge}

We now outline a more general approach to gauge how good is a representation $F$ of a boolean function $f$. Obviously the size of $F$ must be considered, number of variables $n(F)$, number of clauses $c(F)$, number of literal occurrences $\ell(F)$. Currently we do not see a possibility to be more precise than to say that a compromise is to be sought between stronger inference properties of $F$ and the size of $F$. One criterion to judge the inference power of $F$ is GAC, as already explained. This doesn't yield a possibility in case no GAC-representation is feasible, nor is there a possibility for stronger representations. Our approach addresses these concerns as follows. \cite{GwynneKullmann2012SlurSOFSEM,GwynneKullmann2012SlurJ} introduced the measures
\begin{displaymath}
  \hardness, \phardness, \whardness: \Cls \ra \NNZ,
\end{displaymath}
called ``hardness'', ``p-hardness'', and ``asymmetric width'' respectively (indeed $\phardness$ is only mentioned in \cite{GwynneKullmann2012SlurJ}, and properly introduced in this \Schrift), where $\Cls$ is the set of all clause-sets (interpreted as CNFs).\footnote{In \cite{GwynneKullmann2012SlurSOFSEM,GwynneKullmann2012Slur,GwynneKullmann2012SlurJ} the notation ``$\mathrm{whd}$'' (for ``width hardness'') was used instead of ``$\whardness$'', but with \cite{BeyersdorffKullmann2014PHP} we changed terminology.} We relativise these hardness measures here to $\hardness^V, \phardness^V, \whardness^V: \Cls \ra \NNZ$, where $V$ is some set of variables. These measures determine the maximal ``effort'' (in some sense) needed to show unsatisfiability of instantiations $\vp * F$ of $F$ for partial assignments $\vp$ with $\var(\vp) \sse V$ in case of $\hardness$ and $\whardness$, resp.\ the maximal ``effort'' to determine all forced assignments over $V$ for $\vp * F$ in case of $\phardness$. The ``effort'' in case of $\hardness$ or $\phardness$ is the maximal level of generalised unit-clause propagation needed, that is the maximal $k$ for reductions $\rk_k$ introduced in \cite{Ku99b,Ku00g}, where $\ro$ is unit-clause propagation and $\rk_2$ is (complete) elimination of failed literals (as first considered, in an incomplete version, by \cite{Free95} in a SAT solver). While for $\whardness$ the effort is the maximal $k$ needed for asymmetric width-bounded resolution, i.e., for each resolution step one of the parent clauses must have length at most $k$.\footnote{Symmetric width-bounded resolution requires \emph{both} parent clauses to have length at most $k$, which for arbitrary clause-length is not appropriate as complexity measure, since already unsatisfiable Horn clause-sets need unbounded symmetric width; see \cite{BeyersdorffKullmann2014PHP} for the use of asymmetric width in the context of resolution and/or space lower bounds.}

Now we have that $F$ is a GAC-representation of $f$ iff $\phardness^{\var(f)}(F) \le 1$, while $\phardness^{\var(f)}(F) \le k$ would allow higher levels of generalised unit-clause propagation (allowing potentially shorter $F$). Weaker is the requirement $\hardness^{\var(f)}(F) \le 1$, which is precisely an UR-representation --- now not every forced assignment (concerning the variables of $f$) is necessarily detected by unit-clause propagation, but only unsatisfiability. Similarly, $\hardness^{\var(f)}(F) \le k$ would allow higher levels of generalised unit-clause propagation. In general, we call the restriction to variables from $f$ the ``relative condition/measure/hardness'', while without the restriction we speak of ``absolute condition/measure/hardness''.

If we only consider ``relative (w/p-)hardness'', that is, $V = \var(f)$, then, as shown in \cite{Kullmann2014Collapse}, regarding polysize representations for (fixed) $k \ge 1$ all conditions $\hardness^V(F) \le k$, $\phardness^V(F) \le k$, and $\whardness^V(F) \le k$ are equivalent to GAC ($\phardness^V(F) \le 1$), that is, the representations can be transformed in polynomial time into GAC-representations. The (natural) transformations of \cite{Kullmann2014Collapse} produce large representations (and very likely they are not fixed-parameter tractable in $k$), and so higher $k$ can yield smaller representations, however these savings can not be captured by the notion of polynomial size.

This situation changes, as we show in \cite{GwynneKullmann2013GoodRepresentationsI}, when we do not allow auxiliary variables, that is, we require $\var(F) = \var(f)$: Now higher $k$ for each of these measures allows short representations which otherwise require exponential size. We conjecture, that this strictness of hierarchies also holds in the presence of auxiliary variables, when using the \emph{absolute} condition, i.e., $V = \var(F)$ (all variables are included in the worst-case determinations for (w/p)-hardness).

The measurements in case of $V = \var(F)$ are just written as $\hardness, \phardness, \whardness: \Cls \ra \NNZ$. In this way we capture the classes $\Propc$ and $\Urefc$, namely $\Propc = \set{F \in \Cls : \phardness(F) \le 1}$ and $\Urefc = \set{F \in \Cls : \hardness(F) \le 1}$. More generally we have $\Urefc_k = \set{F \in \Cls : \hardness(F) \le k}$, $\Propc_k = \set{F \in \Cls : \phardness(F) \le k}$ and $\Wrefc_k = \set{F \in \Cls : \whardness(F) \le k}$. The basic relations between these classes are $\Wrefc_k = \Urefc_k$ for $k \le 1$, $\Urefc_k \sse \Wrefc_k$ for $k \ge 2$ (see \cite{BeyersdorffKullmann2014PHP}), and $\Propc_k \subset \Urefc_k \subset \Propc_{k+1}$ for $k \ge 0$ (see Corollary \ref{cor:sandwichUP}).

\subsubsection{What happens for satisfying assignments?}
\label{sec:introsata}

Consider a representation $F$ of a boolean function $f$. It seems that until now it has not been considered in general what happens if a partial assignment $\vp$ satisfies $f$; for example even for a GAC-representation and a total assignment $\vp$ for $f$ it could happen that $\vp * F$ is a hard satisfiable problem. The basic criterion here we call ``UP-representation'', which means that for every total assignment $\vp$ for $f$ via unit-clause propagation we obtain the result from $\vp * F$, i.e., $\ro(\vp * F) \in \set{\top, \set{\bot}}$. It is easy to see that UP-representations of boolean functions are up to linear-time transformations the same as representations by boolean circuits.\footnote{In Subsection 9.4.1 of \cite{GwynneKullmann2013GoodRepresentationsI} this class of representations is called $\ex\mc{UP}$.}

To complete GAC-representations on the satisfying side, we require this criterion additionally to the GAC-criterion. Such representations we call \textbf{forcing representations}; and if we also have the absolute condition, then we speak of \textbf{absolute forcing representations}.\footnote{The terminology ``forcing'' was suggested by Donald Knuth.} All our ``good representations'' fulfil these forcing conditions, and in general we show that every UR-representation can be transformed in polynomial time into a forcing representation (Theorem \ref{thm:stronglyforcing}).

This is now the place for some comments on \emph{backdoors} as introduced in \cite{WilliamsGomesSelman2003Backdoors}; for a recent overview see \cite{GaspersSzeider2012Backdoors}. Recall that a weak $\mc{C}$-backdoor for $F \in \Cls$, using the terminology of \cite{GaspersSzeider2012Backdoors}, where $\mc{C} \sse \Cls$ is polytime- decidable and SAT-decidable, is a set $V \sse \var(F)$ of variables such that a partial assignment $\vp$ with $\var(\vp) = V$ and $\vp * F \in \Sat \cap \mc{C}$ exists. As already discussed in \cite{WilliamsGomesSelman2003Backdoors} and further elaborated in \cite{JarvisaloJunttila2009LimitRestrictedLearning}, if we consider the Tseitin-translation $T \in \Cls$ of circuits representing a boolean function $f \ne 0$, then the input-variables, i.e., $\var(f)$, are a weak backdoor for $T$ for the class $\mc{C}_1 := \set{F \in \Cls : \ro(F) \in \set{\top,\set{\bot}}}$; in \cite{WilliamsGomesSelman2003Backdoors} (end of Section 2 there) this is called a ``backdoor for unit propagation'' (given by the ``independent'' variables). We see that the notion of ``forcing representation'' adds this special backdoor-requirement to GAC-representations, for general representations. Indeed, something stronger is added, namely there is the notion of ``strong $\mc{C}$-backdoors'', where the criterion is that for \emph{all} $\vp$ with $\var(\vp) = V$ holds $\vp * F \in \mc{C}$, and a UP-representation $F$ of $f$ is precisely a representation $F$ of $f$, such that $\var(f)$ is a strong $\mc{C}_1$-backdoor for $F$.

So there are some relations of our approach to backdoors, but there are substantial differences. If we use $\mc{C}$ for the level zero of the general hierarchy developed in \cite{Ku99b,Ku00g}, then the resulting hardness-measure is a lower bound for the size of a weak as well as a strong $\mc{C}$-backdoor, while we note that on satisfiable instances the approach of \cite{Ku99b,Ku00g} is different from the approach of this \Schrift, namely there we just ask to find some satisfying assignment in the satisfiable case. So, on the one hand in this \Schrift{} we ask for much less than in backdoors, i.e., we use stronger algorithmic means (as in \cite{Ku99b,Ku00g}), so that already $\hardness = 0,1,2$ has a strong meaning. And on the other hand we ask for much more (for $f \ne 0$), namely we ask for handling of all \emph{partial} assignments to $V$, even in case of just relative hardness.

\subsubsection{Reflection on partial versus total assignments}
\label{sec:prelimrefl}

We conclude our general discussion of ``good representations'' by contrasting UP-representations, which typically arise from some procedural handling of constraints, with the stronger representations considered in this \Schrift. At the first level we view a linear system $S$ as a constraint on $\var(S)$, and encode evaluation via Tseitin's translation, obtaining a UP-representation $F$; this is basically the standard representation $X_1$. However for \emph{partial} assignments applied to $F$ we know nothing, and as our lower bound shows (Theorem \ref{thm:xorclsrel}), there is indeed no polysize representation which handles all partial assignments.

But at the second level, what about writing an algorithm (a special ``constraint propagator'', using Gaussian elimination), which handles all partial assignments in polynomial time (detects unsatisfiability of $\vp * F$ for all partial assignments $\vp$)? This algorithm yields a UP-representation, which should solve our problem? The point is that this can not be integrated into the CNF formalism (by using auxiliary variables and clauses), since algorithms always need \emph{total} assignments (the input must be fully specified), and so partial assignments $\vp$ would need to be \emph{encoded} --- the information ``variable $v$ not assigned'' (i.e., $v \notin \var(\vp)$) needs to be represented by \emph{setting} some auxiliary variable, and this must happen by a mechanism outside of the CNF formalism.

In \cite{Bacchus2007GAC} the difference between using propagators and using CNF-representations is understood as CNF-representations being declarative, while propagators are procedural. It is an essential strength of the CNF formalism (declarations by CNFs) to allow partial instantiation, that is, partial information on the input is encoded by a partial assignment $\vp$, and the application $\vp * F$ represents the original representation $F$ plus the additional information $\vp$ \emph{again} as a CNF-representation.

If we now want to have a ``strong meaning'' of $\vp * F$, that is, a criterion like GAC, or in other words, if we want these partial instantiations also to be easily understandable by a SAT solver, then the results of \cite{BKNW2009CircuitComplexity} and our results show that there are restrictions. Yet there is little understanding of these restrictions. There are many examples where GAC and stronger representations are possible, while the current non-representability results, one in \cite{BKNW2009CircuitComplexity}, one in this \Schrift{} and a variation on \cite{BKNW2009CircuitComplexity} in \cite{BeyersdorffGwynneKullmann2013PHP}, rely on non-trivial lower bounds on monotone circuit complexity; in fact Theorem \ref{thm:otherdir} shows that there is a polysize GAC-representation of a boolean function $f$ if and only if the monotonisation $\widehat{f}$, which encodes partial assignments to $f$, has polysize monotone circuits.

\subsection{Related work}
\label{sec:prelimrelated}

Since the field of ``translating'' or ``encoding'' problems into SAT has various non-trivial aspects, much of the literature discussion is placed later, where then some background is available. In Subsection \ref{sec:disenc} we reflect on terminology, concerning ``encoding'' versus ``representation'', and in Subsection \ref{sec:introAC} we discuss in more depth the CSP-literature on this topic, especially regarding ``arc-consistency'' (in this context). Of special importance to our work is \cite{BKNW2009CircuitComplexity}, and we discuss the relations to our results (in Section \ref{sec:characurmon}) in Subsection \ref{sec:ComparisonBessiere}. Finally in Section \ref{sec:litrev} we review the literature on the handling of XOR-constraints for SAT solving. Here now we give a short overview on our own work on the subject of ``good representations''.

The basics on hardness of unsatisfiable clause-sets, via various forms of resolution complexity measured in a robust way, have been developed in \cite{Ku99b,Ku00g}. An early application of hardness measurements to improved SAT representations is \cite{GwynneKullmann2011TranslationsPrelim}. Our investigations into the classes $\Urefc_k, \Propc_k, \Wrefc_k$ started with the conference paper \cite{GwynneKullmann2012SlurSOFSEM} and its journal version \cite{GwynneKullmann2012SlurJ}, centred around the theorem $\Urefc_k = \Slur_k$ (while $\Propc_k, \Wrefc_k$ are only discussed in the outlook). A proper treatment of $\Wrefc_k$ one finds in \cite{BeyersdorffKullmann2014PHP}, in a proof-complexity setting. The conference version of the current \Schrift{} is \cite{GwynneKullmann2013GoodRepresentationsIILata} (containing Sections \ref{sec:propc}, \ref{sec:xorclausesets}, \ref{sec:transx0}, \ref{sec:transx1}, \ref{sec:norepmsp}, \ref{sec:transarbxor}, \ref{sec:transtxor} of this \Schrift{} in shortened and preliminary form). Finally we have the following relevant work in progress: In \cite{GwynneKullmann2013GoodRepresentationsI} we show that without auxiliary variables, the hierarchies $\Urefc_k$, $\Propc_k$ and $\Wrefc_k$ are strict regarding polysize representations of boolean functions. On the other hand, with auxiliary variables and the \emph{relative} condition, all three hierarchies $\Urefc_k$, $\Propc_k$ and $\Wrefc_k$ collapse to their first level (regarding polysize representations), as we show in \cite{Kullmann2014Collapse}. The predecessor of \cite{GwynneKullmann2013GoodRepresentationsI} and the current \Schrift{} is \cite{GwynneKullmann2013GoodRepresentations}.

Concerning XOR, Section 1.5 of \cite{GwynneKullmann2012SlurJ} discusses the translation of the so-called ``Schaefer classes'' into the $\Urefc_k$ hierarchy; see Section 12.2 in \cite{DH09HBSAT} for an introduction, and see \cite{CreignouKolaitisVollmer2008ComplexityConstraints} for an in-depth overview on recent developments. All Schaefer classes except affine equations have natural translations into either $\Urefc_1$ or $\Urefc_2$. The open question was whether systems of XOR-clauses (i.e., affine equations) can be translated into $\Urefc_k$ for some fixed $k$; the current \Schrift{} answers this question negatively.

\subsection{Overview on results}
\label{sec:overview}

The general structure of this \Schrift{} is as follows:
\begin{enumerate}
\item Section \ref{sec:prelim} is about general preliminaries, Section \ref{sec:measurerepcomp} about preliminaries regarding hardness.
\item Section \ref{sec:propc} introduces propagation-hardness, and proves basic results.
\item Sections \ref{sec:cnfrep}, \ref{sec:characurmon} discuss the general theory of ``SAT representations''.
\item Sections \ref{sec:xorclausesets}, \ref{sec:transx0}, \ref{sec:transx1}, \ref{sec:litrev} present the basics for representations of XOR-constraints.
\item Sections \ref{sec:norepmsp}, \ref{sec:transarbxor} show general lower and upper bounds for ``good'' representations of XOR-constraints.
\item As a starting point for more detailed investigations on XOR-constraints, in Section \ref{sec:transtxor} systems of two XOR-constraints are represented in various ways.
\item Finally Section \ref{sec:open} is the conclusion.
\end{enumerate}

The content in more details is as follows.
After having established in Section \ref{sec:prelim} the basic definitions related to clause-sets, partial assignments, forced assignments, prime implicates and boolean functions, we review in Section \ref{sec:measurerepcomp} the basic concepts and notions related to hardness, asymmetric width, and the classes $\Urefc_k$ and $\Wrefc_k$.

Section \ref{sec:propc} introduces p-hardness and the classes $\Propc_k$, and gives basic criteria for $\bc_{i \in I} F_i \in \Propc_k$ assuming clause-sets $F_i \in \Propc_k$: the main result Theorem \ref{thm:acylcprop} shows that the ``incidence graph'' being acyclic is sufficient.

Section \ref{sec:cnfrep} thoroughly discusses concepts of ``representations'' of boolean functions in the SAT context, and how to measure their strength (Subsection \ref{sec:cnfreprep} -- \ref{sec:measqurep}). In Subsection \ref{sec:introAC} we discuss the literature on translating constraint problems (CSPs) into SAT such that ``local consistencies'' are preserved/detected. We complete our catalogue of representation conditions by the notions of ``forcing representations'' in Subsection \ref{sec:cnfrepforcing}, and finally apply these notions in Subsection \ref{sec:remTseit} to the well-known Tseitin translation, whose representation strength is given in Theorem \ref{thm:extTs}.

Section \ref{sec:characurmon} establishes a close relation between UR-/GAC-representations and monotone boolean circuits. Strengthening \cite{BKNW2009CircuitComplexity} (for the boolean case), in Theorem \ref{thm:acmono} we show that from a UR-representation of a boolean function $f$ we obtain in polynomial time a monotone circuit computing the monotonisation $\widehat{f}$, which extends $f$ by allowing partial assignments to the inputs. The inverse direction is established in Lemma \ref{lem:otherdir}, and thus, in Theorem \ref{thm:otherdir} we get an equivalent characterisation of UR-representations of sequences of boolean functions in terms of monotone circuits. As an application we obtain in Theorem \ref{thm:stronglyforcing}, that an UR-representation of a boolean function can be transformed in polynomial time into a forcing representation. The relations of these results to \cite{BKNW2009CircuitComplexity} finally are discussed in Subsection \ref{sec:ComparisonBessiere}.

After these general preparations on ``good SAT representations'', we turn to the representation of sets of XOR-constraints. In Section \ref{sec:xorclausesets} we introduce ``XOR-clause-sets'' and their CNF-representations, and state in Lemma \ref{lem:characimplxor}, that the sum of XOR-clauses is the (easier) counterpart to the resolution operation for (ordinary) clauses. The fundamental translation $X_0$ of XOR-clause-sets (using the unique equivalent CNF for every XOR-clause) is studied in Section \ref{sec:transx0}, with Lemma \ref{lem:suffx0pc} showing that if the XOR-clause-set $F$ is acyclic, then $X_0(F)$ is an absolute forcing representation. Furthermore the Tseitin formulas are discussed. The standard translation, called $X_1$, uses $X_0$, but breaks up long clauses first (to avoid the exponential size-explosion), and is studied in Section \ref{sec:transx1}. Theorem \ref{thm:suffx1pc} show that if $F$ is acyclic, then $X_1(F)$ is an absolute forcing representation.  In Section \ref{sec:litrev} we provide an overview on the literature on CNF-representation of XOR-clause-sets.

In Section \ref{sec:norepmsp} we present the lower bound on ``good'' representations of XOR-clause-sets. Theorem \ref{thm:xorclsrel} shows that there are no short UR-representations of arbitrary XOR-clause-sets $F$, with Corollary \ref{cor:xorcls} generalising this to arbitrary relative asymmetric width. On the other hand, in Section \ref{sec:transarbxor} we show that $X_1(F^*)$, where $F^*$ is obtained from $F$ by adding all implied XOR-clauses, achieves a forcing representation with time-complexity fixed-parameter tractable in the number of XOR-clauses (Theorem \ref{thm:relxorcnfp}).

In Section \ref{sec:transtxor} we turn to the question of two XOR-clauses $C, D$ and $F = \set{C,D}$. In Theorem \ref{thm:2xorshared} we show how to obtain an absolute forcing representation $X_2(C,D)$. Then we discuss $X_1(F)$ and $X_1(F^*)$ and show, that all three cases can be distinguished here regarding their complexity measures; the worst representation is $X_1(F)$, which still yields an acceptable translation regarding asymmetric width, but not regarding hardness (Theorem \ref{thm:2xor}).

Finally in Section \ref{sec:open} we present the conclusions and open problems.

\section{Preliminaries}
\label{sec:prelim}

We follow the general notations and definitions as outlined in \cite{Kullmann2007HandbuchMU}. We use $\NN = \set{1,2,\dots}$ and $\NNZ = \NN \cup \set{0}$. We apply standard set-theoretic concepts, like that of a map as a set of pairs, and standard set-theoretic notations, like $f(S) = \set{f(x) : x \in S}$ for maps $f$ and $S \sse \dom(f)$, and ``$\subset$'' for the strict subset-relation.

\subsection{Clause-sets}
\label{sec:prelimcls}

Let $\Va$ be the set of variables, and let $\Lit = \Va \cup \set{\ol{v} : v \in \Va}$ be the set of literals, the disjoint union of variables as positive literals and complemented variables as negative literals. The complementation operation is extended to a (fixed point) free involution on $\Lit$, that is, for all $x \in \Lit$ we have $\ol{\ol{x}} = x$. We assume $\NN \sse \Va$, with $\ol{n} = -n$ for $n \in \NN$ (whence $\ZZ \sm \set{0} \sse \Lit$).\footnote{This yields a convenient way of writing down examples for cases, where we do not have to distinguish between different types of variables, and thus can just use natural numbers as variables. Furthermore the set of variables is infinite, and thus is never exhausted by a clause-set.}

We use $\ol{L} := \set{\ol{x} : x \in L}$ to complement a set $L$ of literals, and $\lit(L) := L \cup \ol{L}$ to close a set of literals under complementation. A clause is a finite subset $C \subset \Lit$ which is complement-free, i.e., $C \cap \ol{C} = \es$; the set of all clauses is denoted by $\Cl$. A clause-set is a finite set of clauses, the set of all clause-sets is $\Cls$. A special clause-set is $\top := \es \in \Cls$, the empty clause-set, and a special clause is $\bot := \es \in \Cl$, the empty clause.

For $p \in \NNZ$ let $\Pcls{p} := \set{F \in \Cls \mb \fa\, C \in F : \abs{C} \le p}$ denote the set of all clause-sets with clause-''width'' (i.e., length) at most $p$. Clauses containing at most one positive literal are called Horn clauses, and Horn clause-sets are clause-sets containing only Horn clauses; the set of all Horn clause-sets is denoted by $\Ho := \set{F \in \Cls \mb \fa\, C \in F : \abs{C \cap \Va} \le 1} \subset \Cls$.

By $\var(x) \in \Va$ we denote the underlying variable of a literal $x \in \Lit$, and we extend this via $\var(C) := \set{\var(x) : x \in C} \subset \Va$ for clauses $C$, and via $\var(F) := \bc_{C \in F} \var(C)$ for clause-sets $F$. The ``possible'' literals for a clause-set $F$ are denoted by $\lit(F) := \lit(\var(F))$, while the actually occurring literals are given by union $\bc F \subset \Lit$. A literal $x$ is pure for $F \in \Cls$ iff $\ol{x} \notin \bc F$. For the set of pure literals of $F$ actually occurring in $F$ we use $\purec(F) := \ol{\lit(F) \sm \bc F} \in \Cl$ (the ``pure clause'' of $F$).

The basic measures for clause-sets are $n(F) := \abs{\var(F)} \in \NNZ$ for the number of variables, $c(F) := \abs{F} \in \NNZ$ for the number of clauses, and $\ell(F) := \sum_{C \in F} \abs{C} \in \NNZ$ for the number of literal occurrences.

A basic reduction is $\rsub: \Cls \ra \Cls$, denoting elimination of subsumed clauses, that is, $\rsub(F) := \set{C \in F \mb \neg \ex \, D \in F : D \subset C}$ is the set of all inclusion-minimal clauses of $F$.

\begin{examp}\label{exp:cls0}
  Consider $F := \set{\set{1}, \set{-1,2}, \set{-1,-2,3}} \in \Ho$:
  \begin{enumerate}
  \item $\var(F) = \set{1,2,3}$, $\lit(F) = \set{1,2,3,-1,-2,-3}$, $\bc F = \set{1,2,3,-1,-2}$; the (only) pure literal of $F$ is $3$, i.e., $\purec(F) = \set{3}$.
  \item $n(F) = 3$, $c(F) = 3$, $\ell(F) = 1+2+3 = 6$.
  \item We have $\rsub(F) = F$; if we consider $F' := F \cup \set{\bot} \in \Ho$, then $\rsub(F') = \set{\bot}$.
  \end{enumerate}
\end{examp}

\subsection{Partial assignments}
\label{sec:prelimpass}

A partial assignment is a map $\vp: V \ra \set{0,1}$ for some finite $V \subset \Va$, where we set $\var(\vp) := V$ and $n(\vp) := \abs{\var(\vp)}$. The set of all partial assignments is $\Pass$, while for $V \sse \Va$ we use $\Pass(V) := \set{\vp \in \Pass : \var(\vp) \sse V}$; furthermore we denote by $\Tass(V) := \set{\vp \in \Pass : \var(\vp) = V}$ the set of total assignments on $V$ (for finite $V$). For $v \in \var(\vp)$ let $\vp(\ol{v}) := \ol{\vp(v)}$ (with $\ol{0} = 1$ and $\ol{1} = 0$). We construct partial assignments by terms $\pab{x_1 \ra \ve_1, \dots, x_n \ra \ve_n} \in \Pass$ (in the obvious way) for literals $x_1, \dots, x_n$ with different underlying variables and $\ve_i \in \set{0,1}$. Another construction negates the literals in a clause, that is, for a given $C \in \Cl$ we obtain $\vp_C \in \Pass$ as $\vp_C := \pab{x \ra 0 : x \in C}$, setting precisely the literals of $C$ to $0$.

For $\vp \in \Pass$ and $F \in \Cls$ we denote the result of applying $\vp$ to $F$ by $\vp * F$, removing clauses $C \in F$ containing $x \in C$ with $\vp(x) = 1$, and removing literals $x$ with $\vp(x) = 0$ from the remaining clauses. By $\Sat := \set{F \in \Cls \mb \ex\, \vp \in \Pass : \vp * F = \top}$ the set of satisfiable clause-sets is denoted, and by $\Usat := \Cls \sm \Sat$ the set of unsatisfiable clause-sets.

\begin{examp}\label{exp:cls1}
  For $F$ from Example \ref{exp:cls0} and $\vp := \pab{1 \ra 1, 2 \ra 1, 3 \ra 1}$ we have $\vp * F = \top$, and thus $F \in \Sat$, while for example $\pab{-1 \ra 0, 2 \ra 0} * F = \set{\bot}$ and $\pao 11 * F = \set{\set{2},\set{-2,3}}$.
\end{examp}

\subsection{Forced assignments}
\label{sec:prelimforced}

A fundamental inference mechanism for clause-sets is unit-clause propagation, which we denote by $\bmm{\ro}: \Cls \ra \Cls$, and which is defined recursively via:
\begin{itemize}
\item $\ro(F) := \set{\bot}$ if $\bot \in F$,
\item $\ro(F) := F$ if $F$ contains only clauses of length at least $2$,
\item while otherwise a unit-clause $\set{x} \in F$ is chosen, and $\ro(F) := \ro(\pao x1 * F)$.
\end{itemize}
It is easy to see that the final result $\ro(F)$ does not depend on the choices of the unit-clauses. In \cite{Ku99b,Ku00g} the theory of generalised unit-clause propagation $\bmm{\rk_k}: \Cls \ra \Cls$ for $k \in \NNZ$ was developed (reviewed in \cite[Section 4]{GwynneKullmann2012SlurJ}), where the basic idea should become clear by considering $\rk_2(F)$, which is complete ``failed literal elimination'' (see Section 5.2.1 in \cite{HvM09HBSAT} for the usage of failed literals in SAT solvers): If there is a literal $x \in \lit(F)$ such that $\ro(\pao x0 * F) = \set{\bot}$, then we have to set $x$ to $1$, and $\rk_2(F) := \rk_2(\pao x1 * F)$, while otherwise $\rk_2(F) := F$. The general definition for $k \in \NNZ$ and $F \in \Cls$ is as follows (using recursion in $k$ and $n(F)$):
\begin{itemize}
\item $\rk_0(F) := F$ if $\bot \notin F$, while otherwise $\rk_0(F) := \set{\bot}$.
\item If there is $x \in \lit(F)$ with $\rk_k(\pao x0 * F) = \set{\bot}$, then  choose such an $x$ and let $\rk_{k+1}(F) := \rk_{k+1}(\pao x1 * F)$.
\item If there is no such $x$, then $\rk_{k+1}(F) := F$.
\end{itemize}
It is easy to see that all maps $\rk_k$ are well-defined (do not depend on the choices made for the literals $x$). By definition of $\rk_k(F)$, for every $F \in \Cls$ there is a partial assignment $\vp$ with $\rk_k(F) = \vp * F$:
\begin{enumerate}
\item If $\rk_k(F) = \set{\bot}$, then one can choose for example any $\vp \in \Tass(\var(F))$.
\item However, if $\rk_k(F) \ne \set{\bot}$, then the collection of assignments performed in the process of calculating $\rk_k(F)$ (according to the definition) is unique (does not depend on the choices), and we call this $\vp \in \Pass(\var(F))$ the \emph{associated partial assignment}.
\end{enumerate}
The associated partial assignments consists of certain \textbf{forced assignments} for $F$, which are assignments $\pao x1$ such that the opposite assignment yields an unsatisfiable clause-set, that is, where $\pao x0 * F \in \Usat$; the literal $x$ here is also called a \textbf{forced literal} (in Definition \ref{def:frl} we will introduce a notation for the set of forced literals). The reduction applying all forced assignments is denoted by $\rki: \Cls \ra \Cls$ (so $F \in \Usat \Lra \rki(F) = \set{\bot}$). Forced assignments are also known under other names, for example ``necessary assignments'', or ``backbones'' or ``frozen variables'' concerning the underlying variables (we just speak of forced variables); see \cite{JanotaLycneMarquesSilva2012Backbones} for an overview on algorithms computing all forced assignments.

\begin{examp}\label{exp:forced}
  Some basic examples for $\rk_k$ and forced literals:
  \begin{enumerate}
  \item For $F$ from Example \ref{exp:cls0} we have $\ro(F) = \top$.
  \item If for $F \in \Cls$ we have $\fa\, C \in F : \abs{C} > k$, then $\rk_k(F) = F$.
  \item If $F \in \Usat$, then every literal $x \in \Lit$ is forced for $F$.
  \item A clause-set $F \in \Cls$ is uniquely satisfiable, that is, has exactly one $\vp \in \Tass(\var(F))$ with $\vp * F = \top$, iff $F$ has exactly $n(F)$ forced literals (in other words, all variables of $F$ are forced).
  \item $F$ from Example \ref{exp:cls0} is uniquely satisfiable, since $\ro$ sets all its variables.
  \end{enumerate}
\end{examp}

We conclude this short review on the reductions $\rk_k$ by some remarks on modifications. In \cite{BacchusWinter2003HyperBin} we find the exploration of ``hyper binary resolution'', which goes beyond $\rk_2$ by adding binary resolvents; from our point that yields a reduction $\rk_2': \Cls \ra \Cls$ by forced assignments via all the derived unit clauses. In \cite[Subsection 5.2]{Kullmann2007Uebersicht} the more general reductions $\rk_k': \Cls \ra \Cls$ for $k \ge 2$ are discussed, which are defined as the $\rk_k$, but additionally in case of $\rk_{k-1}(\pao x0 * F) \ne \set{\bot}$ consider the associated partial assignment with $\vp * (\pao x0 * F) = \rk_{k-1}(\pao x0 * F)$, and for all literals $y$ with $\vp(y) = 1$ add the binary clause $\set{x,y}$ (note that this clause is implied by $F$). As shown in \cite[Subsection 2.2]{BacchusWinter2003HyperBin}, for $k=2$ this is the same as above. As remarked in \cite{Kullmann2007Uebersicht}, this is weaker than $\rk_{k+1}$ and stronger than $\rk_k$. For heuristic weakenings of $\rk_2'$ see \cite{KaufmannKottler2011BeyondUCP}. Finally the St\aa{}lmarck approach, when restricted to CNF, is presented in \cite[Subsection 3.5]{Ku99b}, yielding reductions $\rk_k'': \Cls \ra \Cls$ for $k \ge 2$, which are weaker than $\rk_k'$ and stronger than $\rk_k$: if $\rk_{k-1}(\pao x0 * F) \ne \set{\bot}$ and $\rk_{k-1}(\pao x1 * F) \ne \set{\bot}$, then for the associated partial assignments $\vp_0, \vp_1$ the reduction $F \leadsto (\vp_0 \cap \vp_1) * F$ is performed. These modified reductions are of practical interest, but in this \Schrift{} we concentrate on the ``pure forms'' $\rk_k$.

\subsection{Resolution and prime implicates}
\label{sec:prelimresl}

Two clauses $C, D \in \Cl$ are resolvable iff they clash in exactly one literal $x$, that is, $C \cap \ol{D} = \set{x}$, in which case their resolvent is $\bmm{C \res D} := (C \cup D) \sm \set{x,\ol{x}}$ (with resolution literal $x$). A resolution tree is a full binary tree (every non-leaf node has exactly two children) formed by the resolution operation. We write \bmm{T : F \vdash C} if $T$ is a resolution tree with axioms (the clauses at the leaves) all in $F$ and with derived clause (at the root) $C$.

A \emph{prime implicate} of $F \in \Cls$ is a clause $C$ such that a resolution tree $T$ with $T: F \vdash C$ exists, but no $T'$ exists for some $C' \subset C$ with $T': F \vdash C'$; the set of all prime implicates of $F$ is denoted by $\bmm{\primec_0(F)} \in \Cls$. The term ``implicate'' refers to the implicit interpretation of $F$ as a conjunctive normal form (CNF). Considering clauses as combinatorial objects, one can speak of ``\ul{pr}ime \ul{c}lauses'', and the ``$0$'' in our notation reminds of ``unsatisfiability'', due to prime implicates of $F$ corresponding to minimal partial assignments $\vp$ with $\vp * F \in \Usat$.

The underlying semantics of the resolution calculus is denoted by $F \models F'$ for $F, F' \in \Cls$, which is true iff for all partial assignments $\vp$ with $\vp * F = \top$ we also have $\vp * F' = \top$; the clause-sets $F, F'$ are logically equivalent iff $F \models F'$ and $F' \models F$, which in turn is equivalent to $\primec_0(F) = \primec_0(F')$. For clauses $C$ we write $F \models C :\Lra F \models \set{C}$; the implicates of $F$ are precisely the clauses $C$ with $F \models C$, while the prime implicates are characterised by the additional condition, that for no $C' \subset C$ holds $F \models C'$. The number of prime implicates of $F \in \Cls$ can be estimated as follows:
\begin{itemize}
\item $c(\primec_0(F)) \le 2^{c(F)}-1$; for an overview on the history of the inequality (generalised to non-boolean clause-sets in \cite[Corollary 4.6]{Kullmann2007ClausalFormZII}) see \cite{SloanSzoerenyiTuran2005Primimplikanten_1}.
\item $c(\primec_0(F)) \le 3^{n(F)}$; see \cite[Theorem 3.16]{CramaHammer2011BooleanFunctions} for asymptotic sharpness.
\end{itemize}

It is known (though apparently not stated explicitly in the literature), that for a clause-set $F \in \Cls$ the computation of $\primec_0(F)$ is fixed-parameter tractable (fpt) in the number of clauses (using space which is linear in the output size). This follows from \cite[Theorem 3.9]{CramaHammer2011BooleanFunctions}, when using $c(\primec_0(F)) \le 2^{c(F)}-1$, together with a SAT-algorithm, which for inputs $F \in \Cls$ decides SAT in time $O(\ell(F) \cdot 2^{c(F)})$ and linear space.\footnote{The exponential $2^{c(F)}$ can be improved further (see \cite{DH09HBSAT} for an overview on such bounds), but we don't go into this, since it wouldn't improve our stronger bound in Lemma \ref{lem:primcec}.} We obtain run-time $O(\ell(F)^2 \cdot 2^{3 c(F)})$ for the computation of $\primec_0(F)$. See Lemma \ref{lem:primcec} in Appendix \ref{sec:compprimeimpl} for a simple proof (yielding also a better exponent).

Computation of $\primec_0(F)$ is also fpt in the parameter $n(F)$, which also can be obtained from \cite[Theorem 3.9]{CramaHammer2011BooleanFunctions}), again with also linear output-space. This applies $c(\primec_0(F)) \le 3^{n(F)}$, and uses a SAT-algorithm, which for inputs $F \in \Cls$ decides SAT in time $O(\ell(F) \cdot 2^{n(F)})$ and linear space (by the trivial algorithm).\footnote{Here no better exponential is known; see \cite{DH09HBSAT}.} We obtain run-time $O(n \cdot 2^{7 n} \cdot (n \cdot 2^{2 n} + \ell(F)))$, where $n := n(F)$.

Finally, a \emph{boolean function}\footnote{it seems finally adequate to make ``boolean'' a proper adjective} $f$ is a map $f: \Tass(V) \ra \set{0,1}$ for some finite $V \subset \Va$; we use $\var(f) := V$ for the set of variables, $n(f) := \abs{\var(f)}$ for the number of variables, and $\lit(f) := \lit(\var(f))$ for the associated literals. Special boolean functions are $0^V, 1^V$ for finite $V \subset \Va$, which denote the constant functions $f$ with $\var(f) = V$. A boolean function $f$ is \emph{monotone} iff flipping any input variable from $0$ to $1$ never flips the output from $1$ to $0$. Note that $\primec_0(f)$ is well-defined as the set of prime implicates of $f$ (minimal clauses implied by $f$), and for every CNF-clause-set $F$ equivalent to $f$ we have $\primec_0(f) = \primec_0(F)$. A boolean function $f$ is monotone iff $f$ has only positive prime implicates, that is, $\bc \primec_0(f) \subset \Va$.

\begin{examp}\label{exp:boolfres}
  Some simple examples for prime implicates:
  \begin{enumerate}
  \item $\primec_0(0^V) = \set{\bot}$ and $\primec_0(1^V) = \top$.
  \item A clause-set $F \in \Cls$ is unsatisfiable iff $\primec_0(F) = \set{\bot}$, while for satisfiable $F$ a literal $x$ is forced iff $\set{x} \in \primec_0(F)$.
  \item For $F$ from Example \ref{exp:cls0} we have $\primec_0(F) = \set{\set{1},\set{2},\set{3}}$, which are obtained by resolution via $\set{1} \res \set{-1,2} = \set{2}$, $\set{1} \res \set{-1,-2,3} = \set{-2,3}$, $\set{2} \res \set{-2,3} = \set{3}$.
  \item For the boolean function $a \vee b$ we have $\primec_0(a \vee b) = \set{\set{a,b}}$, while for the boolean function $a \wedge b$ we have $\primec_0(a \wedge b) = \set{\set{a},\set{b}}$.
  \end{enumerate}
\end{examp}

\section{Measuring unsatisfiable sub-instances}
\label{sec:measurerepcomp}

In this section we define and discuss the measures $\hardness, \whardness: \Cls \ra \NNZ$ and the corresponding classes $\Urefc_k \sse \Wrefc_k \subset \Cls$. It is mostly of an expository nature, explaining the background from \cite{Ku99b,Ku00g,GwynneKullmann2012SlurSOFSEM,GwynneKullmann2012SlurJ}. For the measure $\phardness: \Cls \ra \NNZ$ and the corresponding classes $\Propc_k$ see Section \ref{sec:propc}. The basic measurement happens on unsatisfiable clause-sets, and is then extended in a generic way to satisfiable clause-sets (see \cite[Section 3]{BeyersdorffKullmann2014PHP} for some discussion).

\subsection{Hardness and $\Urefc_k$}
\label{sec:prelimhdUC}

Hardness for unsatisfiable clause-sets was introduced in \cite{Ku99b,Ku00g}, while the specific generalisation to arbitrary clause-sets used here was first mentioned in \cite{AnsoteguiBonetLevyManya2008Hardness}, and systematically studied in \cite{GwynneKullmann2012SlurSOFSEM,GwynneKullmann2012Slur,GwynneKullmann2012SlurJ}. The most natural approach in our context uses necessary levels of generalised unit-clause propagation $\rk_k$:
\begin{defi}\label{def:charachd}
  For $F \in \Cls$ and $V \sse \Va$ let $\bmm{\hardness^V(F)} \in \NNZ$ (``hardness relative to $V$'') be the minimal $k \in \NNZ$ such that for all $\vp \in \Pass(V)$ and $\vp * F \in \Usat$ holds $\rk_k(\vp * F) = \set{\bot}$, i.e., the minimal $k$ such that $\rk_k$ detects unsatisfiability of any partial instantiation of variables in $V$. Furthermore $\bmm{\hardness(F)} := \hardness^{\var(F)}(F)$ (``absolute hardness'').
\end{defi}
For $F \in \Cls \sm \set{\top}$ we have $\hardness(F) = \max_{C \in \primec_0(F)} \hardness(\vp_C * F)$, since these $\vp_C$ are precisely the minimal $\vp \in \Pass$ with $\vp * F \in \Usat$. An equivalent characterisation uses the Horton-Strahler number $\hts(T)$ (see \cite{Viennot1990Trees,EsparzaLuttenbergerSchlund2014Strahler} for overviews) of resolution trees $T: F \vdash C$ (deriving clause $C$ from $F$). The Horton-Strahler number of a binary tree is the smallest $k \in \NNZ$ such that for every node there exists a path to some leaf of length at most $k$; equivalently, it is the largest $k \in \NNZ$ such that the complete binary tree with $2^k$ leaves can be embedded. Now the hardness $\hardness(F)$ for $F \in \Cls$ is the minimal $k \in \NNZ$ such that for all prime implicates $C$ of $F$ there exists $T : F \vdash C$ with $\hts(T) \le k$.

Recall that for $F \in \Cls$ and $k \in \NNZ$ the associated partial assignment $\vp$ has $\vp * F = \rk_k(F)$, where $\vp$ consists of certain forced assignments $\pao x1 \sse \vp$. A weaker localisation of forced assignments has been considered in \cite{HaanKanjSzeider2013BackbonesC}, namely ``$k$-backbones'', which are forced assignments $\pao x1$ for $F$ such that there is $F' \sse F$ with $c(F') \le k$ and such that $\pao x1$ is forced also for $F'$. It is not hard to see that $\rk_k$ for $k \in \NNZ$ will set all $k$-backbones of $F \in \Cls$, using that for $F \in \Usat$ we have $\hardness(F) < c(F)$ by Lemma 3.18 in \cite{Ku99b}.

Absolute hardness yields the $\Urefc_k$-hierarchy (with ``UC'' for ``unit-refutation complete''):
\begin{defi}\label{def:UC}
  For $k \in \NNZ$ let $\bmm{\Urefc_k} := \set{F \in \Cls : \hardness(F) \le k}$.
\end{defi}
$\Urefc_1 = \Urefc$ is the class of unit-refutation complete clause-sets, as introduced in \cite{Val1994UnitResolutionComplete}. In \cite{GwynneKullmann2012SlurSOFSEM,GwynneKullmann2012Slur,GwynneKullmann2012SlurJ} we show that $\Urefc = \Slur$, where $\Slur$ is the class of clause-sets solvable via Single Lookahead Unit Resolution (see \cite{FrGe98}). Using \cite{CepekKuceraVlcek2012SLUR} we then obtain (\cite{GwynneKullmann2012SlurSOFSEM,GwynneKullmann2012Slur,GwynneKullmann2012SlurJ}) that membership decision for $\Urefc_k$ ($ = \Slur_k$) is coNP-complete for $k \ge 1$. The class $\Urefc_2$ is the class of all clause-sets where unsatisfiability for any partial assignment is detected by complete failed-literal reduction. In Sections 5 and 6 of \cite{GwynneKullmann2012Slur} one finds many examples for clause-sets in $\Urefc_k$.
\begin{examp}\label{ecp:UC}
  Obviously for all $F \in \Cls$ holds $\hardness(F) = \hardness(\rsub(F))$ (hardness is invariant under subsumption elimination). We have $F \in \Urefc_0$ iff $\primec_0(F) = \rsub(F)$. The simplest example of $F \in \Urefc_1 \sm \Urefc_0$ is $\set{\set{1},\set{-1}}$.

  Indeed, all $F \in \Ho$ (Horn clause-sets) are in $\Urefc_1$ (since unit-clause propagation is sufficient to detect unsatisfiability of Horn clause-sets; see \cite{GwynneKullmann2012Slur}). It is well-known that Horn clause-sets can have exponentially many prime implicates; consider for example $F_n := \set{\set{\ol{x_1},\dots,\ol{x_n}}} \cup \set{\set{\ol{x_i},y_i}, \set{x_i,\ol{y_i}}}_{i \in \tb 1n} \in \Ho$ (according to \cite[Theorem 3.17]{CramaHammer2011BooleanFunctions}, there in the language of DNF). Thus already Horn clause-sets can provide an exponential saving in size over $\Urefc_0$, without giving up on inference power (modulo $\ro$). Different from $\Ho$, all class $\Urefc_k$ are functionally complete, that is, can represent all boolean functions.
\end{examp}

Additional characterisations of hardness one finds in \cite{BeyersdorffKullmann2014PHP}: In Subsection 4.1 there a game characterisation of $\hardness(F)$ for $F \in \Cls$ is given (extending the Prover-Delayer game for unsatisfiable $F$ from \cite{PI2000}), while in Subsection 4.2 one finds a characterisation for unsatisfiable $F$ in terms of ``weakly consistent'' sets of partial assignments. We remark that from the variations $\rk_k', \rk_k''$, as discussed at the end of Subsection \ref{sec:prelimforced}, we obtain hardness measures $\hardness', \hardness'': \Cls \ra \NNZ$ with $\hardness -1 \le \hardness' \le \hardness'' \le \hardness$, whose practical potential needs to be explored in future studies.

\subsection{Asymmetric width and $\Wrefc_k$}
\label{sec:prelimwhdWC}

A basic weakness of the standard notion of width-restricted resolution, which demands that \emph{both} parent clauses plus the resolvent must have length at most $k$ for some fixed $k \in \NNZ$ (``width'', denoted by $\wid(F)$ below; see \cite{SW98J}), is that even Horn clause-sets require unbounded width in this sense. A better solution seems, as investigated and discussed in \cite{Ku99b,Ku00g,BeyersdorffKullmann2014PHP}, to use the notion of ``$k$-resolution'' as introduced in \cite{Kl93}, where only \emph{one} parent clause needs to have length at most $k$ (thus properly generalising unit-resolution; the length of the resolvent is also unrestricted).\footnote{In the literature on proof complexity, which makes only asymptotic statements and ignores constant factors, symmetric width is only applied to clause-sets with bounded clause-length, and here everything can be done as well via asymmetric width, as discussed below. For unbounded initial clause-length, asymmetric width is the proper generalisation. Combinatorially, both measures are different, and it is conceivable that symmetric width could have a relevant combinatorial meaning also for unbounded clause-length.} Nested input-resolution (\cite{Ku99b,Ku00g}) is the proof-theoretic basis of hardness, and approximates tree-resolution. In the same vein, $k$-resolution is the proof-theoretic basis of ``asymmetric width'', and approximates dag-resolution (see Theorem 6.12 in \cite{Ku00g}):
\begin{defi}\label{def:whd}
  The \textbf{asymmetric width} $\bmm{\whardness}: \Cls \ra \NNZ$ (``width-hardness'', or ``asymmetric width'') is defined for $F \in \Cls$ as follows:
  \begin{enumerate}
  \item If $F \in \Usat$, then $\whardness(F)$ is the minimum $k \in \NNZ$ such that $k$-resolution refutes $F$, that is, such that $T : F \vdash \bot$ exists where for each resolution step $R = C \res D$ in $T$ we have $\abs{C} \le k$ or $\abs{D} \le k$ (this concept corresponds to Definition 8.2 in \cite{Ku99b}, and is a special case of ``$\mr{wid}_{\mc{U}}$'' as introduced in Subsection 6.1 of \cite{Ku00g}).
  \item If $F = \top$, then $\whardness(F) := 0$.
  \item If $F \in \Sat \sm \set{\top}$, then $\DST \whardness(F) := \max_{\vp \in \Pass} \set{\whardness(\vp * F) : \vp * F \in \Usat}$.
  \end{enumerate}
  For $k \in \NNZ$ let $\bmm{\Wrefc_k} := \set{F \in \Cls : \whardness(F) \le k}$.

  The \textbf{symmetric width} $\bmm{\wid}: \Cls \ra \NNZ$ is defined in the same way, only that for $F \in \Usat$ we define $\wid(F)$ as the minimal $k \in \NNZ$ such that there is $T : F \vdash \bot$, where all clauses of $T$ (axioms and resolvents) have length at most $k$.

  More generally, for $V \sse \Va$ we define the relativisations $\bmm{\whardness^V(F)} := \whardness(F)$ and $\bmm{\wid^V(F)} := \wid(F)$ for unsatisfiable $F$, while for satisfiable $F$ only $\vp \in \Pass$ with $\var(\vp) \sse V$ are considered.
\end{defi}
We remark that although these width-notions are closely related to full or dag-resolution, the easier treatment of resolution via trees is fully sufficient here, since resolution-dags can always be unfolded into trees without affecting the width of clauses. We have $\Wrefc_0 = \Urefc_0$, $\Wrefc_1 = \Urefc_1$, and for all $k \in \NNZ$ holds $\Urefc_k \sse \Wrefc_k$ (this follows by Lemma 6.8 in \cite{Ku00g} for unsatisfiable clause-sets, which extends to satisfiable clause-sets by definition), and, more generally, for all $V \sse \Va$ and $F \in \Cls$ holds $\whardness^V(F) \le \hardness^V(F)$.

\begin{examp}\label{exp:whard}
  A trivial lower bound for $\whardness(F)$ in case of unsatisfiable $F$, where all clauses of $F$ have length $k$ or bigger, is $\whardness(F) \ge k$ . This does not generalise (directly) to satisfiable clause-sets; if for example we consider a singleton clause-set $F := \set{C}$ for $C \in \Cl$, then we have $\whardness(F) = \hardness(F) = 0$.
\end{examp}

We consider now the relation between asymmetric width $\whardness(F)$ and symmetric width $\wid(F)$. By definition we have $\whardness(F) \le \wid(F)$ for all $F \in \Cls$. In the other direction there is no such relation, if the clause-length is unbounded:
\begin{examp}\label{exp:whardHO}
  Consider $F \in \Ho \cap \Usat$ (recall $\Ho$ is the set of Horn clause-sets). The symmetric width $\wid(F)$ is unbounded, and is equal to the maximal clause-length of $F$ in case $F$ is minimally unsatisfiable. But $\whardness(F) \le 1$.
\end{examp}

  So for unbounded clause-length there is an essential difference between symmetric and asymmetric width. On the other hand we have
\begin{displaymath}
  \wid(F) \le \whardness(F) + \max(\whardness(F),p)
\end{displaymath}
for $F \in \Pcls{p}$, $p \in \NNZ$, by Lemma 8.5 in \cite{Ku99b}, or, more generally, Lemma 6.22 in \cite{Ku00g} (also shown in \cite{BeyersdorffKullmann2014PHP,BeyersdorffGwynneKullmann2013PHP}). So for bounded clause-length and considered asymptotically, symmetric and asymmetric width can be considered equivalent.

\section{Propagation-completeness and acyclic clause-sets}
\label{sec:propc}

The measurements $\hardness, \whardness: \Cls \ra \NNZ$ considered on $\Sat$ are concerned with partial assignments making the clause-set unsatisfiable, and the complexity of the refutations (i.e., they are concerned with implicates and the effort of deriving them). In this section we refine $\hardness$ by considering the forced literals after partial assignments, and the effort to determine them. This leads to $\phardness: \Cls \ra \NNZ$ (which on $\Usat$ is equal to $\hardness$), which is defined in Subsection \ref{sec:defpropc}, and where we show the basic properties; in parallel, the corresponding classes $\Propc_k$ are also studied there. In Subsection \ref{sec:Acyclicity} we discuss the concept of an ``acyclic'' family of clause-sets $(F_i)_{i \in I}$, and show in Theorem \ref{thm:acylcprop}, that if $(F_i)_{i \in I}$ is acyclic and fulfils $\fa\, i \in I : F \in \Propc_k$, then also $\bc_{i \in I} F_i \in \Propc_k$.

Forced assignments/literals become now very important, and so we introduce a notation for the set of forced literals of a clause-set (or boolean function):
\begin{defi}\label{def:frl}
  Let $\bmm{\frl(F)} := \set{x \in \Lit : \pao x0 * F \in \Usat}$ for $F \in \Cls$.
\end{defi}
 So $F \in \Usat \Lra \frl(F) = \Lit$ and $F \in \Sat \Lra \frl(F) \sse \lit(F)$. And for $F \in \Sat$ we indeed have $\frl(F) \in \Cl$, and $\rki(F) = \pab{x \ra 1 : x \in \frl(F)} * F$.

\subsection{P-Hardness and $\Propc_k$}
\label{sec:defpropc}

Complementary to ``unit-refutation completeness'', there is the notion of ``pro\-pa\-ga\-tion-com\-ple\-te\-ness'' as investigated in \cite{BordeauxMarquesSilva2012KnowledgeCompilation}, yielding the class $\Propc \subset \Urefc$. This was captured and generalised by a measure $\phardness: \Cls \ra \NNZ$ of ``propagation-hardness'' along with the associated hierarchy, as defined in \cite[Subsection 9.1]{GwynneKullmann2012SlurJ} (in the Outlook). We present a generalised definition, which allows relativisation to a set $V$ of variables; the idea is that $\phardness^V(F)$ is the smallest $k$ such that for any partial assignment $\vp$ to variables in $V$, all forced assignments in $\vp * F$ with variables in $V$ are obtained by $\rk_k$:
\begin{defi}\label{def:phardness}
  For $F \in \Cls$ and $V \sse \Va$ we define the (relative) \textbf{propagation-hardness} (for short ``p-hardness'') $\bmm{\phardness^V(F)} \in \NNZ$ as the minimal $k \in \NNZ$ such that for all partial assignments $\vp \in \Pass$ with $\var(\vp) \sse V$ holds:
  \begin{enumerate}
  \item If $\vp * F \in \Usat$, then $\rk_k(\vp * F) = \set{\bot}$.
  \item If $\vp * F \in \Sat$, then $\frl(\rk_k(\vp * F)) \cap \lit(V) = \es$.
  \end{enumerate}
  Furthermore $\bmm{\phardness(F)} := \phardness^{\var(F)}(F)$.
\end{defi}

\begin{examp}\label{exp:PC}
  Some simple examples for p-hardness (absolute and relative):
  \begin{enumerate}
  \item $\phardness(\top) = \phardness(\set{\bot}) = 0$.
  \item For $C \in \Cl$ with $\abs{C} \ge 1$ holds $\phardness(\set{C}) = 1$.
  \item If $F \in \Cls$ is positive (i.e., $\bc F \subset \Va$), then $\phardness(F) \le 1$ (while $\hardness(F) = 0$).
  \item For $F \in \Cls$ holds $\phardness^{\es}(F) = 0$ iff $\bot \in F$ or $\frl(F) = \es$, and generally we have for $F \in \Usat$ that $\phardness^{\es}(F) = \hardness^{\es}(F) = \hardness(F)$, while for $F \in \Sat$ with $\frl(F) \ne \es$ we have that $\phardness^{\es}(F)$ is the smallest $k \in \NN$ with $\frl(\rk_k(F)) = \es$.
  \end{enumerate}
\end{examp}

The most basic properties of p-hardness are as follows:
\begin{enumerate}
\item By definition we have $\phardness(F) = \hardness(F)$ for $F \in \Usat$.
\item For $F \in \Cls$ the ``absolute'' p-hardness $\phardness(F)$ is the minimal $k \in \NNZ$ such that for all partial assignments $\vp \in \Pass$ we have $\rk_k(\vp * F) = \rki(\vp * F)$.
\item By definition (and composition of partial assignments) we have $\phardness^V(\vp * F) \le \phardness^V(F)$ for all $F \in \Cls$, $V \subset \Va$ and $\vp \in \Pass$.
\end{enumerate}

\begin{defi}\label{def:PC}
  For $k \in \NNZ$ let $\bmm{\Propc_k} := \set{F \in \Cls : \phardness(F) \le k}$ (the class of \textbf{propagation-complete clause-sets of level $k$}).
\end{defi}
By definition holds $\Propc_1 = \Propc$ for the class $\Propc$ as introduced in \cite[Definition 1]{BordeauxMarquesSilva2012KnowledgeCompilation}. The most basic properties of the classes $\Propc_k$ are as follows:
\begin{enumerate}
\item $\Propc_k \cap \Usat = \Urefc_k \cap \Usat$ for $k \in \NNZ$.
\item All classes $\Propc_k$ are stable under application of partial assignments.
\item We have $F \in \Propc_k$ iff for all $\vp \in \Pass$ the clause-set $F' := \rk_k(\vp * F)$ in case of $F' \ne \set{\bot}$ has no forced literals. In other words for $F \in \Cls$ holds $F \in \Propc_k \Lra \fa\, \vp \in \Pass : \frl(\rk_k(\vp * F)) \in \set{\es, \Lit}$.
\end{enumerate}
As shown in \cite[Proposition 1]{BordeauxMarquesSilva2012KnowledgeCompilation}, extracting the notion of \emph{empowering clause} from \cite{DarwichePipatsrisawat2011ClauseLearnRes}, for $F \in \Cls$ holds $F \in \Propc$ iff $F$ contains (in a sense) all empowering clauses. And starting from $F \in \Cls \sm \Propc$, one can compute an equivalent $F' \in \Propc$ by adding empowering clauses (\cite[Section 4]{BordeauxMarquesSilva2012KnowledgeCompilation}); this process must create exponentially many clauses for certain examples, as shown in \cite[Theorem 5.8]{BBCGKV2013Propc}.

The underlying theme of (relative) p-hardness $\phardness^V(F)$ and the classes $\Propc_k$ can be seen in ``maintaining (generalised) arc-consistency'', that is, viewing $F$ as a (global) constraint and enforcing, that to every variable every value can be assigned without causing inconsistency, and this after every instantiation; see \cite{Bessiere2006Propagation} for general information on this central topic from the CSP world. Adopting a remark from \cite{Bacchus2007GAC}, the difference is that p-hardness takes a declarative point of view, by using standardised algorithms $\rk_k$ for enforcing arc-consistency, not the procedural point of view of constraint programming (with dedicated algorithms, operating on top of the constraints); see Subsection \ref{sec:Acyclicity} and especially Subsection \ref{sec:introAC} for more on this.

The definition of relative p-hardness contains a somewhat subtle point:
\begin{examp}\label{exp:relphd}
  In \cite{GwynneKullmann2013GoodRepresentationsIILata} by mistake relative p-hardness $\phardness^V(F)$ was defined as the minimal $k \in \NNZ$ such that for all partial assignments $\vp \in \Pass$ with $\var(\vp) \sse V$ we have $\rk_k(\vp * F) = \rki(\vp * F)$. Thus the reduction $\rk_k$ had to eliminate also forced literals outside of $V$ in $\vp * F$. But the concept of relative p-hardness is motivated by representations $F \in \Cls$ of boolean functions $f$ (or ``constraints''), where $V := \var(f)$ and $\phardness^V(F)$ is considered, generalising the notions of ``maintaining arc-consistency by unit-clause propagation'' as discussed in Subsection \ref{sec:introAC}, and in this context the auxiliary variables $\var(F) \sm V$ are excluded from inference considerations.\footnote{The exclusion of auxiliary variables is not explicitly stated in \cite{Een2006Translating}, since there the auxiliary (``introduced'') variables are not part of the picture, but only the given constraint.}

  A trivial example showing the difference is given by considering $V := \es$: Now for all $F \in \Sat$ we have $\phardness^V(F) = 0$, while $k$ with $\rki(F) = \rk_k(F)$ is unbounded.
\end{examp}

\subsection{Basic properties}
\label{sec:phdbprop}

The class $\Propc_0$ is decidable in polynomial time, as is its superclass $\Urefc_0$ (\cite[Lemma 6.10]{GwynneKullmann2012SlurJ}), but different from $\Urefc_0$, the class $\Propc_0$ is not functionally complete, but contains only the most trivial clause-sets:
\begin{lem}\label{lem:characPropc0}
  $\Propc_0 = \set{\top} \cup \set{F \in \Cls : \bot \in F}$.
\end{lem}
\begin{prf}
Clearly $\set{\top} \cup \set{F \in \Cls : \bot \in F} \sse \Propc_0$. We show that for all $F \in \Cls$ with $F \ne \top$ and $\bot \notin F$ we have $\phardness(F) \ge 1$, by induction on $n(F)$: The assertion holds trivially for $n(F) = 0$, so consider $n(F) \ge 1$. If there is a unit-clause $\set{x} \in F$, then $F$ has the forced literal $x$, while $\rk_0(F) = F$, and thus $\phardness(F) \ge 1$. Otherwise all clauses of $F$ have length at least $2$. Choose $v \in \var(F)$ and $\ve \in \set{0,1}$ such that $F' := \pao{v}{\ve} * F \ne \top$ (note that for a literal $x$ and a clause-set $F$ holds $\pao x1 * F = \top \Lra x \in \bca F$). Now the induction hypothesis can be applied to $F'$, and we obtain $\phardness(F') \ge 1$, while $\phardness(F) \ge \phardness(F')$. \Qed
\end{prf}

P-hardness and (ordinary) hardness are close related:
\begin{lem}\label{lem:sandwichUPrel}
  For $F \in \Cls$ and $V \sse \Va$ holds $\hardness^V(F) \le \phardness^V(F) \le \hardness^V(F) + 1$.
\end{lem}
\begin{prf}
$\hardness^V(F) \le \phardness^V(F)$ follows by definition. In order to show $\phardness^V(F) \le \hardness^V(F) + 1$, let $k := \hardness^V(F)$, and assume that there is $\vp \in \Pass(V)$, such that for $F' := \rk_{k+1}(\vp * F) \in \Sat$ there exists a forced literal $x \in \lit(V)$ for $F'$. Thus $\pao x0 * F' \in \Usat$, and so for $\vp' := \vp \cup \pao x0$ (note $\var(\vp) \sse V$) we have $\rk_k(\vp' * F) = \set{\bot}$ (due to $\hardness^V(F) \le k$), and it follows by definition of $\rk_{k+1}$ (and the confluence of the computation), that $\rk_{k+1}$ must set $x \ra 1$ in the computation of $\rk_{k+1}(\vp * F)$. \Qed
\end{prf}

Thus the classes $\Urefc_k$ are (strictly) interspersed between the classes $\Propc_k$ (\cite[Lemma 9.3]{GwynneKullmann2012SlurJ}, but without a proof there):
\begin{corol}\label{cor:sandwichUP}
  For $k \in \NNZ$ we have $\Propc_k \subset \Urefc_k \subset \Propc_{k+1}$.
\end{corol}
\begin{prf}
It remains to show strictness of the inclusions. Let $A_k \in \Usat$ be the clause-set containing the $2^k$ full clauses (of length $k$) over the variables $1,\dots, k$. Using \cite[Lemma 6.2]{GwynneKullmann2012SlurJ}, we have $\phardness(A_k) = \hardness(A_k) = k$, and the separation $\Urefc_k \subset \Propc_{k+1}$ (by $A_{k+1}$) follows. Finally, let $A_k' \in \Sat$ be obtained from $A_k$ by adding to all clauses the new positive literals $k+1$; note that the literal $k+1$ is forced for $A_k'$. We have $\hardness(A_k') = \hardness(A_k) = k$, since for every partial assignment $\vp$ with $\vp * A_k' \in \Usat$ we must have $\vp(k+1) = 0$. But $\phardness(A_k') \ge k+1$, since $\rk_k(\epa * A_k') = A_k'$ (note that all clauses of $A_k'$ have length $k+1$). \Qed
\end{prf}

We complete these general considerations on propagation hardness by a general class of examples of clause-sets in $\Propc$, which correspond to the ``support encoding'' of binary constraints, as introduced by \cite{Kasif1990SupportEncoding} and further explored in \cite{Gent2002ArcConsistency} (see Subsection \ref{sec:introAC} for a discussion of translating constraints into SAT):
\begin{examp}\label{exp:suppenc}
  Consider a boolean function $f$, $V := \var(f)$, such that there are $A, B \sse V$ with $A \cap B = \es$ and $A \cup B = V$, and such that for every $\vp \in \Tass(V)$ with $f(\vp) = 1$ precisely one $A$-variable and one $B$-variable is true, that is, there are $a \in A$ and $b \in B$ with $\vp(a) = \vp(b) = 1$, while for all $v \in V \sm \set{a,b}$ holds $\vp(v) = 0$. The following clause-set $F \in \Propc$ with $\ell(F) = O(n(f)^2)$ is logically equivalent to $f$:
  \begin{enumerate}
  \item ALO-clauses (``at least one''): $A, B \in F$.
  \item AMO-clauses (``at most one''): $\set{\ol{a_1},\ol{a_2}}, \set{\ol{b_1},\ol{b_2}} \in F$ for $a_1, a_2 \in A$, $b_1, b_2 \in B$, with $a_1 \ne a_2$ and $b_1 \ne b_2$.
  \item For $a \in A$ let $S$ (the ``support'') be the set of $b \in B$ such that $\vp \in \Tass(V)$ with $f(\vp) = 1$ and $\vp(a) = \vp(b) = 1$ exists; now $\set{\ol{a}} \cup S \in F$.
  \item Similarly, for $b \in B$ let $S$ be the set of $a \in A$ such that $\vp \in \Tass(V)$ with $f(\vp) = 1$ and $\vp(b) = \vp(a) = 1$ exists; now $\set{\ol{b}} \cup S \in F$.
  \end{enumerate}
  Both properties ($F$ is logically equivalent to $f$, and $F \in \Propc$) are easy to verify.
\end{examp}

\subsection{Acyclicity}
\label{sec:Acyclicity}

Recall that a clause-set $F$ has no forced assignments (at all) (i.e., $\frl(F) = \es$) if and only if all prime implicates of $F$ have length at least $2$.\footnote{The ``at all'' is for the case $F = \set{\bot}$, where every literal is forced for $F$, but $F$ has no literals.} Before proving the main lemma (Lemma \ref{lem:acyclpropc}), we need a simple characterisation of clause-sets without forced assignments. Recall that a partial assignment $\vp$ is an autarky for $F \in \Cls$ iff for all $C \in F$ with $\var(\vp) \cap \var(C) \ne \es$ holds $\vp * \set{C} = \top$; for an autarky $\vp$ for $F$ the (sub-)clause-set $\vp * F$ is satisfiable iff $F$ is satisfiable. See \cite{Kullmann2007HandbuchMU} for the general theory of autarkies (but we need only the above definition and basic property).
\begin{lem}\label{lem:characnofa}
  For $F \in \Cls$ and $x \in \Lit$ holds $x \notin \frl(F)$ if and only if $F$ is satisfiable, and there is an autarky $\vp$ for $F$ with $\vp(x)=0$.
\end{lem}
\begin{prf}
If $x$ is not a forced literal for $F$, then $F$ is satisfiable (otherwise every literal is forced), and there is a satisfying assignment $\vp$ for $F$ with $\vp(x)=0$, while satisfying assignments are autarkies. For the other direction let $F$ be satisfiable, and assume there is an autarky $\vp$ for $F$ with $\vp(x)=0$. If $F$ had the forced literal $x$, then $\pao x0 * F$ would be unsatisfiable, while by the autarky condition $\vp * F$ would be satisfiable. \Qed
\end{prf}

In the rest of this section we show that having an ``acyclic incidence graph'' yields a sufficient criterion for $\bc_{i \in I} F_i \in \Propc_k$ for clause-sets $F_i \in \Propc_k$. A ``graph'' $G$ is a pair $G = (V,E)$, where $V$ is the (finite) set of ``vertices'', while $E$, the edge-set, is a set of 2-element subsets of $V$. A ``hypergraph'' $G$ is a pair $G = (V, E)$, where again $V$ is the (finite) set of ``vertices'', while $E$, the hyperedge-set, is an arbitrary set of subsets of $V$.

\begin{defi}\label{def:incgr}
  For a finite family $(F_i)_{i \in I}$ of clause-sets $F_i \in \Cls$ the \textbf{in\-ci\-den\-ce graph} $B((F_i)_{i \in I})$ is the bipartite graph, where the two parts are given by $\bc_{i \in I} \var(F_i)$ and $I$, while there is an edge between $v$ and $i$ if $v \in \var(F_i)$. We say that \textbf{$(F_i)_{i \in I}$ is acyclic} if $B((F_i)_{i \in I})$ is acyclic (has no cycle, i.e., is a forest). A single clause-set \textbf{$F \in \Cls$ is acyclic} if $(\set{C})_{C \in F}$ is acyclic.
\end{defi}

From the family $(F_i)_{i \in I}$ of clause-sets we can derive the hypergraph $G := (\bc_{i \in I} \var(F_i), \set{\var(F_i) : i \in I})$, whose hyperedges are the variable-sets of the $F_i$. Now $(F_i)_{i \in I}$ is acyclic iff $G$ is ``Berge-acyclic'' (which just means that the bipartite incidence graph of $G$ is acyclic). The standard notion of a constraint satisfaction instance being acyclic, as defined in Subsection 2.4 in \cite{GottlobSzeider2008FPT}, is ``$\alpha$-acyclicity'' of the corresponding ``formula hypergraph'' (as with $G$, given by the variable-sets of the constraints), which is a more general notion.

Since the property of the incidence graph being acyclic only depends on the occurrences of variables, if $(F_i)_{i \in I}$ is acyclic, then this is maintained by applying partial assignments and by adding new variables to each $F_i$:
\begin{lem}\label{lem:mainacyc}
  Consider an acyclic family $(F_i)_{i \in I}$ of clause-sets.
  \begin{enumerate}
  \item\label{lem:mainacyc1} For every family $(\vp_i)_{i \in I}$ of partial assignments the family $(\vp_i * F_i)_{i \in I}$ is acyclic.
  \item\label{lem:mainacyc2} Every family $(F'_i)_{i \in I}$ with $\var(F_i') \supseteq \var(F_i)$ and $(\var(F_i') \sm \var(F_i)) \cap (\var(F_j') \sm \var(F_j)) = \es$ for all $i,j \in I$, $i \ne j$, is acyclic.
  \end{enumerate}
\end{lem}

We are ready to prove that an acyclic union $\bc_{i \in I} F_i$ of clause-sets without forced assignments has itself no forced assignments. This is kind of folklore in the CSP-literature, but is (apparently) always stated in terms of search (as an algorithmic property), and never stated explicitly, that is, not stated as a semantical property (which has nothing to do with algorithms). Namely the general ``meta theorem'' is, that search for a solution, when done properly, and the underlying (hyper-)graph is ``sufficiently acyclic'' together with sufficient ``local consistency'' of the constraints (the $F_i$), can proceed without backtracking.\footnote{For example with \href{http://en.wikipedia.org/wiki/Local_consistency}{Wikipedia: Local consistency} we find ``Indeed, if the constraints are binary and form an acyclic graph, values can always be propagated across constraints: for every value of a variable, all variables in a constraint with it have a value satisfying that constraint. As a result, a solution can be found by iteratively choosing an unassigned variable and recursively propagating across constraints. This algorithm never tries to assign a value to a variable that is already assigned, as that would imply the existence of cycles in the network of constraints.''} In \cite{Freuder1982BacktrackFreeTree,Freuder1990KTreeCSP} this is studied for binary CSPs (i.e., $n(F_i) = 2$ for all $i \in I$), while in \cite{JanssenJegouNouguierVilarem1989Filtering,PangGoodwin1997BacktrackFreeSearch} these considerations are generalised to non-binary CSPs; see \cite{Dechter2006Tractable} for an overview. Due to the importance of this basic result, we provide a self-contained (complete) proof. The idea is simple: any assignment to a (single) variable in some $F_i$ can be extended to a satisfying assignment $\vp_i$ of $F_i$ (that is just what $\frl(F_i) = \es$ means), which sets single variables in other $F_j$, which can again be extended to satisfying assignments, and so on, and due to acyclicity never two or more variables are set in some $F_j$.

\begin{lem}\label{lem:acyclpropc}
  For an acyclic family $(F_i)_{i \in I}$ of clause-sets with $\bc_{i \in I} \frl(F_i) = \es$ we have $\frl(\bc_{i \in I} F_i) = \es$.
\end{lem}
\begin{prf}
Let $F := \bc_{i \in I} F_i$. We consider $x \in \lit(F)$ and show that $\pao x0 * F$ can be extended to an autarky for $F$; the assertion then follows by Lemma \ref{lem:characnofa}. We use the following simple property of acyclic graphs $G$: if $V \sse V(G)$ is a connected set of vertices and $v \in V(G) \sm V$, then there is at most one vertex in $V$ adjacent to $v$ (since otherwise there would be a cycle in $G$). Let $G := B((F_i)_{i \in I})$ (so the vertices are $\var(F)$ on the one side and $I$ on the other side); as usual we assume w.l.o.g.\ $\var(F) \cap I = \es$.

Choose $i_0 \in I$ with $\var(x) \in \var(F_{i_0})$. For $J \sse I$ we use $F_J := \bc_{i \in J} F_i$ (the clause-set corresponding to $J$) and $J' := \var(F_J) \cup J \sse V(G)$ (the closed neighbourhood of $J$, i.e., the vertex-set consisting of $J$ and the adjacent variables). Consider a maximal $J \sse I$ with the three properties:
\begin{enumerate}
\item $i_0 \in J$;
\item the vertex set $J'$ is connected in $G$;
\item there is a partial assignment $\vp$ with
  \begin{enumerate}
  \item $\var(\vp) = \var(F_J)$
  \item $\vp(x) = 0$
  \item $\vp * F_J = \top$.
  \end{enumerate}
\end{enumerate}
$\set{i_0}$ fulfils these three properties (since $x$ is not forced for $F_{i_0}$), and so there is such a maximal $J$. If there is no $i \in I \sm J$ adjacent to some variable in $J'$, then $\vp$ is an autarky for $F$ and we are done; so assume there is such an $i \in I \sm J$. According to the above property of the acyclic graph $G$ there is exactly one $v \in J'$ adjacent to $i$, that is, $\var(F_i) \cap \var(F_J) = \set{v}$. Since $F_i$ has no forced assignments, there is a partial assignment $\vp'$ with $\var(\vp') = \var(F_i)$, $\vp'(v) = \vp(v)$ and $\vp' * F_i = \top$. Now $\vp \cup \vp'$ satisfies $F' \cup F_i$, and thus $J \cup \set{i}$ satisfies the three conditions, contradicting the maximality of $J$. \Qed
\end{prf}

Lemma \ref{lem:acyclpropc} only depends on the boolean functions underlying the clause-sets $F_i$, and thus could be formulated more generally for boolean functions $f_i$.

\begin{examp}
  For arbitrary families $(F_i)_{i \in I}$ of clause-sets holds $\bc_{i \in I} \frl(F_i) \sse \frl(\bc_{i \in I} F_i)$, but also for acyclic families we do not have $\frl(\bc_{i \in I} F_i) \sse \bc_{i \in I} \frl(F_i)$, as the following two examples show (in each case an acyclic family with two clause-sets):
  \begin{enumerate}
  \item $\frl(\set{\set{1}}) = \set{1}$ and $\frl(\set{\set{-1}}) = \set{-1}$, but $\frl(\set{\set{1}} \cup \set{\set{-1}}) = \Lit$.
  \item A satisfiable example is given by $\frl(\set{\set{1}}) = \set{1}$ and $\frl(\set{\set{-1,2}}) = \es$, but $\frl(\set{\set{1}} \cup \set{\set{-1,2}}) = \set{1,2}$.
  \end{enumerate}
\end{examp}

We obtain a sufficient criterion for the union of unit-propagation complete clause-sets to be itself unit-propagation complete:
\begin{thm}\label{thm:acylcprop}
  Consider $k \in \NNZ$ and an acyclic family $(F_i)_{i \in I}$ of clause-sets. If for all $i \in I$ we have $F_i \in \Propc_k$, then also $\bc_{i \in I} F_i \in \Propc_k$.
\end{thm}
\begin{prf}
Let $F := \bc_{i \in I} F_i$, and consider a partial assignment $\vp$ with $F' \ne \set{\bot}$ for $F' := \rk_k(\vp * F)$. We have to show that $F'$ has no forced assignments. For all $i \in I$ we have $\rk_k(\vp * F_i) \ne \set{\bot}$, and thus $\rk_k(\vp * F_i)$ has no forced assignments (since $F_i \in \Propc_k$). So $\bc_{i \in I} \rk_k(\vp * F_i)$ has no forced assignments by Lemma \ref{lem:acyclpropc}. Thus $F' = \rk_k(\bc_{i \in I} \vp * F_i) = \rk_k(\bc_{i \in I} \rk_k(\vp * F_i)) = \bc_{i \in I} \rk_k(\vp * F_i)$, whence $F'$ has no forced assignments. \Qed
\end{prf}

We note that Theorem \ref{thm:acylcprop} is similar in spirit to statements from \cite{Freuder1982BacktrackFreeTree,Freuder1990KTreeCSP,JanssenJegouNouguierVilarem1989Filtering,PangGoodwin1997BacktrackFreeSearch} of the sort: ``if the constraints are locally consistent of some degree, and the constraint hypergraph is acyclic, then some basic scheme can find a solution quickly'', but the reductions $\rk_k$ are different from local consistency notions in the CSP-literature (which are closer related to width as investigated in Subsection \ref{sec:prelimwhdWC}), and the ``declarative'' definition of the classes $\Propc_k$ is fundamentally different to the procedural approach related to local consistency notions. Since a singleton-clause-set is in $\Propc$, we obtain:
\begin{corol}\label{cor:acycpropc}
  If $F \in \Cls$ is acyclic, then $F \in \Propc$.
\end{corol}

Theorem \ref{thm:acylcprop} yields an upper bound on the p-hardness of an acyclic union, but in general we do not have equality:
\begin{examp}\label{exp:phdacycl}
  By Theorem \ref{thm:acylcprop} we have for acyclic families $(F_i)_{i \in I}$, $I \ne \es$, the inequality $\phardness(\bc_{i \in I} F_i) \le \max_{i \in I} \phardness(F_i)$, but equality does not hold in general:
  \begin{enumerate}
  \item Let $I := \set{1,2}$, $F_1 := \set{\set{a}}$ and $F_2 := \set{\set{a,b},\set{a,\ol{b}}}$.
  \item $(F_1, F_2)$ is acyclic (the incidence graph has two vertices in each of the two parts, but one edge is missing, since $b \notin \var(F_1)$).
  \item $\frl(F_1) = \frl(F_2) = \set{a}$.
  \item $\phardness(F_1) = 1$, $\phardness(F_2) = 2$, while $\phardness(F_1 \cup F_2) = \phardness(F_1) = 1$.
  \end{enumerate}
\end{examp}

The conditions for $B((F_i)_{i \in I})$ being acyclic, which are relevant to us, are collected in the following lemma; they are in fact pure graph-theoretical statements on the acyclicity of bipartite graphs, but for concreteness we formulate them in terms of families of clause-sets:
\begin{lem}\label{lem:characacycl}
  Consider a family $(F_i)_{i \in I}$ of clause-sets, and let $G := B((F_i)_{i \in I})$.
  \begin{enumerate}
  \item\label{lem:characacycla} If there are $i, j \in I$, $i \ne j$, with $\abs{\var(F_i) \cap \var(F_j)} \ge 2$, then $G$ is not acyclic.
  \item\label{lem:characacyclb} Assume that for all $i, j \in I$, $i \ne j$, holds $\abs{\var(F_i) \cap \var(F_j)} \le 1$. If the ``variable-interaction graph'', with vertex-set $I$, while there is an edge between $i, j \in I$ with $i \ne j$ if $\var(F_i) \cap \var(F_j) \ne \es$, is acyclic, then $G$ is acyclic.
  \item\label{lem:characacyclc} If there is a variable $v$, such that for $i, j \in I$, $i \ne j$, holds $\var(F_i) \cap \var(F_j) \sse \set{v}$, then $G$ is acyclic.
  \end{enumerate}
\end{lem}
\begin{prf}
For Part \ref{lem:characacycla} note that $i, j$ together with $v, w \in \var(F_i) \cap \var(F_j)$, $v \ne w$, yield a cycle (of length $4$) in $G$. For Part \ref{lem:characacyclb} assume $G$ has a cycle $C$ (which must be of even length $m \ge 4$). The case $m = 4$ is not possible, since different clause-sets have at most one common variable, and thus $m \ge 6$. Leaving out the interconnecting variables in $C$, we obtain a cycle of length $m / 2$ in the variable-interaction graph. Finally for Part \ref{lem:characacyclc} it is obvious that $G$ can not have a cycle $C$, since the length of $C$ needed to be at least $4$, which is not possible, since the only possible vertex in it would be $v$. \Qed
\end{prf}

For ease of access, we explicitly state the application of the two acyclicity conditions of Lemma \ref{lem:characacycl} to Theorem \ref{thm:acylcprop}:
\begin{corol}\label{cor:1acylcprop}
  Consider $k \in \NNZ$ and a family $(F_i)_{i \in I}$ of clause-sets with $F_i \in \Propc_k$ for all $i \in I$. Then each of the following conditions implies $\bc_{i \in I} F_i \in \Propc_k$:
  \begin{enumerate}
  \item\label{cor:1acylcprop1} Any two different clause-sets have at most one variable in common, and the variable-interaction graph is acyclic.
  \item\label{cor:1acylcprop2} There is a variable $v \in \Va$ with $\var(F_i) \cap \var(F_j) \sse \set{v}$ for all $i,j \in I$, $i \ne j$.
  \end{enumerate}
\end{corol}

The following examples show that the conditions of Corollary \ref{cor:1acylcprop} can not be improved in general:
\begin{examp}\label{exp:1va}
  An example for three boolean functions without forced assignments, where each pair has exactly one variable in common, while the variable-interaction graph has a cycle, and the union is unsatisfiable, is $a \oplus b = 0, \ a \oplus c = 0, \ b \oplus c = 1$. And if there are two variables in common, then also without a cycle we can obtain unsatisfiability, as $a \oplus b = 0, \ a \oplus b = 1$ shows. The latter family of two boolean functions yields also an example for a family of two clause-sets where none of them has forced assignments, while the union has (is in fact unsatisfiable). Since a hypergraph with two hyperedges is ``$\gamma$-acyclic'', in the fundamental Lemma \ref{lem:acyclpropc} we thus can not use any of the more general notions ``$\alpha$/$\beta$/$\gamma$-acyclicity'' (see \cite{Fagin1983Azyklisch} for these four basic notions of ``acyclic hypergraphs'', and see \cite{OrdyniakPaulusmaSzeider2013Acyclic} for a recent study regarding SAT-decision).
\end{examp}

\section{CNF-representations of boolean functions}
\label{sec:cnfrep}

We have now all the notions and concepts together to discuss the concept of a ``CNF-representation of a boolean function'', in the context of the terminology used in the literature. In Subsection \ref{sec:cnfreprep} we define the fundamental notion of a ``CNF-representation'' of a boolean function. In Subsection \ref{sec:disenc} we discuss why we use ``representation'', and not ``encoding''. Various characterisations of CNF-representations are given in Subsection \ref{sec:characrepr}. In Subsection \ref{sec:measqurep} we apply our hardness-measures to CNF-representations, obtaining definitions of ``GAC-'' and ``UR-''representations. The relations to the constraint-satisfaction literature (and the notion of ``arc-consistency'') are discussed in depth in Subsection \ref{sec:introAC}. In Subsection \ref{sec:cnfrepforcing} we strengthen GAC-representations to ``forcing representations''. Finally in Subsection \ref{sec:remTseit} we review the Tseitin-translation and its properties.

\subsection{Representations}
\label{sec:cnfreprep}

We consider the general task of representing boolean functions $f: \Tass(V) \ra \set{0,1}$ (for some finite $V \subset \Va$). The clause-sets $F$ equivalent to $f$ are characterised by $\primec_0(f) = \primec_0(F)$, and they are the obvious first candidates for representing $f$. The representation $\primec_0(f)$ has optimal inference power, but in most cases it is too big. Even when ignoring inference power and just considering equivalent $F$, in many interesting cases these $F$ too big, for example even a single XOR-constraint requires exponential size without auxiliary variables, as is well-known and we will see in Subsection \ref{sec:transx0}. We now turn to a more general notion of ``representation'' of boolean functions, allowing auxiliary variables:
\begin{defi}\label{def:rep}
  A \textbf{CNF-representation} of a boolean function $f$ is a clause-set $F$ with $\var(F) \supseteq \var(f)$, such that for $\vp \in \Tass(\var(f))$ the clause-set $\vp * F$ is satisfiable if and only if $f(\vp) = 1$. The elements of $\var(f)$ are the \textbf{primary} (or \textbf{original}) variables, the elements of $\var(F) \sm \var(f)$ the \textbf{auxiliary} variables.
\end{defi}
In this \Schrift{} we only speak about CNF-representations, and thus often we will leave out the ``CNF''.  A clause-set $F$ is a CNF-representation of a boolean function $f$ iff the satisfying assignments of $F$ projected to $\var(f)$ are \emph{precisely} the satisfying assignments of $f$ (note that this allows that satisfying assignments for $f$ can have multiple extensions to satisfying assignments of $F$). This is exactly what allows $F$ to be used in the context of representations of other constraints: no satisfying assignments are added or removed. If $F$ is a CNF-representation of $f$, and $F'$ is a CNF-representation of $F$ (i.e., of the boolean function underlying $F$ as a CNF), then $F'$ is also a CNF-representation of $f$.

\begin{examp}\label{exp:repr}
  For $F \in \Cls$ we obtain a representation $F' \in \Pcls{3}$ by repeated applications of the well-known trick of breaking up a clause $C = \set{x_1,\dots,x_m} \in \Cl$, $\abs{C} = m \ge 4$, into two shorter clauses $C', C'' \in \Cl$ by introducing a new variable $v$, and letting $C' := \set{x_1,\dots,x_{\floor{m/2}}, v}$ and $C'' := \set{\ol{v}, x_{\floor{m/2}+1}, \dots, x_m}$.

  More generally we consider ``inverse DP-reduction''. For $v \in \Va$ and $F \in \Cls$ we have $\dpi{v}(F) := \set{C \in F : v \notin \var(C)} \cup \set{C \res D : C, D \in F, C \cap \ol{D} = \set{v}}$ (this is called ``DP-reduction'', since it it the original DP-procedure \cite{DP60} in one step; it is also called ``variable elimination''). It is well-known that $\dpi{v}(F)$ is logically equivalent to the existential quantification of $v$ in $F$. Thus, for $F, F' \in \Cls$ and $v \in \Va$ such that $\dpi{v}(F') = F$, we have that $F'$ is a representation of $F$. The proof of this is also very simple and instructive: By definition we have for every partial assignment $\vp$ with $\vp * F' = \top$ also $\vp * F \ne \top$. That on the other hand for every partial assignment $\vp \in \Pass(\var(F))$ with $\vp * F = \top$ there is $\ve \in \set{0,1}$, such that for $\vp' := \vp \cup \pao v{\ve}$ holds $\vp' * F' = \top$ is left as an easy exercise.

  This general mechanism of obtaining representations by (repeated applications of) inverse DP-reductions covers the above replacement of clause $C$ by clauses $C',C''$ (since $\dpi{v}(\set{C',C''}) = \set{C}$). It also covers addition of blocked clauses (see \cite{Ku96c}, especially the proof of Lemma 4.1 there), which includes ``Tseitin extensions'', i.e., the extension of a clause-set via additions of clause-sets equivalent to ``definitions'' $v \lra b$ , where $b$ is a boolean function with $v \notin \var(b)$ (this is the essential step of an Extended Resolution refutation as introduced in \cite{Ts68}, and discussed in general in \cite{BubeckBuening2010Definitions}). A special case of such extensions can be seen in the Tseitin translation of a circuit to a CNF, as discussed in Subsection \ref{sec:remTseit}.

  More general preprocessing of clause-sets needs more general transformations between satisfying assignments than just projection; see \cite{JaervisaloHeuleBiere2012Inprocessing} for a recent study.
\end{examp}

\subsection{``Encoding'' versus ``representation''}
\label{sec:disenc}

The first terminological problem that needs discussion is that we use ``representation'' instead of the often used ``encoding''. The issue is that ``encoding'' has been used with various different meanings.\footnote{It is curious that in the whole literature of CSP and SAT there seems to be not a single attempt at defining what ``encoding'' precisely could mean. Perhaps the point is that ``encoding'' just means to represent the original problem ``somehow'', ``appropriately''.} On the one hand, there is the issue of translating non-boolean variables into boolean variables, which can be handled in many ways, and which we mostly ignore in this \Schrift{} (we consider only boolean variables; but see Subsection \ref{sec:introAC} for a discussion of non-boolean variables). If we would allow arbitrary re-encodings of boolean variables, then every boolean function $f$ would have a trivial (``wild'') ``encoding'' --- just sort the satisfying assignments of $f$, so that in the ``encoding'' in lexicographical order we first have all satisfying assignments, and then all falsifying.

 But still, even when considering just boolean variables and without changing their meaning, there are various interpretations of ``encoding''. If just a whole problem is to be translated, then ``encoding'' is used to denote just ``satisfiability equivalence'' (or ``equi-satisfiability''). See \cite{HertelHertelUrquhart2007Dangerous} for a general reflection on ``good'' versus ``bad'' encodings from a proof-theoretic perspective. Now such ``encodings'' can not be combined with other ``encodings'' in general (since the ``constraints'' are not translated properly), and thus when translating just (single) ``constraints'', then something close to a representation in our sense must be used. In this sense ``good'' CNF-representations can be considered as a special case of ``Knowledge Compilation'' (KC) --- the \emph{complete} knowledge (boolean function) has to be represented (now in a form ``good'' for SAT solvers).\footnote{\cite{BarahonaJungKatsirelosWalsh2008EncodingDNNF} is an example for this gone wrong, namely \cite{BordeauxJanotaMarquesSilvaMarquis2012UC} use the ``encoding'' of DNNF into CNF from \cite{BarahonaJungKatsirelosWalsh2008EncodingDNNF} in their Proposition 4 for the purpose of KC, while that ``encoding'' actually may loose solutions in general.} For example \cite{Een2006Translating}, and more explicitly \cite{BailleuxBoufkhadRoussel2009PBCNF}, use ``encoding'' in the sense of representation.\footnote{\cite{BailleuxBoufkhadRoussel2009PBCNF} also speaks about non-boolean variables, but since ``extension'' and ``projection'' are not specified, the ``wild'' encoding mentioned above is not excluded} But there are also uses in a more strict sense, where the extensions to the auxiliary variables must be unique, which is needed for example when \emph{counting} satisfying assignments.

So ``encoding'' is treated as a generic term, without precise meaning, while we use ``CNF-representation'' for a special form of ``encoding'', where the primary variables are kept, the satisfying assignments are preserved (modulo the extension to the auxiliary variables), and we have just a relation between a boolean function and a clause-set, without reference to some computation (as it is usually understood when using ``encoding'').

\subsection{Characterising representations}
\label{sec:characrepr}

We note that a boolean function $f$ has a CNF-representation iff $f \ne 1^V$ for some non-empty $V$ (since as soon there is a clause, there is a falsifying assignment). This is a slight weakness of our formalism of handling clause-sets, which do not allow ``formal'' variables, but only variables actually occurring. More generally, since we require $\var(f) \sse \var(F)$, ``inessential'' variables of $f$ (variables $v \in \var(f)$ which never influence $f$) need possibly be added to a representation:
\begin{examp}\label{exp:repreprimec}
   Consider an arbitrary boolean function $f$, and let $F := \primec_0(f)$.
   \begin{enumerate}
   \item Thus we have $\var(F) \sse \var(f)$.
   \item The inessential variables of $f$ are precisely given by $\var(f) \sm \var(F)$.
   \item In case of $F \ne \top$ we obtain a CNF-representation $F'$ of $f$ with $\var(F') = \var(F)$ by choosing some $C \in F$ and defining $F' := F \cup \set{C \cup (\var(f) \sm \var(F))}$.
   \end{enumerate}
\end{examp}

In our context, this problem with handling inessential variables is of no relevance, and does not warrant the effort of dealing with formal clause-sets instead (pairs $(V,F)$ with $F \in \Cls$ and some finite $V \subset \Va$ with $\var(F) \sse V$).

CNF-representations $F$ of boolean functions $f$ correspond precisely to existentially quantified CNFs as a mechanism for representing boolean functions, as studied in \cite{BubeckBuening2010Definitions}, that is, $f = \ex (\var(F) \sm \var(f)) \, F$, using equality of boolean functions in the ordinary sense of equality of maps, while the existential quantifiers ranges over (all) the auxiliary variables (compare Example \ref{exp:repr}). The notion of ``auxiliary variables'' as defined in \cite[Definition 2]{BubeckBuening2010Definitions}, when restricted to existential quantification (there also universal quantification is considered), are precisely our auxiliary variables (modulo small differences in the framework).

How the prime implicates of a representation $F$ relate to the prime implicates of the represented $f$ is determined in the following instructive lemma, namely  the prime implicates of $f$ are precisely those prime implicates of $F$ which do not use auxiliary variables. It is useful to extend the notions ``satisfiable / unsatisfiable'' to boolean functions in the obvious sense, i.e., we say that a boolean function $f$ is \emph{satisfiable for a partial assignment $\vp$}, if, after removal from $\vp$ of variables not in $f$, it is possible to extend the remaining partial assignment to a partial assignment $\vp'$ with $f(\vp') = 1$, while otherwise we say that $f$ is \emph{unsatisfiable for $\vp$}.
\begin{lem}\label{lem:characCNFrep}
  Consider a boolean function $f$ and $F \in \Cls$ with $\var(f) \sse \var(F)$. The following statements are equivalent:
  \begin{enumerate}
  \item\label{lem:characCNFrep1} $F$ is a CNF-representation of $f$.
  \item\label{lem:characCNFrep2} $\fa\, \vp \in \Tass(\var(f)) : f(\vp) = 1 \Lra \vp * F \in \Sat$.
  \item\label{lem:characCNFrep3} $\fa\, \vp \in \Tass(\var(f)) : f(\vp) = 0 \Lra \vp * F \in \Usat$.
  \item\label{lem:characCNFrep4} $\fa\, \vp \in \Pass(\var(f)) : f \text{ is satisfiable for $\vp$} \Lra \vp * F \in \Sat$.
  \item\label{lem:characCNFrep5} $\fa\, \vp \in \Pass(\var(f)) : f \text{ is unsatisfiable for $\vp$} \Lra \vp * F \in \Usat$.
  \item\label{lem:characCNFrep6} $\fa\, C \in \Cl, \var(C) \sse \var(f) : f \models C \Lra F \models C$.
  \item\label{lem:characCNFrep7} $\primec_0(f) = \set{C \in \primec_0(F) : \var(C) \sse \var(f)}$.
  \end{enumerate}
\end{lem}
\begin{prf}
The equivalence of Statements \ref{lem:characCNFrep1} -- \ref{lem:characCNFrep5} follows by definition. The equivalence of Statements \ref{lem:characCNFrep5} and \ref{lem:characCNFrep6} follows by the simple equivalences $f \models C$ iff $f$ is unsatisfiable for $\vp_C$, and $F \models C \Lra \vp_C * F \in \Usat$. Finally the equivalence of Statement \ref{lem:characCNFrep6} and \ref{lem:characCNFrep7} follows by definition again. \Qed
\end{prf}

So, ignoring the issue about nonessential variables, we obtain all representations of $f$ by choosing some $F \in \Cls$ such that the set of minimal clauses $C$ with $\var(C) \sse \var(f)$ obtained from $F$ via resolution is (precisely) $\primec_0(f)$. One aspect of a ``good'' $F$ is its size, i.e., $c(F)$ or $\ell(F)$ should be ``small''. The second aspect is the inference power of $F$, discussed in the subsequent subsection --- the main task is to make the resolution refutations of the clauses of $\primec_0(f)$ as ``simple'' as possible.

\subsection{Measuring inference strength}
\label{sec:measqurep}

We use the measures $\hardness$ (Definition \ref{def:charachd}), $\whardness$ (Definition \ref{def:whd}) and $\phardness$ (Definition \ref{def:phardness}) to measure the inference-strength of representations $F$ of boolean functions $f$, either under the ``absolute'' or the ``relative'' condition:
\begin{description}
\item[Absolute condition] $\whardness(F) \le \hardness(F) \le \phardness(F)$.
\item[Relative condition] $\whardness^V(F) \le \hardness^V(F) \le \phardness^V(F)$ for $V := \var(f)$.
\end{description}
Note that the absolute condition only plays a role when measuring inference power, while for just short representations only the relative condition makes sense. The task of ``good representations'' $F$ is to find a good balance between the size of $F$ and having these measures as small as possible. Yet in the literature only the relative condition has been considered, and for the two most prominent cases we introduce special terminology:
\begin{defi}\label{def:specrep}
  A representation $F$ of a boolean function $f$ is called
  \begin{itemize}
  \item \textbf{GAC-representation} if $\phardness^{\var(f)}(F)) \le 1$;
  \item \textbf{UR-representation} if $\hardness^{\var(f)}(F)) \le 1$.
  \end{itemize}
\end{defi}
``GAC'' reminds of ``generalised arc-consistency'', while ``UR'' reminds of ``unit refutation''. In \cite{GwynneKullmann2013GoodRepresentationsIILata} we used ``AC'' instead of ``GAC'', but this turned out to be too ambiguous. In the language of \cite{BKNW2009CircuitComplexity} one could say that a representation $F$ of $f$ with $\phardness^{\var(f)}(F)) \le k$ achieves complete propagation via $\rk_k$, while in case of $\hardness^{\var(f)}(F)) \le k$ we have a complete consistency-checker (complete detection of dis-entailment) via $\rk_k$, however the concepts of \cite{BKNW2009CircuitComplexity} are in fact somewhat weaker, as discussed in Subsection \ref{sec:ComparisonBessiere}.

A clause-set $F$ represents for every $V \sse \var(F)$ the boolean function obtained by projecting the satisfying assignments of $F$ to $V$, and $F$ is propagation-complete (i.e., $F \in \Propc$) iff $F$ is a GAC-representation for all these boolean functions.

\subsection{Discussion of CSP-literature and ``arc-consistency''}
\label{sec:introAC}

Recall that a constraint is ``hyperarc-consistent'' or ``generalised arc-consistent'' iff for each variable each value in its (current) domain is still available (does not yield an inconsistency); see Chapter 3 of \cite{RBW2006HandbookCSP}. There are various algorithms for achieving generalised arc-consistency of a constraint, and the dynamic version is ``maintaining arc-consistency'' (MAC), where (generalised) arc-consistency is established at the nodes of a search tree of a constraint solver, which uses restriction of variable-domains for branching. In \cite{Gent2002ArcConsistency} the idea of using SAT encodings and unit-clause propagation to maintain arc-consistency has been introduced. If the non-boolean variables are arbitrarily encoded into boolean variables, then it is very difficult in general to establish a precise relation between the actions of the constraint-solver and the actions of the SAT-solver. However when using the so-called ``direct variable-encoding'', then there is a precise correspondence between the two sides, and this variable-encoding is thus used throughout these investigations on the use of SAT-solving for CSP-solving. Before reviewing these investigations, we review the basics of CSP, within our framework.

The semantic aspect of a boolean clause-set is the (underlying) boolean function $f$, and accordingly the semantic aspect of a ``constraint'' is the (underlying) ``non-boolean function'' (using an abuse of language) defined as follows.

To a variable $v \in \Va$ associate the domain $D_v$, a finite non-empty set. For a finite $V \subset \Va$, the set of \emph{non-boolean functions} $f$ for $(D_v)_{v \in V}$ is the set of all maps $f: \Tass((D_v)_{v \in V}) \ra \set{0,1}$, where $\Tass((D_v)_{v \in V}) := \prod_{v \in V} D_v$ is the set of all maps $\vp$ with $\dom(\vp) = V$ such that $\vp(v) \in D_v$ for all $v \in V$; as usual we use $n(f) := \abs{V}$ for the number of variables. To handle the ``direct'' variable-translation, we assume that for $v \in \Va$ and $\ve \in D_v$ we have $(v,\ve) \in \Va$, where the underlying meaning of variable $(v,\ve)$ is ``$v = \ve$''. Then the non-boolean functions $f$ for $(D_v)_{v \in V}$ correspond 1-1 to those boolean functions $f'$ with $\var(f') = \bc_{v \in V} \set{v} \times D_v$ and having the property, that for all total assignments $\vp$ with $f'(\vp) = 1$ and all $v \in V$ there is exactly one $\ve \in D_v$ with $\vp((v,\ve)) = 1$. The boolean functions of Example \ref{exp:suppenc} are precisely the $f'$ corresponding to non-boolean functions $f$ with two variables (``binary constraints'').

``Constraints'' are representations of non-boolean functions $f$ by the set of all $\vp \in \Tass((D_v)_{v \in V})$ with $f(\vp) = 1$.\footnote{More precisely one speaks here about ``extensional constraints'', while ``intensional constraints'' use some other representation.} And what are partial assignments for clause-sets, are ``domain restrictions'' for constraints, assigning to each $v \in V$ a subset $D_v' \sse D_v$. Partial assignments to the original variables in a CNF-representation $F$ of $f'$ correspond 1-1 to domain restrictions for $f$ in the presence of AMO- and ALO-clauses and unit-clause propagation.

When in the literature an ``arc-consistent encoding'' of a constrain is given, then what actually (and more precisely) is achieved, is that for a non-boolean function $f$ a GAC-representation $F$ of $f'$ is presented, where the complexity estimations for the computation of $F$ use the constraint-representation of $f$. This point of view has the advantage to spell out what precisely are the properties, and it separates the five fundamental aspects:
\begin{enumerate}
\item the semantical aspect, the underlying non-boolean function $f$,
\item the CSP-representation of $f$ (the constraint),
\item the variable-encoding (the ``direct variable-encoding''),
\item the boolean function $f'$ obtained from $f$ and the variable-encoding,
\item and finally the representation of $f'$ by a clause-set $F$.
\end{enumerate}
To show ``GAC'' for $F$, in principle we only need to consider $F$ itself, plus the knowledge what is $\var(f')$; if $\var(f') = \var(F)$, then automatically we have a representation in $\Propc$. Based on \cite{Kasif1990SupportEncoding}, in \cite{Gent2002ArcConsistency} the case of binary constraints (i.e., $n(f) = 2$) has been handled: the representation of $f'$ is given in Example \ref{exp:suppenc} (there just called ``$f$''), and since no auxiliary variables are used, we have a representation in $\Propc$. The general case, a GAC-representation for constraints of arbitrary arity, now using auxiliary variables, has been given in \cite{Bacchus2007GAC}, based on \cite{BessiereHebrardWalsh2003LocalEncodings}.

In \cite{Bacchus2007GAC} we furthermore find the following results:
\begin{enumerate}
\item In Section 3 the representation $\primec_0(f') \in \Urefc_0$ is considered, and a special case of $\Urefc_0 \sse \Propc$ is shown.
\item GAC-representations for ``regular constraints'' (expressing regular languages), which as a special case contain cardinality constraints, are given in Subsections 5.1, 5.2.
\item In Subsection 5.3 we find the interesting case of ``generalised sequence constraints'', where the representation $F$ of $f'$ has relative p-hardness $2$ (in our terminology), the only case yet in the literature we are aware of not going for (relative, p-)hardness $1$.
\end{enumerate}

We turn to a discussion of the terminology used in the SAT-literature on translating constraints; here, different from the CSP-literature, the SAT-solving is the main thing (not just an auxiliary device helping to solve CSP-problems). The current terminology in the literature can potentially cause confusion between, and thus some clarifying discussion is needed. Typically the starting point is already a boolean function $f$. Our (new) notion ``GAC-representation $F$ of $f$'' expanded says: ``a representation maintaining hyperarc-consistency (or generalised arc-consistency) via unit-clause propagation for the (single, global) boolean constraint $f$ after (arbitrary) partial assignments to the original variables'', which often is stated just as ``arc-consistent under unit propagation''.

Note that especially for XOR-clause-sets it is tempting to take each XOR-clause as a constraint, but this is not interesting here --- the real interest is in bundling together of XOR-constraints into one XOR-clause-set (a ``single'' boolean function). We also emphasise \emph{global} constraints, which do not need to have bounded arity, as it is assumed for ordinary constraints. That for the forced literals also only the original variables are considered, has been discussed in Example \ref{exp:relphd}.

Apparently the first definition of representations, called ``correct encoding'', and the GAC-condition, called ``efficiency'', is given in \cite{BailleuzBoufkhad2003CardinalityConstraints}, for the special case of cardinality constraints. While the first explicit general definition of ``arc-consistency under unit propagation'' (what we now call GAC-representations) is in \cite{Een2006Translating}, the Definition on Page 5; it is left open there whether the partial assignment $\sigma$ there may also involve the introduced (auxiliary) variables, but it is a kind of automatic assumption to not consider them, since only the variables of the (original) constraint are considered in this context.\footnote{An assumption we challenge by considering $\Propc$ (and in general the ``absolute condition'').} For further examples for pseudo-boolean constraints see Section 22.6.7 in \cite{RM09HBSAT} and \cite{Sinz2005CardinalityConstraints,BailleuxBoufkhadRoussel2009PBCNF}. We prefer to speak of ``GAC-representations'', introducing a partially new terminology, to help precision; it seems superfluous to mention in this context ``unit(-clause) propagation''. One could also say ``GAC-translation'' or``GAC-encoding'', but we reserve ``translation'' for (poly-time) functions computing a representation, and, as discussed in Subsection \ref{sec:disenc}, we use ``(variable-)encoding'' for the translation of non-boolean into boolean variables.

\subsection{Forcing -- considering also satisfying assignments}
\label{sec:cnfrepforcing}

We now turn to what seems the most important concepts regarding \emph{satisfying total} assignments. The motivation is as follows. Consider a representation $F \in \Sat$ of a boolean function $f$. If $F \in \Urefc_k$ for some $k$, then via $\Urefc_k = \Slur_k$ and the generalised SLUR algorithms, as shown in \cite{GwynneKullmann2012SlurJ}, we have an oblivious polytime algorithm (repeating simple steps, in arbitrary fashion, with guaranteed success) for finding a satisfying assignment for $F$. That covers the absolute condition, and we are now concentrating on the relative condition. The strongest guarantee here (of general practical importance) is that $F$ has relative p-hardness $1$, i.e., $F$ is a GAC-representation. We can determine by unit-clause propagation whether $F$ is satisfiable, by $F \in \Sat \Lra \ro(F) \ne \set{\bot}$, but how to find a satisfying assignment? By repeating the process ``perform unit-clause propagation, and assign any free variable from $\var(f)$'', starting from the empty partial assignment, we obtain a total assignment $\vp \in \Tass(\var(f))$ with $f(\vp) = 1$. Now how to get a satisfying assignment for $\vp * F$ ? This ability is not of relevance for general KC, because there $F$ is only an auxiliary device (for computing $f(\vp)$), but a SAT solver will stumble over $\vp * F$, and in general $\vp * F$ can be a hard (satisfiable) problem, since no guarantee is given what happens when assigning to auxiliary variables. So we consider the condition that via some reduction $r$ we can reduce $\vp * F$ to $\top$ (where typically this reduction will yield a satisfying assignment).

\begin{defi}\label{def:extension}
  Consider a clause-set $F \in \Cls$ and a finite $V \subset \Va$.
  \begin{itemize}
  \item \textbf{$F$ has sat-recognition via $r: \Cls \ra \Cls$ w.r.t.\ $V$} if for every $\vp \in \Tass(V)$ with $\vp * F \in \Sat$ we have $r(\vp * F) = \top$.
  \item A CNF-representation $F$ of a boolean function $f$ \textbf{has sat-recognition via $r$} if $F$ has this property w.r.t.\ $\var(f)$.
  \end{itemize}
\end{defi}

We have the following characterisation of the condition $\rki(F) = \top$ for $F \in \Cls$: We say that a clause-set $F \in \Sat$ is \emph{uniquely satisfiable mod(ulo) inessential variables} if there is $W \sse \var(F)$ such that the satisfying assignments $\vp \in \Tass(\var(F))$ of $F$ are fixed on $W$, while they are unrestricted on $\var(F) \sm W$ (so that we have exactly $2^{n(F) - \abs{W}}$ satisfying assignments). It is easy to see that $F$ is uniquely satisfiable modulo inessential variables iff $\rki(F) = \top$.\footnote{Recall that a variable $v \in \Va$ is \emph{essential} for a boolean function $f$ iff $v \in \var(f)$ and there is $\vp \in \Tass(\var(f))$ such that for $\vp'$ obtained from $\vp$ by flipping the value of $v$ we have $f(\vp) \ne f(\vp')$; otherwise $v$ is called \emph{inessential}.} We see that having sat-recognition via $\rki$ is a weakening of having unique extension, appropriate for deciding SAT (but not necessarily for counting).

\begin{examp}\label{exp:uniqsatmod}
  $\set{\set{a,b},\set{a}}$ is uniquely satisfiable mod inessential variables.
\end{examp}

If a representation $F$ has sat-recognition via $\rki$ and we have $F \in \Propc_k$, then $F$ has sat-recognition via $\rk_k$. This is a property of the absolute condition, and in general we do \emph{not} have that if $F$ has sat-recognition via $\rki$ w.r.t.\ $V$, then $F$ has sat-recognition via $\rk_k$ w.r.t.\ $V$ for $k := \phardness^V(F)$, since if we take for example $V = \es$, then we have $\phardness^V(F) = 0$ (recall Example \ref{exp:relphd}).

We now come to the central combination of the (practically) strongest relative hardness-condition together with the (practically) strongest sat-recognition-condition:
\begin{defi}\label{def:forciing}
  A clause-set $F$ is \textbf{forcing w.r.t.\ $V \subset \Va$} if
  \begin{itemize}
  \item $\phardness^V(F) \le 1$
  \item $F$ has sat-recognition via $\ro$ w.r.t.\ $V$.
  \end{itemize}
  A \textbf{forcing representation} of a boolean function $f$ is a CNF-representation $F$ of $f$ which is forcing w.r.t.\ $\var(f)$. If additionally $F \in \Propc$ holds, then $F$ is an \textbf{absolute forcing representation} of $f$.
\end{defi}

In other words, a forcing representation is a GAC-representation with sat-recognition via $\ro$. If $F$ is forcing w.r.t.\ $V$, then the following two conditions are fulfilled for all $\vp \in \Pass(V)$:
\begin{enumerate}
\item $\vp * F \in \Usat$ iff $\ro(\vp * F) = \set{\bot}$;
\item $\vp * F$ is uniquely satisfiable mod inessential variables iff $\ro(\vp * F) = \top$.
\end{enumerate}
The main reason why we think that forcing representations are of fundamental importance for SAT solving, more so than mere relative p-hardness 1, is that every representation of relative hardness $1$ can be transformed in polynomial time into a forcing representation, as we will show in Theorem \ref{thm:stronglyforcing}, and furthermore most of our GAC-representations naturally are forcing representations (or close to it).

\begin{examp}\label{exp:rep01}
  The representations of $0^{\es}$ (the constant-0 function with empty var\-iable-set) are precisely the unsatisfiable clause-sets $F$ (here sat-recognition doesn't play a role):
  \begin{itemize}
  \item $F$ is forcing iff $\hardness(F) \le 1$ (i.e., iff $\ro(F) = \set{\bot}$).
  \end{itemize}
  The representations of $1^{\es}$ are precisely the satisfiable clause-sets $F$ (here unsatisfiable sub-instances don't play a role):
  \begin{itemize}
  \item $F$ is forcing iff $F$ has sat-recognition via $\ro$ (i.e., iff $\ro(F) = \top$).
  \end{itemize}
\end{examp}

\subsection{The Tseitin translation, and UP-representations}
\label{sec:remTseit}

Boolean circuits represent boolean functions, and they are translated into CNF-representations via the Tseitin translation, as we will discuss in this subsection. The idea of the Tseitin translation, first mentioned in \cite[Section 1]{Ts68} and fully worked out (for first-order logic) in \cite{GreenbaumPlaisted1986ClauseFormTrans}, with further refinements in \cite{JacksonSheridan2004CircuitstoSAT,Een2006Translating},  is to introduce new variables $v$ for the nodes of the circuit, and to represent the equivalence $v \lra b$, where $b$ is the boolean function corresponding to the gate, via a CNF --- the union of all these CNFs plus the assertion, that the final gate be true, yields the Tseitin translation. One might use only one of the two directions of the equivalence $a \lra b$, if one can determine the ``polarity'', and we will handle this in a simplified form, where only positive polarities are allowed, as the ``reduced Tseitin translation''. We start these reflections by the definition of a ``general circuit''.

A dag (directed acyclic graph) is a pair $G = (V,E)$, with $V$ the (finite) vertex set and $E \sse V^2$ the arc set, where $V(G) := V$ and $E(G) := E$, such that there are no directed cycles. For $v \in V$ let $\income_G(v) := \set{w \in V : (w,v) \in E}$ be the set of vertices with an arc to $v$ (so $\income(v) = \es$ means that $v$ is a source of $G$). A \textbf{general circuit} is a quadruple $\mc{C} = (G, I, b, o)$, such that
\begin{itemize}
\item $G$ is a dag with $V(G) \subset \Va$ (the vertices are the auxiliary variables);
\item $I \subset \Va$ with $I \cap V(G) = \es$ is the finite set of ``input variables'';
\item $o \in V(G)$ is the ``output node'' (also ``output variable'');
\item $b$ maps every vertex $v \in V(G)$ to a boolean function $b_v$ (the ``gate function'') such that $\income(v) \sse \var(b_v) \sse \income(v) \cup I$ (the boolean function $b_v$ uses the variables of the incoming nodes plus possibly the input variables).
\end{itemize}
The \emph{length} of the circuit is $\ell(\mc{C}) := \abs{V(G)} + \abs{E(G)} + \abs{I}$.
The \emph{underlying boolean function} of the circuit $\mc{C}$ has as variables those $v \in I$ which are essential (not taking the variables on which $f$ does not depends enables us to use CNF-representations), and its value is read off at the output-node $o$ in the usual way (performing the computation from the sources up).
The \textbf{Tseitin translation} $\bmm{\tstr(\mc{C})} \in \Cls$ is
\begin{displaymath}
  \tstr(\mc{C}) := \set{\set{o}} \cup \bc_{v \in V(G)} \primec_0(v \lra b_v),
\end{displaymath}
that is, taking for each vertex the strongest CNF-representation, via the prime implicates, of the boolean function $v \lra b_v$, plus the requirement that the output variable shall be true. Note that $\tstr(\mc{C})$ is computable in time linear in $\ell(\mc{C})$ if the clause-sets $\primec_0(b_v)$ can be computed in linear time in $n(b_v)$ (which is the case if the arity of the $b_v$ (i.e., $n(b_v)$) is bounded, or if we have only ANDs and ORs).

We say that $\mc{C}$ is in \emph{general NNF} (``general negation normal form''), if all $b_v$ are monotone when considered as functions of the literals $I \cup \ol{I}$ (that is, negation of inputs is allowed), which is equivalent to $\bc \primec(b_v) \sm \Va \sse \ol{I}$ (negative literals in the prime implicates of $b_v$ come only from the input variables). For general circuits in NNF the \textbf{reduced Tseitin translation} $\bmm{\tstrr(\mc{C})} \in \Cls$ is defined as
\begin{displaymath}
  \tstrr(\mc{C}) := \set{\set{o}} \cup \bc_{v \in V(G)} \primec_0(v \ra b_v),
\end{displaymath}
that is, only one direction of the original equivalences is used. Note that for an arbitrary boolean function $f$ and $v \in \Va \sm \var(f)$ holds $\primec_0(v \ra f) = \set{\set{\ol{v}} \cup C : C \in \primec_0(f)}$. It is well-known that the Tseitin-translation in both forms yields a representation, and so we only give a terse proof:
\begin{lem}\label{lem:extTs0}
  The Tseitin translation $\tstr(\mc{C})$ of a general boolean circuit $\mc{C}$ with underlying boolean function $f$ is a CNF-representation of $f$. If $\mc{C}$ is in NNF, then also $\tstrr(\mc{C})$ is a CNF-representation of $f$.
\end{lem}
\begin{prf}
The only interesting point is to show that if $\mc{C}$ is in NNF, then $\tstrr(\mc{C})$ does not add new satisfying assignments, and this follows from the fact that the relaxation of the reduced translation additionally only allows the auxiliary variables to take values $0$ instead of $1$, which due to the monotonicity condition (note that all input variables are fixed) can not turn a falsifying (total) assignment for $f$ into a satisfying assignment for $\tstrr(\mc{C})$. \Qed
\end{prf}

Instead of NNF we could handle more general circuits, by distinguishing between ``positive'' and ``negative'' polarities, as in  \cite{GreenbaumPlaisted1986ClauseFormTrans}, but these technical considerations would lead us too far astray, and so we concentrate on the most prominent case of (general) NNF. We want to understand the evaluation of Tseitin translations if values for all input variables are given. Analogously to handling total assignments to the original variables in the satisfiable case via ``sat-recognition'', we consider now ``usat-recognition'':
\begin{defi}\label{def:usatextension}
  Consider a clause-set $F \in \Cls$ and a finite $V \subset \Va$.
  \begin{itemize}
  \item \textbf{$F$ has usat-recognition via $r: \Cls \ra \Cls$ w.r.t.\ $V$} if for every $\vp \in \Tass(V)$ with $\vp * F \in \Usat$ we have $r(\vp * F) = \set{\bot}$.
  \item A CNF-representation $F$ of a boolean function $f$ \textbf{has usat-recognition via $r$} if $F$ has this property w.r.t.\ $\var(f)$.
  \end{itemize}
\end{defi}
Note that usat-recognition is weaker than the notions related to relative or absolute (p/w-)hardness, since these stronger notions consider all \emph{partial} assignments.

\begin{defi}\label{def:UPrep}
  A \textbf{UP-representation} of a boolean function $f$ is a representation $F$ with sat-recognition via $\ro$ and usat-recognition via $\ro$.
\end{defi}
In other words, a UP-representation is a representation $F$ of $f$ such that for all $\vp \in \Tass(\var(f))$ holds $\ro(\vp * F) \in \set{\top, \set{\bot}}$. Every forcing representation is a UP-representation. In order to understand the effect of the reduced Tseitin translation, let $\ropl: \Cls \ra \Cls$ denote the combination of (complete) unit-clause propagation with (complete) elimination of pure literals. This combination is easily seen to be confluent, and can be computed by first applying the reduction $F \leadsto \ro(F)$, followed by repeated applications of $F \leadsto \pab{x \ra 1 : x \in \purec(F)} * F$ until no change happens anymore (note that elimination of pure literals can not create new possibilities for unit-clause propagation). Strengthening Lemma \ref{lem:extTs0}:
\begin{thm}\label{thm:extTs}
  The Tseitin translation of a general boolean circuit with underlying boolean function $f$ is a UP-representation of $f$. If the circuit is in NNF, then the reduced Tseitin translation has usat-recognition via $\ro$ and sat-recognition via $\ropl$.
\end{thm}
\begin{prf}
The only non-trivial assertion is that for a circuit $\mc{C}$ in NNF we have sat-recognition via $\ropl$. Consider a node $v$ of $\mc{C}$ and a partial assignment $\vp$ with $\var(\vp) = \var(b_v)$. Then we have $\vp * \primec_0(v \ra b_v) \in \set{\top, \set{\set{v}}, \set{\set{\ol{v}}}}$. In the latter two cases unit-clause propagation sets $v$, while in the first case the literal $v$ is pure in $\vp * \tstrr(\mc{C})$ (due to the monotonicity condition), and thus is set by elimination of pure literals. So from the sources to the output-sink all variables get set by $\ropl$. \Qed
\end{prf}

In principle the UP-part of Theorem \ref{thm:extTs} is well-known, though not explicitly expressed; see for example \cite[Subsection 4.1]{JarvisaloJunttila2009LimitRestrictedLearning}. We conclude this section with general reflections on (SAT-)representations of boolean functions by the observation that UP-representations are basically the same as representations by circuits (again this is kind of folklore of the field):
\begin{corol}\label{cor:characUP}
  Via the Tseitin translation we see that UP-re\-pre\-sen\-ta\-tions are equivalent modulo linear-time translations to representations of boolean functions by boolean circuits with gates having polysize equivalent CNFs (CNF-rep\-re\-sen\-ta\-tions without auxiliary variables); for example using only gates with bounded arity.
\end{corol}
Thus the problem of separating general CNF-representations from UP-re\-pre\-sen\-ta\-tions, i.e., showing that there is a sequence of boolean functions with polysize CNF-representations but without polysize UP-representations, is precisely the holy grail of the theory of circuit complexity, proving a non-polynomial lower bound for the circuit-complexity for decision problems in NP (see \cite{Jukna2012BooleanFunctionComplexity}). On the other hand, separating UP-representations from GAC-representations (or, equivalently, as we will see, from UR-representations), has been achieved in effect in \cite{BKNW2009CircuitComplexity}, or will be seen by our Theorem \ref{thm:xorclsrel}.

\section{Characterising UR-representations by monotone circuits}
\label{sec:characurmon}

The topic of the section is the close relation between UR-representations and monotone circuits, extending the CSP-approach introduced in \cite{BKNW2009CircuitComplexity} specifically for the boolean context. In Subsection \ref{sec:acmon} we show in Theorem \ref{thm:acmono} that from an UR-representation of a boolean function $f$ we obtain in polynomial time a monotone circuit computing the monotonisation $\widehat{f}$ (which captures evaluation of partial assignments). In Subsection \ref{sec:charACreps} we consider the other direction, and show how from monotone circuits computing $\widehat{f}$ we obtain an UR-representation (which also has sat-recognition via $\ro$). In Subsection \ref{sec:forcrepingen} we show in Theorem \ref{thm:stronglyforcing}, that from an UR-representation we can compute in polynomial time a forcing representation. Finally the precise relation to \cite{BKNW2009CircuitComplexity} is discussed in Subsection \ref{sec:ComparisonBessiere}.

\subsection{UR-representations versus monotone circuits}
\label{sec:acmon}

As a preparation we introduce a natural monotonisation of boolean functions. Recall that a boolean function $f(v_1,\dots,v_n)$ assigns to every total assignment $\vp$ a boolean value $f(\vp)$, and thus is a map $f: \Tass(\set{v_1,\dots,v_n}) \ra \set{0,1}$. $f$ is monotone iff $(\fa\, i \in \tb 1n : v_i \le v_i') \Ra f(v_1,\dots,v_n) \le f(v_1',\dots,v_n')$. We want to extend $f$ to \emph{partial} assignments $\vp$, obtaining the value $0$ iff there is no total assignment $\psi \supseteq \vp$ with $f(\psi) = 1$. Furthermore, we want indeed a \emph{monotone} boolean function $\widehat{f}$, and thus setting more arguments of $\widehat{f}$ to $1$ should mean setting fewer variables of $f$ (at all).
\begin{itemize}
\item For that purpose, every variable $v_i$ is replaced by two new variables $v_i^0, v_i^1$, where $v_i^{\ve} = 1$ means that $v_i \ne \ve$ for $\ve \in \set{0,1}$, that is:
  \begin{itemize}
  \item $v_i^0 = v_i^1 = 1$ means that $v_i$ has not been assigned,
  \item $v_i^0 = 1$, $v_i^1 = 0$ means $v_i = 1$,
  \item $v_i^0 = 0$, $v_i^1 = 1$ means $v_i = 0$,
  \item while $v_i^0 = 0$, $v_i^1 = 0$ means ``contradiction''.
  \end{itemize}
\item $\widehat{f}(v_1^0,v_1^1,\dots,v_n^0,v_n^1) = 0$ iff either
  \begin{enumerate}
  \item there is $i$ with $v_i^0 = v_i^1 = 0$, or
  \item for the corresponding partial assignment $\vp$ (with $\var(\vp) \sse \set{v_1,\dots,v_n}$) there is no total assignment $\psi \supseteq \vp$ with $f(\psi) = 1$.
  \end{enumerate}
\item Accordingly $\widehat{f}(v_1^0,v_1^1,\dots,v_n^0,v_n^1) = 1$ iff there is $\psi \in \Tass(\set{v_1,\dots,v_n})$ with $f(\psi) = 1$, such that for all $i \in \tb 1n$ holds $v_i^{\ol{\psi(v_i)}} = 1$.
\end{itemize}
Obviously $\widehat{f}(v_1^0,v_1^1,\dots,v_n^0,v_n^1)$ is a monotone boolean function. Strengthening \cite[Lemma 4]{BKNW2009CircuitComplexity} for the boolean case, by providing a simpler proof and the details of the polytime construction:

\begin{thm}\label{thm:acmono}
  Consider a boolean function $f(v_1,\dots,v_n)$ and a UR-representation $F \in \Pcls{p}$ for some $p \in \NNZ$, that is,
  \begin{itemize}
  \item $\set{v_1,\dots,v_n} \sse \var(F)$,
  \item for every partial assignment $\vp$ with $\var(\vp) \sse \set{v_1,\dots,v_n}$, such that $f$ is unsatisfiable for $\vp$, we have $\ro(\vp * F) = \set{\bot}$,
  \item while otherwise $\vp * F \in \Sat$.
  \end{itemize}
  From $F$ we can compute in time $O(p \cdot n(F) \cdot \ell(F) )$ a monotone circuit $\mc{C}$ (using only binary ANDs and ORs, with $O(n(F)^2)$ nodes) which computes $\widehat{f}(v_1^0,v_1^1,\dots,v_n^0,v_n^1)$.
\end{thm}
\begin{prf}
In $\bot \in F$, then $\widehat{f}$ is the constant-0 function, if $F = \top$, then $\widehat{f}$ is the constant-1 function; so assume $\bot \notin F$ and $F \ne \top$. Let $N := n(F)$ and $\var(F) = \set{v_1,\dots,v_n,v_{n+1},\dots,v_N}$. Let the nodes of $\mc{C}$ be $v^0_{i,j}, v^1_{i,j}$ for $i \in \tb 1N$, $j \in \tb 0N$, plus one additional output-node $o$. We define $\mc{C}$ via the defining equations for its nodes.

The inputs of $\mc{C}$ are the nodes $v^0_{i,0} = v^0_i$, $v^1_{i,0} = v^1_i$ for $i \in \tb 1n$, while $v^0_{i,0} = v^1_{i,0} = 1$ for $i \in \tb{n+1}{N}$.\footnote{According to our definition of a circuit, the duplication of inputs $v_i^0, v_i^1$ as nodes is not needed, but it simplifies the notation.} The output of $\mc{C}$ is given by
\begin{displaymath}
  o = \bw_{i \in \tb 1N} v^0_{i,N} \vee v^1_{i,N}.
\end{displaymath}
The meaning of $v^{\ve}_{i,j} = 0$ is that $v_i$ got value $\ve$ at stage $j$ or earlier of unit-clause propagation (and thus the equation for $o$ means that a contradiction was derived). For $x \in \lit(F)$ and $j \in \tb 0N$ let
\begin{displaymath}
  l(x,j) :=
  \begin{cases}
    v^0_{i,j} & \text{if } x = v_i\\
    v^1_{i,j} & \text{if } x = \ol{v_i}
  \end{cases}.
\end{displaymath}
For the remaining nodes with $i \in \tb 1N$ and $j \in \tb{0}{N-1}$ we have:
\begin{eqnarray*}
  v^0_{i,j+1} & = & v^0_{i,j} \wedge \bw_{C \in F \atop \ol{v_i} \in C} \bv_{x \in C \sm \set{\ol{v_i}}} l(x,j),\\
  v^!_{i,j+1} & = & v^1_{i,j} \wedge \bw_{C \in F \atop v_i \in C} \bv_{x \in C \sm \set{v_i}} l(x,j).
\end{eqnarray*}
Obviously these equations express unit-clause propagation, and thus yield the desired meaning. $N$ stages (layers) are enough, since at each stage of unit-clause propagation at least one new assignment is created. For each level $j$ of the defining equations we need time $O(p \cdot \ell(F))$ for the constructions, which yields total time $O(N \cdot p \cdot \ell(F))$ for the construction of the circuit. \Qed
\end{prf}

Independently of \cite{BKNW2009CircuitComplexity}, in \cite[Theorem 8]{Bailleux2011ExpressivenessUnitRes}, a similar result has been obtained (couched in a language aiming at the computational content of unit-clause propagation).

\paragraph{Remarks on the monotonisation} Via $f(v_1,\dots,v_n) \mapsto \widehat{f}(v_1^0,v_1^1, \dots, v_n^0,v_n^1)$ every bool\-ean function with $n$ arguments is embedded into a monotone boolean function with $2 n$ arguments. If $f$ is given via the full truth-table, then $\widehat{f}$ can be computed in polynomial time, while if $f$ is given via an equivalent CNF $F$, then decision of $\widehat{f}(1,\dots,1)=1$ is NP-complete (since $\widehat{f}(1,\dots,1)=1$ iff $F$ is satisfiable). As pointed out by George Katsirelos\footnote{personal communication, October 2013}, there is another, related and simpler monotonisation $f'(v_1^0,v_1^1, \dots, v_n^0,v_n^1)$, as used in \cite{Goldschlager1977MCVP}, but where now $f'$ depends on a representation of $f$ (while $\widehat{f}$ is semantically defined). Namely a deMorgan-circuit $\mc{C}$ for $f$ is taken, a monotone circuit with inputs $v_i$ and $\ol{v_i}$, and then $v_i$ is renamed to $v_i^0$ and $\ol{v_i}$ to $v_i^1$. In order to compare $f'$ to $\widehat{f}$, where $v_i^0=v_i^1=0$ means $\widehat{f}=0$, we also apply to the result additionally the conjunction of all $v_i^0 \vee v_i^1$. We now have $\widehat{f} \le f'$, but not equality in general: Take $f = v \wedge \ol{v}$. So $f$ is constant $0$, and so is $\widehat{f}$. But $f' = (v^0 \wedge v^1)$ (using the given circuit for $f$), and thus $f'$ is not constant $0$ (since $f'(1,1) = 1$).

\subsection{Characterising UR-representations}
\label{sec:charACreps}

Also the other direction of Theorem \ref{thm:acmono} holds, and we indeed obtain a characterisation of UR (strengthening \cite[Corollary 1]{BKNW2009CircuitComplexity}):
\begin{lem}\label{lem:otherdir}
  Consider a boolean function $f$ and a monotone circuit $\mc{C}$ for $\widehat{f}$. We can construct in linear time from $\mc{C}$ a UR-representation $F$ of $f$ with sat-recognition via $\ro$.
\end{lem}
\begin{prf}
Let $f = f(v_1,\dots,v_n)$ and $\widehat{f} = \widehat{f}(v_1^0,v_1^1,\dots,v_n^0,v_n^1)$. Let $F_0 := \tstr(\mc{C})$ (recall Subsection \ref{sec:remTseit}), that is, use the Tseitin translation of $\mc{C}$, translating an or/and-node $w$ with inputs $w_1,\dots,w_m$ via the equivalence $w \lra (w_1 \vee \dots \vee w_m)$ resp.\ $w \lra (w_1 \wedge \dots \wedge w_m)$. We have $\set{v_1^0,v_1^1,\dots,v_n^0,v_n^1} \subset \var(F_0)$, and now replace $v_i^0$ by $v_i$ and $v_i^1$ by $\ol{v_i}$ in $F_0$, remove (pseudo-)clauses containing clashing literals, and obtain $F$. By Lemma \ref{lem:extTs0}, $F$ is a CNF-representation of $f$, and by Theorem \ref{thm:extTs} (sat-recognition) we get sat-recognition via $\ro$.

It remains to show that for a partial assignment $\vp$ with $\var(\vp) \sse \set{v_1,\dots,v_n}$ and $\vp * F \in \Usat$ we have $\ro(\vp * F) = \set{\bot}$. Consider the corresponding $\vp'$ with $\var(\vp) = \set{v_1^0,v_1^1,\dots,v_n^0,v_n^1}$; we get $\widehat{f}(\vp') = 0$. For the computation in $\mc{C}$, due to monotonicity, variables $v_i^0, v_j^1$ set to $1$ do not contribute, and so we get $\ro(\vp * F) = \set{\bot}$ by Theorem \ref{thm:extTs} (usat-recognition). \Qed
\end{prf}

Instead of the (full) Tseitin translation one can also use the reduced translation in the proof of Lemma \ref{lem:otherdir}, that is, using $F_0 := \tstrr(\mc{C})$; the only change is that sat-recognition then needs $\ropl$. So then the implications $w \ra (w_1 \vee \dots \vee w_m)$ and $w \ra (w_1 \wedge \dots \wedge w_m)$ are used instead of the equivalences. That leads to clauses of the type $\set{\ol{w},w_1,\dots,w_m}$ resp.\ $\set{\ol{w},w_i}$, and thus $F_0$ here is a dual Horn clause-set (has at most one negative literal in each clause; indeed except of the one unit-clause with the output-variable we have a pure dual Horn clause-set, with exactly one negative literal in each clause). Corollary 3 in \cite{BKNW2009CircuitComplexity} makes a similar statement: ``Let $C_C$ be a CNF decomposition of a consistency checker $f_C$. The variables of $C_C$ can be renamed to that each clause has exactly one negative literal.'' In our different context (where we require proper representations; see Subsection \ref{sec:ComparisonBessiere}), this renamability to dual Horn clause-sets does not hold for arbitrary representations $F$ of boolean functions $f$ of relative hardness $1$, but only after first constructing a monotone circuit $\mc{C}$ from $F$ by Theorem \ref{thm:acmono}, and then transforming $\mc{C}$ into a representation of relative hardness $1$ by Lemma \ref{lem:otherdir}, and finally ignoring negative literals for variables in $f$. Since flipping the signs of auxiliary variables in a representation still yields a representation, we get:

\begin{corol}\label{cor:URrepHorn}
  From a UR-representation $F \in \Pcls{p}$ of a boolean function $f$, in time $O(p \cdot n(F) \cdot \ell(F))$ a UR-representation $F' \in \Cls$ can be computed, which has sat-recognition via $\ropl$, and where after removal of positive literals in $\var(f)$ from $F'$ we have a Horn clause-set (i.e., $\set{C \sm \var(f) : C \in F'} \in \Ho$).
\end{corol}

In Theorem \ref{thm:stronglyforcing} we obtain a stronger ``standardisation'' of UR-representations.

\begin{examp}\label{exp:monohorn}
  We consider the boolean function $f = (a \vee b \vee c) \wedge (\neg a \vee \neg b \vee \neg c)$ and its representation $F_0 := \primec_0(f) = \set{\set{a,b,c},\set{\ol{a},\ol{b},\ol{c}}}$. This representation can not be renamed into a (dual) Horn clause-set, but of course after removal of negative literals it is trivially dual Horn. We note that $F_0$ itself can be used here to obtain a monotone circuit $\mc{C}$ computing $\widehat{f}$, since $F_0$ is of hardness $0$. Furthermore we can use gates of arbitrary fan-in, since we allow clauses of arbitrary size. So the monotone circuit (in fact, a monotone formula here) for $\widehat{f}$ is
  \begin{displaymath}
    \mc{C} = ((a^0 \vee b^0 \vee c^0) \wedge (a^1 \vee b^1 \vee c^1)) \wedge (a^0 \vee a^1) \wedge (b^0 \vee b^1) \wedge (c^0 \vee c^1).
  \end{displaymath}
  Using the reduced Tseitin translation, we obtain $w_1 \ra (a^0 \vee b^0 \vee c^0)$, $w_2 \ra (a^1 \vee b^1 \vee c^1)$, $w_3 \ra (a^0 \vee a^1)$, $w_4 \ra (b^0 \vee b^1)$, $w_5 \ra (c^0 \vee c^1)$, and $o \ra w_1 \wedge w_2 \wedge w_3 \wedge w_4 \wedge w_5$ from $\mc{C}$, and thus we get the pure dual Horn clause-set $F_0' := F_0 \sm \set{\set{o}}$
  \begin{multline*}
    F_0' = \set{\set{\ol{w_1},a^0,b^0,c^0},\set{\ol{w_2},a^1,b^1,c^1},\\
      \set{\ol{w_3},a^0,a^1},\set{\ol{w_4},b^0,b^1},\set{\ol{w_5},c^0,c^1},\\
      \set{\ol{o},w_1},\dots,\set{\ol{o},w_5}}.
  \end{multline*}
  Finally we get $F = \set{\set{\ol{w_1},a,b,c}, \set{\ol{w_2},\ol{a},\ol{b},\ol{c}}, \set{\ol{o},w_1},\dots,\set{\ol{o},w_5}, \set{o}}$.
\end{examp}

We conclude by characterising the expressive power of UR-representations in terms of monotone circuits (similar to \cite[Theorem 2]{BKNW2009CircuitComplexity}):
\begin{thm}\label{thm:otherdir}
  A sequence $(f_n)_{n \in \NN}$ of boolean functions has a CNF-representation $(F_n)_{n \in \NN}$ with $\hardness^{\var(f_n)}(F_n) \le 1$ and $\ell(F_n) = n^{O(1)}$ (a polysize UR-representation) if and only if the sequence $(\widehat{f_n})_{n \in \NN}$ can be computed by monotone circuits of size polynomial in $n$.
\end{thm}

By \cite{Kullmann2014Collapse} the condition ``$\hardness^{\var(f_n)}(F_n) \le 1$'' in Theorem \ref{thm:otherdir} can be replaced by ``$h^{\var(f_n)}(F_n) \le k$'' for any $h \in \set{\hardness,\phardness,\whardness}$ and any fixed $k \in \NN$.

We remark that the computation by circuits in Theorem \ref{thm:otherdir} is non-uniform --- to spell out the uniformity conditions for their computation (and the corresponding computation of the sequence $(F_n)_{n \in \NN}$) would lead us to far astray here, and must be left for future work.

If in Theorem \ref{thm:otherdir} we drop the requirements on relative hardness, but just use UP-representations, then by Corollary \ref{cor:characUP} we have the equivalence to arbitrary (non-uniform, poly-size) boolean circuits computing $(f_n)_{n \in \NN}$. If we also consider arbitrary boolean circuits, but use the stronger requirement, that they compute $(\widehat{f_n})_{n \in \NN}$, then we had the most general form of a non-uniform Knowledge Compilation mechanism for the representation of the boolean functions $(f_n)_{n \in \NN}$ such that Clausal Entailment queries can be answered in non-uniform polynomial time (see \cite{DarwicheMarquis2002KCmap} for an overview). It is apparently not known to what class of CNF-representations this corresponds.

\subsection{Forcing representations}
\label{sec:forcrepingen}

In Corollary \ref{cor:URrepHorn} we have already seen an example of strengthening UR-re\-pre\-sen\-ta\-tions. We now show that UR-representations can be transformed into forcing representations (recall Definition \ref{def:forciing}). First we strengthen the construction of Lemma \ref{lem:otherdir}, by actually producing a forcing representation, through adding $2 n$ versions of the translation, each responsible for one potential forced literal (this is similar to the proof of the direction ``$\la$'' of \cite[Theorem 1]{BKNW2009CircuitComplexity}, but the details are different):
\begin{lem}\label{lem:achieveforcing}
  Consider a boolean function $f$ and a monotone circuit $\mc{C}$ for $\widehat{f}$. We can construct in time $O(n(f) \cdot \ell(\mc{C}))$ from $\mc{C}$ a forcing representation $F$ of $f$.
\end{lem}
\begin{prf}
Consider $F' := F$ from Lemma \ref{lem:otherdir}, and additionally construct clause-sets $F^x$ for $x \in \lit(f)$ as follows (this clause-set shall produce the unit-clause $\set{x}$ by $\ro$ iff $x$ is forced): The (monotone) circuit $\mc{C}_x$ is obtained from $\mc{C}$ by substituting $0$ for $x$, that is, setting the two corresponding variables to $(0,1)$ (if $x$ is positive) resp.\ $(1,0)$, simplifying this accordingly,  and renaming all nodes so that they are all new, but call the output-node $x$ (in a slight generalisation of our treatment of circuits, here we allow them to be also negative literals). $\mc{C}_x$ produces $0$ (at node $x$) iff literal $x$ is forced. Let $F_0^x$ be obtained by the Tseitin translation from $\mc{C}_x$, but without the additional unit clause for the output, and using the reduced Tseitin translation at the output node $x$. As before replace $v_i^0$ by $v_i$ and $v_i^1$ by $\ol{v_i}$ in $F_0^x$, remove (pseudo-)clauses containing clashing literals, and obtain $F^x$. Note that literal $\ol{x}$ is pure in $F^x$, and $\pao x0 * F_0^x$ is satisfiable for any total assignment for $\var(f)$ applied to it. Finally $F := F' \cup \bc_{x \in \lit(f)} F^x$. By Theorem \ref{thm:extTs}, applied to the NNFs $\mc{C}_x$, the outputs $0$ of $\mc{C}_x$ are propagated via $\ro$ in $F^x$, and thus we get that $F$ is a GAC-representation of $f$. While for the sat-recognition we don't need $\ropl$, but $\ro$ is enough, since a total assignment for $\var(f)$ sets the output-variables anyway (these are the only places where the reduced Tseitin translation was applied). \Qed
\end{prf}

Now we can transform UR-representations into forcing representations, by applying first the translation from Theorem \ref{thm:acmono} and then the translation (back) of Lemma \ref{lem:achieveforcing}, strengthening Corollary \ref{cor:URrepHorn}. Compared to \cite[Theorem 1]{BKNW2009CircuitComplexity}, we provide the constructive details, and by starting from a representation we also obtain a representation (according to our definition), which furthermore additionally to GAC has sat-recognition by $\ro$:
\begin{thm}\label{thm:stronglyforcing}
  From a UR-representation $F \in \Pcls{p}$ of a boolean function $f$ a forcing representation of $f$ can be computed in time $O(n(f) \cdot p \cdot n(F) \cdot \ell(F))$.
\end{thm}

\subsection{Comparison with Bessiere et al}
\label{sec:ComparisonBessiere}

The main result of \cite{BKNW2009CircuitComplexity} is Theorem 2: ``A consistency checker $f_C$ can be decomposed to a CNF of polynomial size if and only if it can be computed by a monotone circuit of polynomial size.'' The direction from left to right is expressed more precisely by Lemma 4 there: ``Let $C_C$ be a CNF decomposition of a consistency checker $f_C$. Then, there exists a monotone circuit $S_C$ of size $O(n \abs{C_C})$ that computes $f_C$.'', and corresponds to our Theorem \ref{thm:acmono} (though in \cite{BKNW2009CircuitComplexity} the CNF $C_C$ is assumed to have maximal clause-length $3$; see Footnote 2 there). The other direction corresponds to our Lemma \ref{lem:otherdir}. We now discuss the similarities and differences.

\cite{BKNW2009CircuitComplexity} uses, in our terminology, the translation of non-boolean functions into boolean functions via the direct variable-translation, as discussed in Subsection \ref{sec:introAC}. A ``propagator'' for \cite{BKNW2009CircuitComplexity} is a function, which for given domains $D_v$ restricts them, correctly but possibly incompletely, to some sub-domains, while a ``consistency checker'' is (basically) the special case which detects only (some) cases of unsatisfiability (and thus is similar to representations of relative hardness $1$). If we start directly with a boolean function $f$, and use a complete consistency checker, then the monotone boolean function corresponding to the consistency checker is up to a flipping (and renaming) of the variables precisely $\widehat{f}$: the doubling of the variables comes from the direct variable encoding, which for a variable $v \in \var(f)$ introduces two new variables $v_0, v_1$ where $v_{\ve} = 0$ means $v \ne \ve$; we use the negated version in order to get $\widehat{f}$ monotone, that is, the more inputs are $1$ (``unassigned''), the more ``likely'' the output is to be $1$ (``satisfiable''). In this sense Theorem 2 of \cite{BKNW2009CircuitComplexity} is more general than our results, since non-boolean variables are used, and the consistency checker can be incomplete.

However \cite{BKNW2009CircuitComplexity} does not treat ``representations''. More precisely, \cite{BKNW2009CircuitComplexity} does not treat representations of (non-)boolean functions for the sake of SAT-solving, but the main motivation is to ``decompose'' a constraint-propagator into a CNF, as a tool to replace the propagator (compare our Subsection \ref{sec:introAC}). First of all, if we start with a boolean function, then every variable is doubled due to the direct variable-encoding. Then the definition of a ``decomposition'' of a propagator  (\cite[Definition 4]{BKNW2009CircuitComplexity}) considers only setting the variables of the direct encoding to $0$, which corresponds to removing values from the (current) domain of a variable, while setting them to $1$ is irrelevant here. Accordingly Lemma 2 in \cite{BKNW2009CircuitComplexity} states that negative literals on the original variables can be removed from the decomposition; there are some remarks after the proof how to circumvent this in practice, but clearly the (full) representation of boolean functions is not the focus of \cite{BKNW2009CircuitComplexity}. So our Lemma \ref{lem:otherdir} makes a stronger assertion, namely that we obtain a CNF-representation (additionally we also have sat-recognition), and that for the original boolean function (without doubled variables).

These differences could be handled via additional translations, but this would obscure the picture --- the proof of our Theorem \ref{thm:acmono} is simpler than the proof in \cite{BKNW2009CircuitComplexity}, which uses various normalisation steps, while we directly translate the CNF-representation into a boolean circuit. There are actually two further normalisation steps in the definition of a ``consistency checker'', namely an output-variable is used there, and there is an additional condition, the second bullet point in Definition 5 of \cite{BKNW2009CircuitComplexity}, which essentially states that unit-clause propagation run on a decomposition of a consistency checker does not touch the original variables. Again, these normalisations could be handled, but we do not need any of them.

Finally, for the proof of Theorem \ref{thm:stronglyforcing} we need that the translation of Theorem \ref{thm:acmono} happens in polynomial time; the translation from \cite{BKNW2009CircuitComplexity} to monotone circuits is ``constructive'' and ``polysize'', but it is not stated whether it is polynomial time.

\section{Systems of XOR-constraints}
\label{sec:xorclausesets}

We now review the concepts of ``XOR-constraints'' and their representations via CNF-clause-sets. In Subsection \ref{sec:xorclausesetsI} we model XOR-constraints via ``XOR-clauses'' (and sets of XOR-constraints via ``XOR-clause-sets''), and we define their semantics. In Subsection \ref{sec:gencnfrepxor} we define ``CNF-representations'' of XOR-clause-sets, and show in Lemma \ref{lem:characimplxor} that all XOR-clauses following from an XOR-clause-set are obtained by summing up some XOR-clauses.

\subsection{XOR-clause-sets}
\label{sec:xorclausesetsI}

An \textbf{XOR-constraint} (also known as ``parity constraint'') is a (boolean) constraint of the form $x_1 \oplus \dots \oplus x_n = \ve$ for literals $x_1, \dots, x_n$ and $\ve \in \set{0,1}$, where $\oplus$ is the addition in the 2-element field $\ZZ_2 = \set{0,1}$. Note that $x_1 \oplus \dots \oplus x_n = y$ is equivalent to $x_1 \oplus \dots \oplus x_n \oplus y = 0$, while $x \oplus x = 0$ and $x \oplus \ol{x} = 1$, and $0 \oplus x = x$ and $1 \oplus x = \ol{x}$. Two XOR-constraints are \emph{equivalent}, if they have exactly the same set of solutions. In this \Schrift{} we prefer a lightweight approach, and so we do not present a full framework for working with XOR-constraints, but we use a representation by \textbf{XOR-clauses}. These are just ordinary clauses $C \in \Cl$, but under a different interpretation, namely implicitly interpreting $C$ as the XOR-constraints $\oplus_{x \in C} = 0$. And instead of systems of XOR-constraints we just handle \textbf{XOR-clause-sets} $F$, which are sets of XOR-clauses, that is, ordinary clause-sets $F \in \Cls$ with a different interpretation. So two XOR-clauses $C, D$ are equivalent iff $\var(C) = \var(D)$ and the number of complements in $C$ has the same parity as the number of complements in $D$. That clauses are sets is justified by the commutativity of XOR, while repetition of literals is not needed due to $x \oplus x = 0$. Clashing literal pairs can be removed by $x \oplus \ol{x} = 1$ and $1 \oplus y = \ol{y}$, as long as there is still a literal left. So every XOR-constraint can be represented by an XOR-clause except of inconsistent XOR-constraints, where the simplest form is $0=1$; we can represent this by \emph{two} XOR-clauses $\set{v}, \set{\ol{v}}$. In our theoretical study me might even assume that the case of an inconsistent XOR-clause-set is filtered out by preprocessing.

The appropriate theoretical background for (systems of) XOR-constraints is the theory of systems of linear equations over a field (here the two-element field). To an XOR-clause-set $F$ corresponds a system $A(F) \cdot \vec{v} = b(F)$, using ordinary matrix notation. To make this correspondence explicit we use $n := n(F)$, $m := c(F)$, $\var(F) = \set{v_1, \dots, v_n}$, and $F = \set{C_1,\dots,C_m}$. Now $F$ yields an $m \times n$ matrix $A(F)$ over $\ZZ_2$ together with a vector $b(F) \in \set{0,1}^m$, where the rows $A(F)_{i,-}$ of $A(F)$ correspond to the clauses $C_i \in F$, such that a coefficient $A(F)_{i,j}$ of $v_j$ is $0$ iff $v_j \notin \var(C_i)$, while $b_i = 0$ iff the number of complementations in $C_i$ is even.
\begin{examp}\label{exp:xorcls}
  Consider $F = \set{\set{v_1,\ol{v_2}},\set{\ol{v_2},\ol{v_3}},\set{v_1,v_3}}$, where the clauses are taken in this order. Then
  \begin{displaymath}
    A(F) =
    \begin{pmatrix}
      1 & 1 & 0\\
      0 & 1 & 1\\
      1 & 0 & 1
    \end{pmatrix}, \quad b(F) =
    \begin{pmatrix}
      1\\ 0\\ 0
    \end{pmatrix}.
  \end{displaymath}
\end{examp}

\subsection{Semantical aspects}
\label{sec:gencnfrepxor}

A partial assignment $\vp \in \Pass$ satisfies an XOR-clause-set $F$ iff $\var(\vp) \supseteq \var(F)$ and for every $C \in F$ the number of $x \in C$ with $\vp(x) = 1$ is even. An XOR-clause-set $F$ implies an XOR-clause $C$ if every satisfying partial assignment $\vp$ for $F$ is also a satisfying assignment for $\set{C}$. The satisfying total assignments for an XOR-clause-set $F$ correspond 1-1 to the solutions of $A(F) \cdot \vec{v} = b$ (as elements of $\set{0,1}^n$), while implication of XOR-clauses $C$ by $F$ correspond to single equations $c \cdot \vec{v} = d$, which follow from the system, where $c$ is an $1 \times n$-matrix over $\ZZ_2$, and $d \in \ZZ_2$. Note that for every satisfiable XOR-clause-set $F$ we can compute, via computation of a row basis of $A(F)$, an equivalent XOR-clause-set $F'$ with $c(F') \le c(F)$, $n(F') \le n(F)$, and $c(F') \le n(F')$.

A \textbf{CNF-representation} of an XOR-clause-set $F \in \Cls$ is a clause-set $F' \in \Cls$ with $\var(F) \sse \var(F')$, such that the projections of the satisfying total assignments for $F'$ (as CNF-clause-set) to $\var(F)$ are precisely the satisfying (total) assignments for $F$ (as XOR-clause-set). The central question of representing XOR-clause-sets $F$ is how to obtain implied XOR-clauses $C$ from the representation $F'$; using resolution and $F'$ without auxiliary variables, it is costly in general to obtain even just one $C$ (see Section \ref{sec:transx0} for more details). How to derive any single $C$ cheaply from $F$ at the XOR-level is now discussed.

What is for (resolvable) CNF-clauses $C, D \in \Cl$ the resolution operation $C \res D \in \Cl$, is for (arbitrary) XOR-clauses $C, D \in \Cl$ the addition of clauses, which corresponds to symmetric difference, that is, from two XOR-clauses $C, D$ follows the pseudo-clause (possible containing clashing literals) $C \symdif D := (C \sm D) \cup (D \sm C) = (C \cup D) \sm (C \cap D)$ (literals $x \in C \cap D$ are cancelled due to $x \oplus x = 0$). Since we do not allow clashing literals, some rule is supposed here to translate $C \symdif D$ into an equivalent $E \in \Cl$ in case of $C \cap \ol{D} \ne \es$; such a translation is possible iff $C \symdif D$ is not precisely the disjoint union of an odd number of clashing literals $v, \ol{v}$. More generally, for an arbitrary XOR-clause-set $F$ we can consider the \emph{sum}, written as $\oplus F \in \Cl$, which is defined as the reduction of $\symdif_{C \in F} C$ (note that the symmetric difference is associative and commutative) to some clause $\oplus F := E \in \Cl$, assuming that the reduction does not end up in the situation $E = \set{v,\ol{v}}$ for some variable $v$ --- in this case we say that $\oplus F$ is inconsistent (which is only possible for $c(F) \ge 2$). More precisely, the reduction removes all quadruples $v,\ol{v},w,\ol{w}$ for variables $v \ne w$ (due to $x \oplus \ol{x} = 1$ and $1 \oplus 1 = 0$); if a single pair $v, \ol{v}$ remains, and there is another literal $x$ left, then $v, \ol{v}$ is removed, and according to some choice-rule one such remaining literal $x$ is chosen and replaced by $\ol{x}$ (due to $1 \oplus x = \ol{x}$), while if no other literal $x$ is left, then we have the situation that $\oplus F$ is inconsistent.

\begin{examp}\label{exp:xorcls2}
  For the following computation we consider only variables from $\NN$, and assume that the chosen literal $x$ is the one with minimal $\var(x)$:
  \begin{enumerate}
  \item $\oplus \top = \oplus(\set{\bot}) = \bot$ (note that as an XOR-clause, $\bot$ is a tautology).
  \item $\oplus \set{\set{1,2},\set{2,3}} = \set{1,3}$.
  \item $\oplus \set{\set{1,2,-3},\set{-1,2,3}} = \bot$.
  \item $\oplus \set{\set{1,2},\set{-1,2}}$ is inconsistent.
  \item $\oplus \set{\set{1,2,-3,4},\set{-1,2,3,-4,5,6}} = \set{-5,6}$.
  \end{enumerate}
\end{examp}

The following fundamental lemma translates witnessing of unsatisfiable systems of linear equations and derivation of implied equations into the language of XOR-clause-sets; it is basically a result of linear algebra, but since it might not be available in this form, we provide a proof in Appendix \ref{sec:Proofsla} (which is instructive anyway).
\begin{lem}\label{lem:characimplxor}
  Consider an XOR-clause-set $F \in \Cls$.
  \begin{enumerate}
  \item\label{lem:characimplxor1} $F$ is unsatisfiable if and only if there is $F' \sse F$ such that $\oplus F'$ is inconsistent.
  \item\label{lem:characimplxor2} Assume that $F$ is satisfiable. Then for all $F' \sse F$ the sum $\oplus F'$ is defined, and the set of all these clauses is modulo equivalence precisely the set of all XOR-clauses which follow from $F$.
  \end{enumerate}
\end{lem}

\section{The most basic translation $X_0$}
\label{sec:transx0}

In this section we investigate the basic building block of any representation of XOR-clause-sets, the translation $X_0(F)$, which translates single XOR-clause-sets, by the unique equivalent CNF-clause-set.

\subsection{The unique equivalent clause-set}

There is precisely one CNF-clause-set equivalent to the XOR-clause-set $\set{C}$, i.e., there is exactly one representation without auxiliary variables, namely
\begin{displaymath}
  \bmm{X_0(C)} := \primec_0(x_1 \oplus \dots \oplus x_n = 0) \in \Urefc_0,
\end{displaymath}
the set of prime implicates of the underlying boolean function, which is unique since the prime implicates are not resolvable (and are full, so that not even subsumptions are possible). $X_0(C)$ has $2^{n-1}$ clauses for $n \ge 1$ (while for $n=0$ we have $X_0(C) = \top$), namely the full clauses (containing all variables) over $\set{\var(x_1),\dots,\var(x_n)}$, where the parity of the number of complementations is different from the parity of the number of complementations in $C$.

\begin{examp}\label{exp:X0}
  $X_0(\set{1,2}) = \set{\set{-1,2},\set{1,-2}}$, $X_0(\set{1,-2}) = \set{\set{1,2},\set{-1,-2}}$.
\end{examp}

Note that for two XOR-clauses $C, D$ we have $X_0(C) = X_0(D)$ iff $C, D$ are equivalent. By definition we have $X_0(C) \in \Urefc_0$.

More generally, we define $X_0: \Cls \ra \Cls$, where the input is interpreted as XOR-clause-set and the output as CNF-clause-set, by $\bmm{X_0(F)} := \bc_{C \in F} X_0(C)$.
\begin{lem}\label{lem:suffx0pc}
  If $F \in \Cls$ is acyclic, then $X_0(F)$ is an absolute forcing representation of the XOR-clause-set $F$.
\end{lem}
\begin{prf}
 By Theorem \ref{thm:acylcprop} and Lemma \ref{lem:mainacyc}, Part \ref{lem:mainacyc2} we obtain $X_0(F) \in \Propc$. Since $X_0(F)$ as CNF-clause-set is equivalent to the XOR-clause-set $F$ (yields the same underlying boolean function), trivially we have that the representation $X_0(F)$ has sat-recognition by $\ro$. \Qed
\end{prf}

In the rest of this section we consider $X_0(F)$ for unsatisfiable XOR-clause-sets $F$. These cases can be handled by preprocessing, but nevertheless they are instructive, and they have been at the heart of lower bounds for the resolution calculus from the beginnings. Ignoring the size of the obtained representation, the following simple example shows that $X_0(\set{C,D})$ for $C, D \in \Cl$ in general has high asymmetric width.
\begin{examp}\label{exp:2xor0}
  For $n \in \NN$ and (different) variables $v_1,\dots,v_n$ consider the system
  \begin{eqnarray*}
    v_1 \oplus v_2 \oplus \dots \oplus v_n &=& 0 \\
    v_1 \oplus v_2 \oplus \dots \oplus \ol{v_n} &=& 0,
  \end{eqnarray*}
  that is, consider the XOR-clauses $C_1 := \set{v_1,\dots,v_n}$ and $C_2 := \set{v_1,\dots,v_{n-1},\ol{v_n}}$. Then $X_0(\set{C_1,C_2})$ is the clause-set with all $2^n$ full clauses over $\set{v_1,\dots,v_n}$, and thus $\hardness(X_0(\set{C_1,C_2})) = \whardness(X_0(\set{C_1,C_2})) = n$ (due to the minimal clause-length $n$ we have $n \le \whardness(X_0(\set{C_1,C_2}))$, while due to the variable-number $n$ we have $\hardness(X_0(\set{C_1,C_2})) \le n$).
\end{examp}

\subsection{The Tseitin formulas}
\label{sec:Tseitinformulas}

An early and very influential example of (hard) unsatisfiable clause-sets are the ``Tseitin formulas'' introduced in \cite{Ts68}, which are defined as follows. Consider a general graph $G = (V,E,\eta)$, that is, $V$ is the set of vertices, $E$ is the set of edge-labels, while $\eta: E \ra \set{e \sse V : 1 \le \abs{e} \le 2}$ maps every edge-label $x$ to some edge $\eta(x)$ (so parallel edges and loops are allowed). The special conditions are:
\begin{itemize}
\item $E$ is a clause (i.e., $E \in \Cl$; alternatively one could say that every edge is labelled by a literal with pairwise distinct underlying variables and no clashes);
\item there is additionally a ``charge'' $\rho: V \ra \set{0,1}$.
\end{itemize}
In order that we only have to deal with XOR-clauses, we forbid the case that an isolated vertex can have charge $1$ (that would lead to ``$0 = 1$''; otherwise $\rho$ is arbitrary). For every vertex $w \in V(G)$ the XOR-clause $C_w \in \Cl$ is defined via the equation
\begin{displaymath}
  \oplus_{x \in E(G), w \in \eta(x)} \; x = \rho(w),
\end{displaymath}
that is, the XOR over all literal-edges incident with $w$ is $\rho(w)$. Let
\begin{displaymath}
  T_0(G,\rho) := \set{C_w : w \in V(G)} \in \Cls
\end{displaymath}
be the XOR-clause-set derived from $G$. Then $T_0(G,\rho)$ is unsatisfiable if $G$ has no loops and  $\oplus_{w \in V(G)} \rho(w) = 1$, since $\oplus_{w \in V(G)} \oplus_{x \in E(G), w \in \eta(x)} x = 0$, due to every edge occurring precisely twice in the sum. Finally the \textbf{Tseitin clause-set} is $T(G,\rho) := X_0(T_0(G,\rho))$; typically we just use ``$T(G)$''.

\begin{examp}\label{exp:2xor0T}
  For an XOR-clause $C$ we have $X_0(C) = T(B_C)$, where $B_C$ is the ``bouquet'' (a general graph with one vertex) with the (single) vertex $C$, which has charge $0$, and the literals of $C$ as edges (loops).

  To obtain the translation of the two XOR-clauses $C_1, C_2$ from Example \ref{exp:2xor0}, we consider the dipole $D_n$, which is the general graph with two vertices and the variables $v_1,\dots,v_n$ as edges connecting these two vertices, where the first vertex gets charge $0$ and the second gets charge $1$. We have $X_0(\set{C_1,C_2}) = T(D_n)$.
\end{examp}

In \cite{Ts68} an exponential lower bound for regular resolution refutations of (special) Tseitin clause-sets was shown, and thus unsatisfiable Tseitin clause-sets in general have high hardness. This was extended in \cite{Urq87} to full resolution, and thus unsatisfiable Tseitin clause-sets in general also have high asymmetric width. In the following we refine $X_0: \Cls \ra \Cls$ in various ways, by first transforming an XOR-clause-set $F$ into another XOR-clause-set $F'$ representing $F$, and then using $X_0(F')$.

\section{The standard translation $X_1$}
\label{sec:transx1}

If the XOR-clause-set $F$ contains long clauses, then $X_0(F)$ is not feasible, and the XOR-clauses of $F$ have to be broken up into short clauses, which we consider now. As we have defined how a CNF-clause-set can represent an XOR-clause-set, so we can define that an XOR-clause-set $F'$ represents an XOR-clause-set $F$, namely if the satisfying assignments of $F'$ projected to the variables of $F$ are precisely the satisfying assignments of $F$.

\begin{defi}\label{def:1softxor}
  Consider a linear order $\le$ on $\Va$ and an XOR-clause $C \in \Cl$, where $C = \set{x_1,\dots,x_n}$ is the ordering via $\le$. The \textbf{natural splitting} of $C$ w.r.t.\ $\le$ is the XOR-clause-set $F'$ obtained as follows, using $n := \abs{C}$:
  \begin{itemize}
  \item If $n \le 2$, then $F' := \set{C}$.
  \item Otherwise choose pairwise different new variables $y_2, \dots, y_{n-1} \in \Va \sm \var(C)$, and let
    \begin{displaymath}
      F' := \set{x_1 \oplus x_2 = y_2} \cup \set{y_{i-1} \oplus x_i = y_i}_{i \in \tb 3{n-1}} \cup \set{y_{n-1} \oplus x_n = 0}
    \end{displaymath}
    (i.e., $F' = \set{\set{x_1,x_2,y_2}} \cup \set{\set{y_{i-1},x_i,y_{i}}}_{i \in \tb 3{n-1}} \cup \set{\set{y_{n-1},x_n}}$).
  \end{itemize}
  Then $F'$ as XOR-clause-set is a representation of $\set{C}$. Let $\bmm{X_1^{\le}(C)} := X_0(F')$.
\end{defi}
In the following the underlying linear order on $\Va$ is mentioned in general results only in case it matters. For $C \in \Cl$ and $n := \abs{C} \in \NNZ$ we have for $F := X_1(C)$:
\begin{itemize}
\item If $n \le 2$, then $n(F) = c(F) = n$, and $\ell(F) = 2^{n-1} \cdot n = n^2$.
\item Otherwise $n(F)=2n-2$, $c(F)=4n-6$ and $\ell(F) = 12n-20$.
\end{itemize}

\begin{examp}\label{exp:xortrans}
  For $n = 3$ we get
  \begin{displaymath}
    X_1(C) = \setb{ \underbrace{\set{x_1,x_2,\ol{y_2}}, \set{x_1,\ol{x_2},y_2}, \set{\ol{x_1},x_2, y_2}, \set{\ol{x_1},\ol{x_2},\ol{y_2}}}_{\bmm{x_1 \oplus x_2 = y_2}},\underbrace{\set{y_2,\ol{x_3}}, \set{\ol{y_2},x_3} }_{\bmm{y_2 \oplus x_3 = 0}} }.
  \end{displaymath}
\end{examp}

Computing general XORs with binary XORs, the translation $X_1(C)$ is what we get from the Tseitin-translation (recall Subsection \ref{sec:remTseit}); more precisely if
\begin{itemize}
\item we negate the last input-variable (since we want the sum to be equal $0$),
\item and split the $n$-ary XOR into binary XORs, in the form of a tree with Horton-Strahler number $1$, where these nodes use variables $y_2,\dots,y_{n-1}, o$ (using the additional (new) output-variable $o$),
\end{itemize}
then for the general circuit $\mc{C}$ obtained we have $\pao o1 * \tstr(\mc{C}) = X_1(C)$ (note that we process unit-clause propagation (only) on the output variable $o$).

In Example \ref{exp:2xor0T} we have seen how to obtain $X_0(C)$ as a Tseitin clause-sets (via a bouquet). Also $X_1(C)$ can be obtained as a Tseitin clause-set:

\begin{examp}\label{exp:x1ts}
  Consider a clause $C = \set{x_1,\dots,x_n}$. As we have seen in Example \ref{exp:2xor0T}, $X_1(C)$ for $n \le 2$ is the Tseitin clause-set for the bouquet given by the literals $x_1,\dots,x_n$ (in both cases having one vertex $v_1$). Now the general graph $G_n$ for $n \ge 3$, such that $X_1(C)$ is the Tseitin clause-set for $G_n$, is obtained recursively from $G_{n-1}$ by adding one new vertex $v_n$, which has two incident edges, namely $y_{n-1}$ connected to the vertex $v_{n-1}$ last added (where $v_2 := v_1$) and the loop $x_n$ (recall that edges are literals here); all charges are $0$. For example for $n = 4$ we get (the charges not shown):
  \begin{displaymath}
    G_4 = \quad \xymatrix {
      v_1 \aru@(dl,dr)_{x_1} \aru@(ur,ul)_{x_2} \aru[r]_{y_2} & v_3 \aru@(ur,ul)_{x_3} \aru[r]_{y_3} & v_4 \aru@(ur,ul)_{x_4}
    }
  \end{displaymath}
\end{examp}

Corollary \ref{cor:1acylcprop}, Part \ref{cor:1acylcprop2}, applies to $F'$ from Definition \ref{def:1softxor}, and thus we obtain $X_1(C) \in \Propc$, as was first shown \cite[Proposition 5]{BordeauxMarquesSilva2012KnowledgeCompilation}:
\begin{lem}\label{lem:1softxor}
  $X_1(C)$ is an absolute forcing representation for XOR-clause $C \in \Cl$.
\end{lem}
\begin{prf}
It only remains to show sat-recognition by $\ro$, and this is easy to see directly, and follows also by Theorem \ref{thm:extTs}. \Qed
\end{prf}

We define $X_1: \Cls \ra \Pcls{3}$, where the input is interpreted as XOR-clause-set and the output as CNF-clause-set, by $X_1(F) := \bc_{C \in F} X_1(C)$ for $F \in \Cls$, where some choice for the new variables is used, so that the new variables for different XOR-clauses do not overlap, and for each clause some ordering is chosen. By Lemma \ref{lem:1softxor} we get:
\begin{lem}\label{lem:X1UP}
  $X_1(F)$ is a UP-representation of the XOR-clause-set $F \in \Cls$.
\end{lem}

A simple example shows that we do not have more than UP-representations:
\begin{examp}\label{exp:notevenur}
  The simplest example for an XOR-clause-set $F \in \Cls$ such that $X_1(F)$ is not even UR is $F := \set{\set{1,2},\set{-1,2}}$, which is unsatisfiable (as XOR-clause-set), while $X_1(F) = X_0(F)$ has all clauses of length $2$, and thus $\hardness(X_0(F)) = 2$ (this is also the relative hardness, since the representation here doesn't use auxiliary variables). Note that $F$ is not acyclic. In Lemma \ref{lem:hdtwoxor} we see that indeed $X_1$ for just two XOR-clauses can have arbitrary high (relative) hardness.
\end{examp}

However for acyclic $F$ we can generalise Lemma \ref{lem:1softxor}:
\begin{thm}\label{thm:suffx1pc}
  If $F \in \Cls$ is acyclic, then $X_1(F)$ is an absolute forcing representation of the XOR-clause-set $F$.
\end{thm}
\begin{prf}
By Theorem \ref{thm:acylcprop}, Lemma \ref{lem:1softxor}, Lemma \ref{lem:mainacyc}, Part \ref{lem:mainacyc2}, and Lemma \ref{lem:X1UP} \Qed
\end{prf}

A precursor to Theorem \ref{thm:suffx1pc} is found in Theorem 1 of \cite{LaitinenJunttilaNiemelae2012Parity}, where it is stated that tree-like XOR clause-sets are ``UP-deducible'' (not to be confused with our use of ``UP''), which is precisely the assertion that for acyclic $F \in \Cls$ the representation $X_1(F)$ is GAC. As mentioned in \cite{LaitinenJunttilaNiemelae2012Parity}, such XOR clause-sets have good applications, with 61 out of 474 SAT benchmarks from SAT competition 2005 to 2011 containing only tree-like systems of XOR equations.

The question is now how much Theorem \ref{thm:suffx1pc} can be extended. In Section \ref{sec:transtxor} we will see that $X_0(F)$ and $X_1(F)$ have high hardness in general, even for $c(F) = 2$. But we will also see that appropriate preprocessing of the XOR-clause-set improves the yield of $X_1$. In general we will see by Corollary \ref{cor:xorcls}, that even under the relative condition there is no general (polysize) solution.

\section{Literature review on XOR-constraints for SAT}
\label{sec:litrev}

If we do not specify the representation in the following, then essentially $X_1$ is used (up to small variations), that is, breaking up long XOR-constraints into short ones and using $X_0$ for short constraints (where ``short'' and ``long'' depend somewhat on the context).

\subsection{Applications of XOR-constraints}

XOR-constraints are a typical part of cryptographic schemes, and accordingly it is important to have ``good'' representations for them. The earliest application of SAT to cryptanalysis is \cite{MassacciMarraro2000DESSAT}, translating DES to SAT and then considering finding a key. In \cite{BardCourtois2007AlgebraicDES}, DES is encoded to ANF (``algebraic normal form'', that is, XORs of conjunctions), and then translated. \cite{JovanovicKreuzer2010AlgAttackSAT} attacks DES, AES and the Courtois Toy Cipher via translation to SAT. Each cipher is  first translated to equations over $\mr{GF}(2)$ and then to CNF. A key contribution is a specialised translation of certain forms of polynomials, designed to reduce the number of variables and clauses. The size for breaking up long XOR-constraints is called the ``cutting length'', and has apparently some effect on solver times. \cite{MironovZhang2006SATHash} translates MD5 to SAT and finds collisions. MD5 is translated by modelling it as a circuit (including XORs) and applying the Tseitin translation.

\cite{MarquesSilvaSakallah2000EDA} provides an overview of SAT-based methods in Electronic Design Automation, and suggests keeping track of circuit information (fan in/fan out of gates etc.) in the SAT solver when solving such instances. XOR is relevant here due to the use of XOR gates in the underlying circuit being checked (and translated).

A potential application area is the translation of pseudo-boolean constraints, as investigated by \cite{Een2006Translating}. Translations via ``full-adders'' introduce XORs via translation of the full-adder circuit. It is shown that this translation does not produce a GAC-representation, and the presence of XOR and the log-encoding is blamed for this (in Section 5.5). Experiments conclude that sorting network and BDD methods perform better, as long as their translations are not too large.

\subsection{Hard examples via XORs}

It is well-known that using $X_0$ for unsatisfiable systems can result in hard (unsatisfiable) instances for resolution. This goes back to the ``Tseitin formulas'' introduced in \cite{Ts68} (recall Subsection \ref{sec:Tseitinformulas}), which were proven hard for full resolution in \cite{Urq87}, and modified to (empirically) hard satisfiable instances in \cite{HaanpaaJarvisaloKaskiNiemela2006HardSATBench}.

A well-known (satisfiable) benchmark is based on \cite{CrawfordKearns1994MinimalParityDisagreement}, which considers the ``Minimal Disagreement Parity'' problem, an NP-complete problem, and presents a SAT-translation. Randomly generated instances became the \texttt{parity32} benchmarks in the SAT2002 competition.  Given $m$ vectors $\vec{x}_i \in \set{0,1}^n$, further $m$ bits $y_i \in \set{0,1}$, and $k \in \NNZ$, the computational problem is to find a vector $\vec{a} \in \set{0,1}^n$, such that $\abs{\set{i : \vec{a} \cdot \vec{x}_i \not= y_i}} \le k$, where $\vec{a} \cdot \vec{x}_i$ is the scalar product. The chosen CNF-representation is the union of the default representation of $m$ XOR-clauses
\begin{displaymath}
  X_1 \big ( r_i \oplus y_i = (\vec{a}_1 \wedge (\vec{x}_i)_1) \oplus \dots \oplus (\vec{a}_n \wedge (\vec{x}_i)_n) \big )
\end{displaymath}
for $i = 1, \dots, m$ (note $r_i = 0$ iff $y_i = \vec{a} \cdot \vec{x}_i$), together with the cardinality constraint ``$\sum_{1 \le i \le m} r_i \le k$'', translated by using full-adders. So the XORs occur both in the summations and in the cardinality constraint. These benchmarks were first solved by the solver \texttt{EqSatz} (\cite{Li2000Equivalency}). The general form of these problems is a system of XOR-constraints plus one cardinality constraint. An (empirically) improved translation for the \texttt{parity32} instances is presented in \cite{BailleuzBoufkhad2003CardinalityConstraints}, where the XOR-constraints are simplified by variable-elimination (in the linear-algebra sense), and where for the cardinality constraint the translation from \cite{BailleuzBoufkhad2003CardinalityConstraints}, based on unary addition, is used.

\subsection{Special reasoning}
\label{sec:introspecreas}

It is natural to consider extensions of resolution and/or SAT techniques to handle XOR-constraints more directly. The earliest theoretical approach seems \cite{BaumgartnerMassacci2000TamingXOR}, integrating a proof calculus for Gaussian elimination with an abstract proof calculus modelling DPLL (without clause learning). It is argued that such a system should offer improvements over just DPLL/resolution in handling XORs. \cite{Heule2004Diplom} points out a simple algorithm for extracting ``equivalence constraints''. The earliest SAT solver with special reasoning is \texttt{EqSatz} (\cite{Li2000Equivalency}), extracting XOR-clauses from its input (produced by $X_0(C)$ for $C \in \Cl$ with $2 \le \abs{C} \le 3$), and applying DP-resolution plus incomplete XOR reasoning rules. Further work on the integration of such ``equivalence reasoning'', using ``conjunctions of equivalences'' instead of (equivalent) collections of XOR-constraints, into look-ahead solvers (see \cite{HvM09HBSAT} for a general overview) one finds in \cite{HeulevanMaaren2004Equivalences}.

More recently, conflict-driven solvers are considered (``CDCL''; see \cite{MSLM09HBSAT} for an overview). \texttt{CryptoMiniSAT} (\cite{CryptoSAT2009,Soos2010SATGauss}) integrates Gaussian elimination during search, allowing both explicitly specified XOR-clauses and also XOR-clauses extracted from CNF input. However in the newest version 3.3 the XOR handling during search is removed, since it is deemed too expensive.\footnote{See \url{http://www.msoos.org/2013/08/why-cryptominisat-3-3-doesnt-have-xors/}.} Further approaches for hybrid solvers one finds in \cite{Chen2009HybridXOR} and \cite{HanJiang2012GaussianSAT} (which also computes interpolants).

A systematic study of the integration of XOR-reasoning and SAT-techniques has been started with \cite{Laitinen2010DPLLParity}, by introducing the ``DPLL(XOR)'' framework, similar to SMT. These techniques have also been integrated into \texttt{MiniSat}. \cite{LaitinenJunttilaNiemela2011EqParReasDPLLXOR} expands on this by reasoning about equivalence classes of literals created by binary XORs, while \cite{LaitinenJunttilaNiemela2012ConflictXORLearn} learns conflicts in terms of ``parity (XOR) explanations''. The latest paper \cite{Laitinen2012DPLLParity} (with underlying report \cite{Laitinen2012DPLLParityE}) extends the reasoning from ``Gau\ss{} elimination'' to ``Gau\ss{}-Jordan elimination'', which corresponds to moving from relative hardness to relative p-hardness, i.e., also detecting forced literals, not just inconsistency.\footnote{We say ``relative'' here, since the reasoning mechanism is placed outside of SAT solving, different from the ``absolute'' condition, where also the reasoning itself is made accessible to SAT solving (that is, one can (feasibly!) split in some sense on the higher-level reasoning).} Theorem 4 in \cite{Laitinen2012DPLLParity} is similar in spirit to Corollary \ref{cor:1acylcprop}, Part \ref{cor:1acylcprop2}, considering conditions when strong reasoning only needs to be applied to ``components''.

Altogether we see a mixed picture regarding special reasoning in SAT solvers. The first phase of expanding SAT solvers could be seen as having ended in some disappointment regarding XOR reasoning, but with \cite{Laitinen2010DPLLParity} a systematic approach towards integrating special reasoning has been re-opened. A second approach for handling XOR-constraints, also the approach of the current \Schrift, is by using intelligent translations (possibly combined with special reasoning).

\subsection{Translations to CNF}

Switching now to translations of XORs to CNF, \cite{LaitinenJunttilaNiemelae2012Parity} identifies the subsets of ``tree-like'' systems of XOR constraints, where the standard translation delivers a GAC-representation (our Theorem \ref{thm:suffx1pc} strengthens this, showing that indeed an absolute forcing representation is obtained):
\begin{itemize}
\item \cite{LaitinenJunttilaNiemelae2012Parity} also considered equivalence reasoning, where for ``cycle-partitionable'' systems of XOR constraints this reasoning suffices to derive all conclusions.
\item Furthermore \cite{LaitinenJunttilaNiemelae2012Parity} showed how to eliminate the need for such special equivalence reasoning by another GAC-representation.
\item In general, the idea is to only use Gaussian elimination for such parts of XOR systems which the SAT solver is otherwise incapable of propagating on. Existing propagation mechanisms, especially unit-clause propagation, and to a lesser degree equivalence reasoning, are very fast, while Gaussian elimination is much slower (although still poly-time).
\end{itemize}
Experimental evaluation on SAT 2005 benchmarks instances showed that, when ``not too large'', such CNF translations outperform dedicated XOR reasoning modules. The successor \cite{LaitinenJunttilaNiemelae2013Parity} provides several comparisons of special-reasoning machinery with resolution-based methods, and in Theorem 4 there we find a general GAC-translation; our Theorem \ref{thm:relxorcnfp} yields a better upper bound, but the heuristic reasoning of \cite{Laitinen2012DPLLParity,LaitinenJunttilaNiemelae2013Parity} seems valuable, and should be explored further.

\section{No short UR-representations for general XOR-clause-sets}
\label{sec:norepmsp}

We now prove that there are no short UR-representations of general XOR-systems (recall Subsection \ref{sec:introlowb} for an overview on the proof idea). More precisely, we show in Theorem \ref{thm:xorclsrel}, that if there were polysize UR-representations of all XOR-clause-sets, then we could translate ``monotone span programs'' (MSPs) with only a polynomial size blow-up into monotone boolean circuits, which is not possible by \cite{BabaiGalWigderson1999MonoteSpanPrograms}.\footnote{See Chapter 8 of \cite{Jukna2012BooleanFunctionComplexity} for a recent introduction and overview on span programs.} First we define (within our framework) MSPs. These are representations of monotone boolean functions
\begin{displaymath}
  f: \set{0,1}^n \ra \set{0,1}
\end{displaymath}
by systems of linear equations of $\ZZ_2$ in the following way: Each $x_i$ acts as a switch for its associated system, where $x_i = 0$ means ``on'' (for monotonicity reasons). The total value for $(x_1,\dots,x_n) \in \set{0,1}^n$ is $0$ iff all active systems together are unsatisfiable. More precisely:
\begin{itemize}
\item The input variables are given by $x_1,\dots,x_n$.
\item Additionally $m \in \NNZ$ boolean variables $y_1,\dots,y_m$ can be used, where $m$ is the dimension, which we can also be taken as the size of the span program.
\item For each $i \in \tb 1n$ there is a linear system $A_i \cdot y = b_i$ over $\ZZ_2$, where $A_i$ is an $m_i \times m$ matrix with $m_i \le m$, and $b_i \in \set{0,1}^{m_i}$.
\item For a total assignment $\vp$, i.e., $\vp \in \Pass$ with $\var(\vp) = \set{x_1,\dots,x_n}$, the value $f(\vp)$ is $0$ if and only if the linear systems given by $\vp(x_i) = 0$ together are unsatisfiable, that is,
  \begin{displaymath}
    f(\vp) = 0 \iff \setb{y \in \set{0,1}^m \mb \fa\, i \in \tb 1n : \vp(x_i)=0 \Ra A_i \cdot y = b_i} = \es.
  \end{displaymath}
\end{itemize}
W.l.o.g.\ we assume that each system $A_i \cdot y = b_i$ is satisfiable.

\begin{examp}\label{exp:mspproof}
  Consider $f(x_1, x_2, x_3) = x_1 \oder x_2 \oder x_3$ ($n=3$), which can be represented by an MSP with $m=2$ (with $m_1=m_2=m_3=1$, thus $N=3$), where $x_1 = 0$ activates $y_1 \oplus y_2 = 1$, while $x_2 = 0$ activates $y_1 = 0$ and $x_3 = 0$ activates $y_2 = 0$. If $x_1 = x_2 = x_3 = 0$, then the combined system is unsatisfiable, otherwise it is satisfiable. The relaxation process used in the proof of Theorem \ref{thm:xorclsrel} applied to $M$ yields linear equations $y_1 \oplus y_2 \oplus z_1 = 1$, $y_1 \oplus z_2 = 0$ and $y_2 \oplus z_3 = 0$.
\end{examp}

\begin{thm}\label{thm:xorclsrel}
  There is no polynomial $p$ such that for all XOR-clause-sets $F \in \Cls$ there is a UR-representation $F' \in \Cls$ with $\ell(F') \le p(\ell(F))$.
\end{thm}
\begin{prf}
Consider a monotone boolean function $f$ and its representation by an MSP as above, where we use $N := m_1 + \dots + m_n$. Consider for each $i \in \tb 1n$ an XOR-clause-set $A_i' \in \Cls$ representing $A_i \cdot y = b_i$; so $\var(A_i') \supseteq \set{y_1,\dots,y_m}$, where, as always, new variables for different $A_i'$ are used, that is, for $i \ne j$ we have $(\var(A_i') \cap \var(A_j')) \sm \set{y_1,\dots,y_m} = \es$. Let $A_i'' \in \Cls$ be obtained from $A_i'$ by adding a new variable to each clause; we denote these ``relaxation variables'' (altogether) by $z_1,\dots,z_N$. Let $F := \bc_{i=1}^n A_i''$. Consider a CNF-representation $F'$ of the XOR-clause-set $F$. We have
\begin{displaymath}
  f(\vp)=0 \iff \vp' * F' \in \Usat,
\end{displaymath}
where $\vp'$ is a partial assignment with $\vp'$ assigning only relaxation variables $z_j$, namely if $\vp(x_i) = 0$, then all the relaxation variables used in $A_i''$ are set to $0$, while if $\vp(x_i) = 1$, then nothing is assigned here. The reason is that by setting the relaxation variables to $0$ we obtain the original system $A_i \cdot y = b_i$, while by leaving them in, this system becomes satisfiable whatever the assignments to the $y$-variables are.

Now assume that we have $\hardness^{\set{z_1,\dots,z_N}}(F') \le 1$. By Theorem \ref{thm:acmono} we obtain from $F'$ a monotone circuit $\mc{C}$ (using only ANDs and ORs) of size polynomial in $\ell(F')$ with input variables $z_1^0,z_1^1,\dots,z_N^0,z_N^1$, where
\begin{itemize}
\item $z_j^0 = z_j^1 = 1$ means that $z_j$ has not been assigned,
\item $z_j^0 = 0$, $z_j^1 = 1$ means $z_j = 0$,
\item $z_j^0 = 1$, $z_j^1 = 0$ means $z_j = 1$,
\item while $z_j^0 = z_j^1 = 0$ means ``contradiction'' (where the output of $\mc{C}$ is $0$).
\end{itemize}
The value of $\mc{C}$ is $0$ iff the corresponding partial assignment applied to $F'$ yields an unsatisfiable clause-set. In $\mc{C}$ we now replace the inputs $z_j^0,z_j^1$ by inputs $x_i$, which in case of $x_i = 0$ sets $z_j^0=0$, $z_j^1 = 1$ for all related $j$, while in case of $x_i = 1$ all related $z_j^0,z_j^1$ are set to $1$.\footnote{In other words, all $z_j^1$ are set to $1$, while $z_j^0 = x_i$ for the $j$ related to $i$.} This is now a monotone circuit computing $f$. By \cite{BabaiGalWigderson1999MonoteSpanPrograms}, Theorem 1.1, thus it is not possible that $F'$ is of polynomial size in $F$. \Qed
\end{prf}

In \cite{Kullmann2014Collapse} we show that in the (unrestricted) presence of auxiliary variables even asymmetric width boils down, modulo polytime computations, to $\Propc_1 = \Propc$ under the relative condition:
\begin{corol}\label{cor:xorcls}
  XOR-clause-sets do not have good representations with bounded asymmetric width, not even when using relative asymmetric width. That is, there is no $k \in \NNZ$ and no polynomial $p(x)$ such that for all XOR-clause-sets $F \in \Cls$ there is a CNF-representation $F' \in \Cls$ with $\ell(F') \le p(\ell(F))$ and $\whardness^{\var(F)}(F') \le k$.
\end{corol}

\section{GAC for XOR-clause-sets is fpt in the number of equations}
\label{sec:transarbxor}

As discussed in Subsection \ref{sec:prelimresl}, the computation of $\primec_0(F)$ for a (CNF-)clause-set $F$ is fixed-parameter tractable (fpt) in the number $n(F)$ of variables. Now consider the ``optimal'' (without auxiliary variables, of hardness $0$) CNF-representation $F' := \primec_0(X_0(F))$ of an XOR-clause-set $F \in \Cls$, consisting of precisely the CNF-prime-implicates of the XOR-clause-set $F$; recall $c(F') \le 3^{n(F)}$. For fixed clause-length $p$, i.e., $F \in \Pcls{p}$ for some constant $p$, we thus obtain that computation of $F'$ is fpt in $n(F)$ (due to $n(X_0(F)) = n(F)$. However for unrestricted clause-length this does not work, since $X_0(F)$ is then of exponential size.

Since for an (XOR-)clause-set $F \in \Pcls{p}$ holds $n(F) \le p \cdot c(F)$, we also obtain that the computation of $F'$ is fpt in $c(F)$, the number of XOR-constraints (again, fixed clause-length). It is interesting to note here, that even the computation of $F'' := \primec_0(X_1(F))$ is fpt in $c(F)$ (with principally the same complexity), which is shown in Lemma \ref{lem:fptkc} in Appendix \ref{sec:compprimeimpl}.

So computing a CNF-representation with (absolute) hardness $0$ of an XOR-clause-set $F \in \Pcls{p}$ is fpt in $c(F)$ for each fixed $p$. When allowing CNF-representations with relative p-hardness $1$ (a GAC-representation), then we obtain fpt in the parameter $c(F)$ for arbitrary $F \in \Cls$ (where now also the constants involved are small):
\begin{thm}\label{thm:relxorcnfp}
  Consider a satisfiable XOR-clause-set $F \in \Cls$. Let $F^* := \set{\oplus F' : F' \sse F} \in \Cls$ (recall Lemma \ref{lem:characimplxor}); $F^*$ is computable in time $O(\ell(F) \cdot 2^{c(F)})$ (while $c(F^*) \le 2^{c(F)}$). Then $\bmm{X^*(F)} := X_1(F^*)$ is a forcing representation of $F$.
\end{thm}
\begin{prf}
First we show GAC. Consider some partial assignment $\vp$ with $\var(\vp) \sse \var(F)$, let $F' := \ro(\vp * F^*)$, and assume there is a forced literal $x \in \lit(F')$ for $F'$. Then the XOR-clause $C := \set{y \in \Lit : \vp(y) = 0} \cup \set{\ol{x}}$ follows from $F$. By Lemma \ref{lem:characimplxor} there is $F' \sse F$ with $\oplus F' = C$ modulo equivalence of XOR-clauses. So we have (modulo equivalence) $X_1(C) \sse F^*$, where due to $X_1(C) \in \Propc$ (Lemma \ref{lem:1softxor}) the forced literal $x$ for $\vp * X_1(C)$ is set by $\ro$, contradicting the assumption. Finally by Lemma \ref{lem:X1UP} we obtain sat-recognition by $\ro$. \Qed
\end{prf}

Theorem 4 in \cite{LaitinenJunttilaNiemelae2013Parity} yields the weaker bound  $O(4^{n(F)})$ for the number of clauses in a GAC-representation of $F$ (note that w.l.o.g.\ $c(F) \le n(F)$). The following example shows that the representation of Theorem \ref{thm:relxorcnfp} is not in $\Propc$ (i.e., considering the absolute condition now):

\begin{examp}\label{exp:notpropc}
  Consider the XOR-clauses $C := \set{a,b,c,d}$ and $D := \set{c,d}$, where $a,\dots,d$ are different variables. Then $F^* =  \set{C,D,\set{a,b},\bot}$ (where $\bot$ can be removed, also in general). Then we have $\phardness(X^*(F)) \ge 2$. We assume here that the literals of the clauses are ordered as shown. Now let $y_2$ be the first auxiliary variable for $C$ (so, semantically, $y_2 = a \oplus b$ ). Then $y_2 \ra 0$ is forced for $F$ (and thus also for $F^*$), but all clauses of $F^*$ have length at least two, whence $\phardness(X^*(F)) \ge 2$.
\end{examp}

While the simple Example \ref{exp:notpropc} might be considered as a trivial case, the following example shows in a nutshell that (absolute) hardness of $X^*(F)$ actually can be arbitrary high, just for two XOR-clauses:
\begin{examp}\label{exp:arccfptnpc}
  Consider the XOR-clauses $C := \set{a,b,c,d}$ and $D := \set{a,b,c,e}$, and let $F := \set{C,D}$. Then we have $F^* = \set{C,D,\set{d,e},\bot}$, and $\hardness(X^*(F)) = 2$ (using the given ordering).

Let $y^C_2,y^C_3$ and $y^D_2,y^D_3$ be the new variables in $C$ resp.\ $D$ (so, semantically, $y^C_2 = a \oplus b = y^D_2$ and $y^C_3 = y^C_2 \oplus c$,  $y^D_3 = y^D_2 \oplus c$). To see that $\hardness(X^*(F)) \ge 2$, observe that $F' := \pab{y^C_2 \ra 0, y^D_2 \ra 1} * X^*(F)$ is unsatisfiable (that is, forcing $a \oplus b = 0 \und a \oplus b = 1$), but all clauses in $F'$ are of size $2$ (no unit-clauses).

Considering the upper-bound, the only way to make $X^*(F)$ unsatisfiable without immediately yielding the empty-clause, is (essentially) to set one of the new variables $y_2, y_3$ in each $X_1(C), X_1(D)$ to contradictory values. So consider the different possibilities. If $y^C_2$ is set to $b \in \set{0,1}$ and $y^D_2$ is set to $1 - b$, then $F' := \pab{y^C_2 \ra b, y^D_2 \ra 1 - b} * X^*(F) \in \Pcls{2}$, and hence $\hardness(F') \le 2$. Otherwise, if $y^C_3$ is set to $b \in \set{0,1}$ and $y^D_3$ is set to $1-b$, then unit-clause propagation forces $d$ and $e$ in $F' := \pab{y^C_3 \ra b, y^D_3 \ra 1 - b} * X^*(F)$ to opposing values, and hence creates the empty-clause in $X_1(\set{d,e})$.
\end{examp}

In Lemma \ref{lem:unbhd2xc} we will see that in fact absolute hardness $\hardness(X^*(F))$ even just for $c(F)=2$ is unbounded. On the other hand, in Conjecture \ref{con:relxorcnfp} we state our belief that we can strengthen Theorem \ref{thm:relxorcnfp} by also establishing absolute (p-)hardness $1$. We now turn to the problem of understanding and refining the basic translation $X_1$ for two clauses.

\section{Translating two XOR-clauses}
\label{sec:transtxor}

For an XOR-clause-set $F$ with $c(F) \le 1$ we have $X_1(F) \in \Propc$, which is a perfect representation. We are now considering in detail the case of $c(F) = 2$. By Theorem \ref{thm:relxorcnfp} we can consider $F^* = \set{C,D,\oplus\set{C,D}}$, and obtain the CNF-representation $X^*(F)$ of relative p-hardness $1$. But as Example \ref{exp:arccfptnpc} shows, absolute p-hardness is larger than $1$, and Lemma \ref{lem:unbhd2xc} indeed shows that (absolute) p-hardness is unbounded.

\subsection{In $\Propc$}
\label{sec:X2propc}

With more intelligence, we can provide a representation in $\Propc$ as follows; note that an XOR-clause-set $\set{C,D}$ is unsatisfiable iff $\abs{C \cap \ol{D}}$ is odd and $\var(C) = \var(D)$.
\begin{thm}\label{thm:2xorshared}
  Consider two XOR-clauses $C, D \in \Cl$. Let $I := \var(C) \cap \var(D)$. We assume (to simplify the presentation) $\abs{I} \ge 2$, $\abs{C} > \abs{I}$ and $\abs{D} > \abs{I}$; thus w.l.o.g.\ $C \cap D = I$.
  \begin{enumerate}
  \item Choose $s \in \Va \sm \var(\set{C,D})$, and let $I' := I \cup \set{s}$.
  \item Let $C' := (C \sm I) \cup \set{s}$ and $D' := (D \sm I) \cup \set{s}$.
  \end{enumerate}
  Now $\set{I',C',D'}$ is an XOR-clause-set which represents the XOR-clause-set $\set{C,D}$. Let $\bmm{X_2(C,D)} := X_1(\set{I', C', D'})$.
  Then $X_2(C,D)$ is an absolute forcing representation of the XOR-clause-set $\set{C,D}$.
\end{thm}
\begin{prf}
That $\set{I',C',D'}$ represents $\set{C,D}$ is obvious, since $s$ is the sum of the common part. Corollary \ref{cor:1acylcprop}, Part \ref{cor:1acylcprop2}, applies to $\set{I',C',D'}$ (the only common variable is $s$), and thus we get $X_2(C,D) \in \Propc$. By Lemma \ref{lem:X1UP} we obtain sat-recognition by $\ro$. \Qed
\end{prf}

We believe that this method can be generalised to more than two clauses:
\begin{conj}\label{con:relxorcnfp}
  We can combine a generalisation of Theorem \ref{thm:2xorshared} with Theorem \ref{thm:relxorcnfp} and obtain $X_*: \Cls \ra \Propc$, which computes for an XOR-clause-set $F \in \Cls$ an absolute forcing representation $X_*(F)$ in time $2^{O(c(F))} \cdot \ell(F)^{O(1)}$.
\end{conj}
The stronger Conjecture \ref{con:relxorcnfptw} replaces $c(F)$ by the treewidth $\twidthin(F) \in \NNZ$ of the incidence graph of $F$; since for the complete bipartite graphs $K_{m,n}$, $m,n \in \NNZ$, we have $\twidth(K_{m,n}) = \min(m,n)$, and removing edges does not increase the treewidth, we have $\twidthin(F) \le c(F)$.

\subsection{In $\Wrefc_3$}
\label{sec:2xwc3}

We now turn to the analysis of the ``naked'' translation $X_1(\set{C,D})$ for two XOR-clauses $C, D$. First we consider the worst-case, the unsatisfiable case where $C, D$ coincide except of one flipped literal; this case can be produced from the general case by application of partial assignments. Though we won't use it here, it is instructive to obtain these clause-sets via the Tseitin method:
\begin{examp}\label{exp:twoxorts}
  Consider the XOR-clauses $C_1, C_2$ from Example \ref{exp:2xor0}. The realisation of $X_0(\set{C_1,C_2})$ we have seen in Example \ref{exp:2xor0T}, while for a single clause we have seen the realisation of $X_1(C)$ in Example \ref{exp:x1ts}. Let $T_n := X_1(\set{C_1,C_2})$. We can obtain $T_n$ also as a Tseitin clause-set, where the principle of construction of the underlying general graph should become clear from the following example for $n=4$:
  
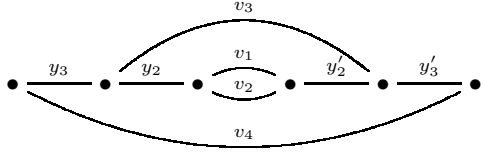
\begin{displaymath}
    \xymatrix {
      {\bullet} \aru[r]^{y_3} \aru@/_2pc/[rrrrr]^{v_4} & {\bullet} \aru[r]^{y_2} \aru@/^2pc/[rrr]^{v_3} & {\bullet} \aru@/^/[r]^{v_1} \aru@/_/[r]^{v_2} & {\bullet} \aru[r]^{y_2'} & {\bullet} \aru[r]^{y_3'} & {\bullet} &
    }
  \end{displaymath}
  All vertices have charge $0$ except of the rightmost vertex.
\end{examp}
In \cite{BeyersdorffGwynneKullmann2013PHP} we show $\hardness(T_n) = n$ (a special case of the following Lemma \ref{lem:hdtwoxor}), and thus these clause-sets are very hard regarding tree-resolution (namely every resolution tree refuting $T_n$ has at least $2^n$ leaves). For the general case we show in \cite{BeyersdorffGwynneKullmann2013PHP}, that $X_1(\set{C,D})$ has even high relative hardness:
\begin{lem}[\cite{BeyersdorffGwynneKullmann2013PHP}]\label{lem:hdtwoxor}
  For two XOR-clauses $C, D \in \Cl \sm \set{\bot}$, where either the XOR-clause-set $\set{C,D}$ is unsatisfiable, or $\var(C) \cup \var(D) \supset \var(C) \cap \var(D)$, holds
  \begin{displaymath}
    \hardness(X_1(\set{C,D})) =
    \hardness^{\var(\set{C,D})}(X_1(\set{C,D})) = \max(1,\abs{\var(C) \cap \var(D)}).
  \end{displaymath}
\end{lem}
It follows that the addition of derived clauses is not sufficient to keep (absolute) hardness low, since the problematic cases can be created by instantiating the auxiliary variables:
\begin{lem}[\cite{BeyersdorffGwynneKullmann2013PHP}]\label{lem:unbhd2xc}
  For two XOR-clauses $C, D$ and  $X^*$ as defined in Theorem \ref{thm:relxorcnfp}, $\hardness(X^*(\set{C,D}))$ is arbitrarily large.
\end{lem}

By Lemma \ref{lem:hdtwoxor} the distance of $X_1(\set{C,D})$ from GAC is as large as possible, and this is provably the worst translation from the three considered. However, still it has merits, namely (absolute) asymmetric width is in fact low, which we show now (and furthermore for dag-resolution the refutation is really easy):
\begin{thm}\label{thm:2xor}
  Consider $T_n$ from Example \ref{exp:twoxorts}. We have:
  \begin{enumerate}
  \item $n(T_n) = 2 \cdot (2n-2) - n = 3n - 4$ for $n \ge 2$.
  \item $c(T_n) = 8 n - 12$ for $n \ge 2$.
  \item $\ell(T_n) = 24 n - 40$ for $n \ge 2$.
  \item $T_n \in \Usat \cap \Pcls{3}$.
  \item For $n \ge 3$ holds $\whardness(T_n) = \wid(T_n) = 3$.
  \item There exists a resolution refutation using altogether $18 n - 29$ clauses.
  \end{enumerate}
\end{thm}
\begin{prf}
To show the lower bound for $\whardness$, consider the closure $\widehat{T_n}$ of $T_n$ under $2$-resolution. The binary clauses in $T_n$ are exactly $\primec_0(y_{n-1} = x_n \und \ol{y'_{n-1}} = x_n)$. The resolution of these binary clauses with ternary clauses in $T_n$ allows the corresponding substitutions ($y_{n-1} = x_n = \ol{y'_{n-1}}$) to be made in (other) clauses containing those variables, but this does not introduce any further clauses of size $\le 2$. Therefore, $\widehat{T_n}$ contains only clauses of size $\ge 2$, so $\whardness(T_n) \ge 3$. To show $\wid(T_n) \le 3$, we construct a resolution refutation.

From
\begin{eqnarray*}
  \primec_0(y_{n-1} \oplus x_n = 0) &=& \set{\set{\ol{y_{n-1}},x_n}, \set{y_{n-1},\ol{x_n}}} \\
  \primec_0(y'_{n-1} \oplus \ol{x_n} = 0) &=& \set{\set{\ol{y'_{n-1}},\ol{x_n}},\set{y'_{n-1},x_n}},
\end{eqnarray*}
via $2$-resolution we derive $\primec_0(y_{n-1} = \ol{y'_{n-1}}) = \set{\set{\ol{y_{n-1}},\ol{y'_{n-1}}}, \set{y_{n-1},y'_{n-1}}}$:
\begin{prooftree}
  \AxiomC{$\set{\ol{y_{n-1}},x_n}$}
  \AxiomC{$\set{\ol{y'_{n-1}},\ol{x_n}}$}
  \BinaryInfC{$\set{\ol{y_{n-1}},\ol{y'_{n-1}}}$}
  \AxiomC{$\set{y_{n-1},\ol{x_n}}$}
  \AxiomC{$\set{y'_{n-1},x_n}$}
  \BinaryInfC{$\set{y_{n-1},y'_{n-1}}$}
  \alwaysNoLine\BinaryInfC{~}
\end{prooftree}

From
\begin{eqnarray*}
  \primec_0(y_{i-1} \oplus x_i = y_i) &=& \set{\underbrace{\set{\ol{y_{i-1}},\ol{x_i},\ol{y_i}}}_{\bmm{C_1}}, \underbrace{\set{\ol{y_{i-1}},x_i,y_i}}_{\bmm{C_2}}, \underbrace{\set{y_{i-1},\ol{x_i}, y_i}}_{\bmm{C_3}}, \underbrace{\set{y_{i-1},x_i,\ol{y_i}}}_{\bmm{C_4}}} \\
  \primec_0(y'_{i-1} \oplus x_i = y'_i) &=& \{ \underbrace{\set{\ol{y'_{i-1}}, \ol{x_i}, \ol{y'_i}}}_{\bmm{D_1}}, \underbrace{\set{\ol{y'_{i-1}}, x_i, y'_i}}_{\bmm{D_2}}, \underbrace{\set{y'_{i-1}, \ol{x_i}, y'_i}}_{\bmm{D_3}}, \underbrace{\set{y'_{i-1}, x_i, \ol{y'_i}}}_{\bmm{D_4}} \}\\
  \primec_0(y_i = \ol{y'_i}) &=& \set{\underbrace{\set{\ol{y_i},\ol{y'_i}}}_{\bmm{E_1}},\underbrace{\set{y_i,y'_i}}_{\bmm{E_2}}}
\end{eqnarray*}
we derive $\primec_0(y_{i-1} = \ol{y'_{i-1}}) = \set{\set{\ol{y'_{i-1}}, \ol{y_{i-1}}}, \set{y'_{i-1}, y_{i-1}}}$:
\begin{prooftree}\small
  \def\defaultHypSeparation{\hskip .09in}
  \AxiomC{$C_1$}
  \AxiomC{$E_2$}
  \BinaryInfC{$\set{\ol{y_{i-1}},\ol{x_i},y'_i}$}
  \AxiomC{$D_2$}
  \BinaryInfC{$\set{\ol{y'_{i-1}},\ol{y_{i-1}},y'_i}$}
  \AxiomC{$C_2$}
  \AxiomC{$E_1$}
  \BinaryInfC{$\set{\ol{y_{i-1}},x_i,\ol{y'_i}}$}
  \AxiomC{$D_1$}
  \BinaryInfC{$\set{\ol{y'_{i-1}},\ol{y_{i-1}},\ol{y'_i}}$}
  \BinaryInfC{\bmm{\set{\ol{y'_{i-1}}, \ol{y_{i-1}}}}}
  \AxiomC{$C_3$}
  \AxiomC{$E_1$}
  \BinaryInfC{$\set{y_{i-1},\ol{x_i},\ol{y'_i}}$}
  \AxiomC{$D_4$}
  \BinaryInfC{$\set{y'_{i-1},y_{i-1},\ol{y'_i}}$}
  \AxiomC{$C_4$}
  \AxiomC{$E_2$}
  \BinaryInfC{$\set{y_{i-1},x_i,y'_i}$}
  \AxiomC{$D_3$}
  \BinaryInfC{$\set{y'_{i-1},y_{i-1},y'_i}$}
  \BinaryInfC{\bmm{\set{y'_{i-1}, y_{i-1}}}}
  \alwaysNoLine\BinaryInfC{~}
\end{prooftree}

Hence, by induction on $n$, we derive $\primec_0(y_2 = \ol{y'_2})$. We conclude: From
\begin{eqnarray*}
  \primec_0(x_1 \oplus x_2 = y_2) &=& \set{\underbrace{\set{\ol{x_1},\ol{x_2},\ol{y_2}}}_{\bmm{C_1}}, \underbrace{\set{\ol{x_1},x_2,y_2}}_{\bmm{C_2}}, \underbrace{\set{x_1,\ol{x_2}, y_2}}_{\bmm{C_3}}, \underbrace{\set{x_1,x_2,\ol{y_2}}}_{\bmm{C_4}}} \\
  \primec_0(x_1 \oplus x_2 = y'_2) &=& \set{\underbrace{\set{\ol{x_1},\ol{x_2},\ol{y'_2}}}_{\bmm{D_1}}, \underbrace{\set{\ol{x_1},x_2,y'_2}}_{\bmm{D_2}},\underbrace{\set{x_1,\ol{x_2}, y'_2}}_{\bmm{D_3}}, \underbrace{\set{x_1,x_2,\ol{y'_2}}}_{\bmm{D_4}}} \\
  \primec_0(y_2 = \ol{y'_2}) &=& \set{\underbrace{\set{\ol{y_2},\ol{y'_2}}}_{\bmm{E_1}},\underbrace{\set{y_2,y'_2}}_{\bmm{E_2}}}
\end{eqnarray*}
we derive $\bot$:
\begin{prooftree}
  \def\defaultHypSeparation{\hskip .1in}
  \AxiomC{$C_1$}
  \AxiomC{$E_2$}
  \BinaryInfC{$\set{\ol{x_1},\ol{x_2},y'_2}$}
  \AxiomC{$D_1$}
  \BinaryInfC{$\set{\ol{x_1},\ol{x_2}}$}
  \AxiomC{$C_2$}
  \AxiomC{$E_1$}
  \BinaryInfC{$\set{\ol{x_1},x_2,\ol{y'_2}}$}
  \AxiomC{$D_2$}
  \BinaryInfC{$\set{\ol{x_1},x_2}$}
  \BinaryInfC{$\set{\ol{x_1}}$}
  \AxiomC{$D_3$}
  \AxiomC{$C_3$}
  \AxiomC{$E_1$}
  \BinaryInfC{$\set{x_1,\ol{x_2},\ol{y'_2}}$}
  \BinaryInfC{$\set{x_1,\ol{x_2}}$}
  \AxiomC{$D_4$}
  \AxiomC{$C_4$}
  \AxiomC{$E_2$}
  \BinaryInfC{$\set{x_1,x_2,y'_2}$}
  \BinaryInfC{$\set{x_1,x_2}$}
  \BinaryInfC{$\set{x_1}$}
  \BinaryInfC{$\bot$}
\end{prooftree}

The number of clauses in this refutation (which uses only clauses of length at most $3$) altogether is
\begin{enumerate}
\item $8 n - 12$ clauses from $T_n$.
\item $2$ clauses from the derivation of $\primec_0(y_{n-1} = \ol{y'_{n-1}})$.
\item $(n-3) \cdot 10$ clauses from $(n-3)$ induction steps.
\item $11$ clauses in the final refutation in step.
\end{enumerate}
So in total, the resolution proof is of size $18 n - 29$. \Qed
\end{prf}

For arbitrary XOR-clauses $C, D$ the worst-case for instantiation of $X_1(\set{C,D})$ happens when we get the situation of $T_n$ above, and thus:
\begin{corol}\label{cor:whd2xor}
  For an XOR-clause-set $F$ with $c(F) \le 2$ holds $\wid(X_1(F)) \le 3$.
\end{corol}

\subsection{Discussion}
\label{sec:2dis}

To summarise, there are three levels of representing two XOR-clauses $C, D$:
\begin{description}
\item[Low asymmetric width] For $F_1 := X_1(\set{C,D})$ we have low width (thus low asymmetric width), namely $\wid(F_1) \le 3$, but high relative hardness.
\item[Relative p-hardness 1] For $F_2 := X_1(C,D,\oplus\set{C,D})$ we have relative p-hardness $1$, but high absolute hardness.
\item[P-hardness 1] For $F_3 := X_2(C,D)$ we have absolute p-hardness $1$.
\end{description}

The most drastic cure to handle more than two XOR-clauses is to resolve Conjecture \ref{con:relxorcnfp} positively, so that for a constant number of clauses we can reach p-hardness $1$ in polynomial time. While the most lazy approach is to do nothing, relying on the asymmetric width not growing too much:
\begin{conj}\label{con:whardc}
  There is a function $\alpha: \NNZ \ra \NNZ$ such that for all XOR-clause-sets $F \in \Cls$ holds $\whardness(X_1(F)) \le \alpha(c(F))$.
\end{conj}
We know $\alpha(0) = 0$, $\alpha(1) = 1$ and $\alpha(2) = 3$. Since $X_1(F)$ has clause-length at most $3$, we have $\wid(X_1(F)) \le \whardness(X_1(F)) + \max(\whardness(X_1(F)),3)$, and so we could have required as well $\wid(X_1(F)) \le \alpha(c(F))$ in Conjecture \ref{con:whardc}. See Conjecture \ref{conj:whdtwX1} for a strengthening.

\section{Conclusion and open problems}
\label{sec:open}

In the first part of this \Schrift{} we gave a framework for the representation of boolean functions $f$, for the purpose of SAT solving. We clarified the notion ``CNF-representation'', which has been naturally used at many places, but without giving it a proper name (besides speaking in general about ``encodings''). The most prominent condition for ``good'' representation  is ``GAC'', and we discussed its definition thoroughly, together with its ``absolute'' form, the class $\Propc$ of unit-propagation complete clause-sets. A weakening of this condition we called ``UR'', with its absolute form, the well-known class $\Urefc$ of unit-refutation complete clause-sets (especially known in the form $\Urefc = \Slur$). We introduced the new condition ``UP'', which regarding detection of unsatisfiability is weaker than the above condition, but which also handles detection of satisfying assignments; it is the outer limit of efficient CNF-representation, equivalent to the power of boolean circuits. UP together with GAC yields the ``forcing'' condition, while UP together with $\Propc$ yields the ``absolute forcing'' condition. We have shown that every UR-representation can be transformed in polynomial time into a forcing representation. This transformation is based on a general characterisation of UR- and GAC-representations via monotone circuits. The characterisations of GAC and $\Propc$ are generalised by the p-hardness measure $\phardness^V(F)$, where for the relative condition we have $V = \var(f)$, while for the absolute condition we have $V = \var(F)$. The clause-sets of absolute p-hardness at most $k$ are collected in the class $\Propc_k$. Similarly, the characterisations of UR and $\Urefc$ are generalised by the hardness measure $\hardness^V(F)$, with the same treatment of relative and absolute condition, yielding for the absolute condition the classes $\Urefc_k = \Slur_k$. We also established tools for obtaining UP-representations, via Tseitin translations, and for obtaining clause-sets in $\Propc_k$, via acyclic unions of clause-sets in $\Propc_k$. Additionally we also treated the weakest measure, asymmetric width $\whardness^V(F)$, and the corresponding largest classes $\Wrefc_k$.

In the second part of this \Schrift{} we investigated ``good'' SAT representations $F'$ of systems of linear equations over $\set{0,1}$, handled via XOR-clause-sets $F$. We showed that even under the most generous measurement of quality of $F'$, relative asymmetric width, i.e., $\whardness^{\var(F)}(F') \le k$ for some constant $k$, in general there are no $F'$ of polynomial size. Then we considered the possibilities of computing $F'$ with $\phardness^{\var(F)}(F') \le 1$, i.e., GAC, or with $\phardness(F') \le 1$, that is, $F' \in \Propc$. The methodology in general is to transform $F$ into another XOR-clause-set $G$, and then to use $F' = X_0(G)$ (translating every XOR-clause into the unique equivalent CNF-clause-set). By adding to $F$ all sums of subsets of $F$ (as XOR-clauses) we obtain GAC, where the computation is fixed-parameter tractable in the number of clauses of $F$. Our remaining endeavours are about obtaining $F' \in \Propc$. We achieved this for two cases, acyclic $F$ and $c(F) \le 2$. In the first case $X_1(F)$ does the job, where $X_1$ just splits clauses up, so that $X_0$ only has to handle XOR-clauses of length at most $3$. While the second case is handled by $X_2(F)$, which additionally factors out the common part of the two XOR-clauses.

The case $c(F) = 2$ we considered more closely, and showed that even with just using $X_1(F)$ we get low asymmetric width, however relative hardness is high. Using the general method to obtain GAC, we then obtain relative p-hardness $1$, however absolute hardness is still high. Finally, via the translation $X_2$ we get absolute p-hardness $1$.

\subsection{Open problems and future research directions}
\label{sec:openprob}

Theorem \ref{thm:acylcprop} (with applications in Lemma \ref{lem:suffx0pc}, Theorem \ref{thm:suffx1pc}, and Theorem \ref{thm:2xorshared}) is a basic general tool for obtaining clause-sets in $\Propc$, based on acyclic graphs. This should be generalised by considering treewidth and related notions, and we discuss now such approaches.

Conjecture \ref{con:relxorcnfp} says, that computing a representation in $\Propc$ should be fixed-parameter tractable in the number of XOR-clauses. More generally, we conjecture to have fixed-parameter tractability in the treewidth of the incidence graph:
\begin{conj}\label{con:relxorcnfptw}
  There exists $X^*: \Cls \ra \Propc$, which computes for an XOR-clause-set $F \in \Cls$ an absolute forcing representation $X^*(F)$ in time $2^{O(\twidthin(F))} \cdot \ell(F)^{O(1)}$, where $\twidthin(F)$ is the treewidth of the incidence graph of $F$.
\end{conj}
Note that $F \in \Cls$ is acyclic iff $\twidthin(F) \le 1$. As we already remarked, Conjecture \ref{con:relxorcnfptw} strengthens Conjecture \ref{con:relxorcnfp}. See Subsection 13.5 in \cite{SS09HBSAT} for an overview on treewidth in the context of SAT. Theorem 7 in \cite{LaitinenJunttilaNiemelae2013ParityE} shows a weaker form of Conjecture \ref{con:relxorcnfptw}, where $\twidthpr(F)$, the treewidth of the variable-interaction graph (or ``primal graph''; recall Lemma \ref{lem:characacycl}) is used instead of the treewidth of the incidence graph, and where instead of (absolute) propagation-completeness only GAC is achieved. For $F \in \Cls$ holds $\twidthin(F) \le \twidthpr(F) + 1$. On the other hand, just for a single clause $C \in \Cl$ with $n := \abs{C} \ge 1$ holds $\twidthin(\set{C}) = 1$, while $\twidthpr(\set{C}) = \twidth(K_n) = n - 1$ ($K_n$ is the complete graph with $n$ vertices).

Considering the general lower bounds for representations of XOR-clause-sets, the main question for Theorem \ref{thm:xorclsrel} and Corollary \ref{cor:xorcls} is to obtain sharp bounds on the size of shortest representations $F'$ with $\phardness^{\var(F)}(F') \le k$ resp.\ $\hardness^{\var(F)}(F') \le k$ resp.\ $\whardness^{\var(F)}(F') \le k$ for fixed $k$.

We already mentioned the possibility, that the asymmetric width of $X_1(F)$ for arbitrary XOR-clause-sets $F \in \Cls$ might not be ``too bad'', which would be encapsulated by the following (motivated by \cite[Section 4]{AlekRaz2011BranchWidth}):
\begin{conj}\label{conj:whdtwX1}
  For $F \in \Cls$ holds $\whardness(X_1(F)) \le \twidthin(F) + 1$.
\end{conj}
Since $\twidthin(F) \le \min(c(F), n(F))$ for $F \in \Cls$, this strengthens Conjecture \ref{con:whardc}. In Subsection \ref{sec:2xwc3} we started a complexity analysis (hardness analysis) of XOR-representations. While in Examples \ref{exp:2xor0T}, \ref{exp:x1ts} and \ref{exp:twoxorts} we gave connections to Tseitin clause-sets. These streams need to be combined in future work. An interesting technical question here is the dependency of $X_1(F)$ on the individual clause-orders for $c(F) \ge 2$ (we sidestepped these issues in this \Schrift; it is easy to see that with different orders we obtain essentially different representations).

We make a few remarks on the general relations between width and treewidth. In \cite[Lemma 6.23]{Ku00g} it is shown that $\wid(F) \le \twidthpr(F) + 1$ holds for $F \in \Usat$ (indeed a slightly stronger version is shown), and thus we get:
\begin{lem}[\cite{Ku00g}]\label{lem:widtwpr}
  For $F \in \Cls$ holds $\wid(F) \le \twidthpr(F) + 1$.
\end{lem}
That the ``$+1$'' is needed here, is shown by the following example:
\begin{examp}\label{exp:tw}
  Let $F := \set{\set{1,3},\set{-1,3},\set{2,-3},\set{-2,-3}} \in \Usat$:
  \begin{enumerate}
  \item $\whardness(F) = \wid(F) = \hardness(F) = 2$.
  \item $\twidthpr(F) = 1$, $\twidthin(F) = 2$.
  \end{enumerate}
\end{examp}

We believe that regarding asymmetric width, the primal graph can be replaced by the incidence graph:
\begin{conj}\label{conj:whdtw}
  For $F \in \Cls$ holds $\whardness(F) \le \twidthin(F)$.
\end{conj}

In this \Schrift{} we have only considered representations of XOR-clause-sets via CNF-clause-sets. Extending the CNF-mechanism however is also a necessary avenue (as shown by Theorem \ref{thm:xorclsrel} and Corollary \ref{cor:xorcls}), and we present a theoretical perspective in the following subsection (based on a semantic perspective, not on a proof-theoretic perspective as the DPLL(XOR)-framework introduced in \cite{Laitinen2010DPLLParity}).

\subsection{Hard boolean functions handled by oracles}
\label{sec:conclhard}

By Corollary \ref{cor:xorcls} we know that systems of XOR-clauses (affine equations) in general have no ``good'' representation, even when just considering GAC. To overcome these limitations, the theory started here can be generalised via the use of oracles as developed in \cite{Ku99b,Ku00g}, and further discussed in Subsection 9.4 of \cite{GwynneKullmann2012Slur,GwynneKullmann2012SlurJ}. The point of these oracles, which are just sets $\mc{U} \sse \Usat$ of unsatisfiable clause-sets stable under application of partial assignments, is to discover hard \emph{unsatisfiable} (sub-)instances (typically in polynomial time); at the end of Subsection \ref{sec:introsata} we already made a remark on them, in the context of \cite{Ku99b,Ku00g}, where the satisfiable instances are handled differently, namely that they yield (much) stronger hierarchies than what is obtained from the underlying classes of backdoors (which count just variables, while we use stronger parameters like the Horton-Strahler number). We obtain relativised hierarchies $\Urefc_k(\mc{U})$, $\Propc_k(\mc{U})$, $\Wrefc_k(\mc{U})$, which are defined as before, with the only change that $\Urefc_0(\mc{U}) \cap \Usat = \Propc_0(\mc{U}) \cap \Usat = \Wrefc_0(\mc{U}) \cap \Usat := \mc{U}$.

The use of such oracles is conceptually simpler than the current integration of SAT solvers and methods from linear algebra (see Subsection \ref{sec:introspecreas}). Recall the simple CNF-representation $X_1: \Cls \ra \Cls$ of XOR-clause-sets. In Theorem \ref{thm:2xor} we have seen that already for two clauses this is a bad translation (at least from the hardness-perspective). Now let $U_{X_1}$ be the set of unsatisfiable $\vp * X_1(F)$ for $F \in \Cls$ and $\vp \in \Pass$ (it is not hard to see that $U_{X_1}$ is decidable in polynomial time). Then we have $X_1: \Cls \ra \Urefc_0(U_{X_1})$.

An important aspect of the theory to be developed must be the usefulness of the representation (with oracles) in context, that is, as a ``constraint'' in a bigger problem: a boolean function $f$ represented by a clause-set $F$ is typically contained in $F^* \supset F$, where $F^*$ is the SAT problem to be solved (containing also other constraints). One approach is to require from the oracle also stability under addition of clauses, as we have it already for the resolution-based reductions like $\rk_k$, so that the (relativised) reductions $\rk_k^{\mc{U}}$ can always run on the whole clause-set (an instantiation of $F^*$). However for example for the oracle mentioned below, based on semidefinite programming, this would be prohibitively expensive. And for some oracles, like detection of minimally unsatisfiable clause-sets of a given deficiency, the problems would turn from polytime to NP-hard in this way (\cite{FKS00,BueningZhao2002MUsubsets}). Furthermore, that we have some (representation of a) constraint which would benefit for example from some XOR-oracle, does not mean that in other parts of the SAT-problems that oracle will also be of help. So in many cases it is better to restrict the application of the oracle $\mc{U}$ to that subset $F \subset F^*$, where the oracle is actually required to achieve the desired hardness.

Another example of a current barrier is given by the satisfiable pigeonhole clause-sets $\php^m_m$, which have variables $p_{i,j}$ for $i, j \in \tb 1m$, and where the satisfying assignments correspond precisely to the permutations of $\tb 1m$ (i.e., the underlying boolean function represents the permutations of a set of size $m$). The question is about ``good'' representations. In \cite{BeyersdorffGwynneKullmann2013PHP} we show $\hardness(\php^m_m) = \whardness(\php^m_m) = m-1$, and so the (standard representation) $\php^m_m \in \Cls$ itself is not a good representation (it is small, but has high asymmetric width). Moreover, as shown in \cite{BeyersdorffGwynneKullmann2013PHP} (closely related to the treatment of all-different constraints in \cite{BKNW2009CircuitComplexity}), from Theorem \ref{thm:acmono} it follows that $\php^m_m$ has no polysize GAC-representation (or, more generally, of bounded relative asymmetric width) at all. So again, oracles could be useful here; see Subsection 9.4 of \cite{GwynneKullmann2012SlurJ} for a proposal of an interesting oracle based on semidefinite programming (with potentially good stability properties).

\paragraph{Acknowledgements}

This work was partially supported (regarding the second author) by National Social Science Foundation of China Grant 13\&ZD186.

\bibliographystyle{elsarticle-num}

\newcommand{\noopsort}[1]{}

\appendix

\section{Proofs of linear algebra theorems}
\label{sec:Proofsla}

This section is devoted to the proof of Lemma \ref{lem:characimplxor}. \Zusatz

\subsection{The four subspaces of a matrix}
\label{sec:relfours}

We need to recall a fundamental theorem of linear algebra. Consider a field $K$ (we only need to consider $K = \ZZ_2$, but it seems that the greater generality adds lucidity here), consider $m, n \in \NNZ$, and an $m \times n$-matrix $A$ over $K$.

The kernel $\ker(A) \sse K^n$ is the set of $x \in K^n$ such that $A \cdot x = 0$. We denote the rows of $A$ by $A_{1,-}, \dots A_{m,-}$, which we consider as $1 \times n$-matrices, which are identified (for convenience) with vectors in $K^n$, while the columns of $A$ are denoted by $A_{-,1}, \dots, A_{-,n}$, which are considered as $m \times 1$-matrices, and which are identified with vectors in $K^m$. The \emph{row space} of $A$ is the linear hull of the rows of $A$, denoted by $\rows(A) \sse K^n$, while the \emph{column space} is the linear hull of the columns of $A$, denoted by $\cols(A) \sse K^m$. Finally the canonical scalarproduct on $K^n$ is defined by $\inprod {x,y} := \sum_{i=1}^n x_i \cdot y_i$, and for a set $X \sse K^n$ the \emph{orthogonal complement} is $X^{\bot} := \set{y \in K^n : \inprod{x,y} = 0} \sse K^n$.

Now we have (denoting transposition of $A$ by $\trans{A}$):
\begin{eqnarray*}
  \ker(A)^{\bot} & = & \rows(A)\\
  \ker(\trans{A})^{\bot} & = & \cols(A).
\end{eqnarray*}
We couldn't find a statement of these relations in the literature for the field $K = \ZZ_2$: At the \href{http://en.wikipedia.org/wiki/Fundamental_theorem_of_linear_algebra}{Wikipedia page} it is only formulated for $K = \RR$, and in \cite{Hogben2007HandbuchLA}, Chapter 5, Section 5.2 ``Orthogonality'', Fact 15, it is stated for $K \in \set{\RR, \CC}$. So we provide the simple proof, where we use basic facts from \cite{Ro1992ZA}.

By definition of the kernel we have for a row $A_{i,-}$ and $x \in \ker(A)$ the equation $\inprod{A_{i,-}, x} = 0$, whence $\rows(A) \sse \ker(A)^{\bot}$. 

We use $\dim S \in \NNZ$ for the dimension of a finite-dimensional (sub-)space. By \cite[Theorem 11.8]{Ro1992ZA} we have $\dim \ker(A)^{\bot} = n - \dim \ker(A)$. By \cite[Theorem 2.8]{Ro1992ZA} we have $\dim \ker(A) = n - \dim \cols(A)$ (using also the equation at \cite[Page 60]{Ro1992ZA}, directly before Theorem 2.11). Finally with \cite[Theorem 1.16]{Ro1992ZA} we have $\dim \cols(A) = \dim \rows(A)$. Altogether this yields $\dim \rows(A) = \dim \ker(A)^{\bot}$, and thus $\rows(A) = \ker(A)^{\bot}$.

Applying this to the transposed matrix gives $\ker(\trans{A})^{\bot} = \rows(\trans{A}) = \cols(A)$. \Qed

\subsection{Proof of Lemma \ref{lem:characimplxor}}
\label{sec:Proofslalem}

Consider an XOR-clause-set $F \in \Cls$. The assertions are:
\begin{enumerate}
\item\label{lem:characimplxor1a} $F$ is unsatisfiable if and only if there is $F' \sse F$ such that $\oplus F'$ is inconsistent.
\item\label{lem:characimplxor2a} Assume that $F$ is satisfiable. Then for all $F' \sse F$ the sum $\oplus F'$ is defined, and the set of all these clauses is modulo equivalence precisely the set of all XOR-clauses which follow from $F$.
\end{enumerate}

Obviously, if for some $F' \sse F$ we have that $\oplus F'$ is inconsistent, then $F$ is unsatisfiable, while if $F$ is satisfiable, then for every $F' \sse F$ we have that $\oplus F'$ as an XOR-clause follows from $F$. It remains to show the other directions from Part \ref{lem:characimplxor1a} resp.\ \ref{lem:characimplxor2a}.

We show the remaining assertions first at the level of linear algebra, and then we show how to translate them to the language of XOR-clause-sets. Let $A := A(F)$ be an $m \times n$ matrix, and let $b := b(F)$ (recall Subsection \ref{sec:xorclausesetsI}). We denote by $A' := (A,b)$ the extended matrix, which is an $m \times (n+1)$-matrix, obtained by appending $b$ as last column.

For the direction from left to right of Part \ref{lem:characimplxor1a} now assume that $A \cdot x = b$ is unsatisfiable; we show that for the vector $e_{n+1} := (0,\dots,0,1) \in K^{n+1}$ we have $e_{n+1} \in \rows(A')$. That $A \cdot x = b$ is unsatisfiable means that $b \notin \cols(A)$, that is, there is $c \in \ker(\trans{A})$ with $\inprod{c,b} \ne 0$. So $c_1 \cdot A'_{1,-} + \dots, c_m \cdot A'_{m,-}$ is a vector in $\rows(A')$, which is $0$ in the first $n$ components and non-zero in the last component (since this is $\inprod{c,b}$); division by the last component yields the desired result.

For the completeness-assertion of Part \ref{lem:characimplxor2a}, assume that $A \cdot x = b$ is satisfiable, and consider an equation $c \cdot x = d$ for some $1 \times n$-matrix $c$ and $d \in K$, which logically follows, that is, such that for all $x \in K^n$ holds $A x = b \Ra c x = d$. We have to show that for the vector $(c;d) \in K^{n+1}$ holds $(c;d) \in \rows(A')$. First we note that this holds for the case of homogeneous systems and conclusions, that is, for cases $b = 0$ and $d=0$, since then we have $c \in \ker(A)^{\bot}$, and thus $c \in \rows(A)$. So, introducing an additional variable $x_{n+1}$ and letting $x' = (x_1,\dots,x_n,x_{n+1})$, if we can show that the system $A' \cdot x' = 0$ implies $(c;d) \cdot x' = 0$, then we are done. So consider some $x'$ with $A' x' = 0$; we have to show that $(c;d) \cdot x' = 0$ holds.

If $x_{n+1} \ne 0$, then $x_1 A_{-,1} + \dots + x_n A_{-,n} + x_{n+1} b = 0$ is equivalent to $\frac{x_1}{-x_{n+1}} A_{-,1} + \dots +  \frac{x_n}{-x_{n+1}} A_{-,n} = b$, thus $c_1 \frac{x_1}{-x_{n+1}} + \dots + c_n \frac{x_n}{-x_{n+1}} = d$, which is equivalent to $c_1 x_1 + \dots + c_n x_n = -x_{n+1} d$, that is, $(c;d) \cdot x' = 0$.

If on the other hand $x_{n+1} = 0$ holds, then we have $A x = 0$. Since $A x = b$ is solvable, there is a solution $A x_0 = b$ (and we have $c x_0 = d$). Then we have $A (x_0+x) = b$, thus $c (x_0+x) = d$, which is equivalent to $ c x_0 + c x = d$, where $c x_0 = d$, whence $c x = 0$, that is, $(c;d) x' = 0$. This concludes the proof of the linear-algebra-formulation.

Coming finally back to the XOR-clause-sets level, we see that the row space of $A'$, considered as XOR-clauses, is precisely the set of all sums $\oplus F'$ for $F' \sse F$ (since linear combinations over $\ZZ_2$ just allow coefficients $0,1$). \Qed

\section{Computing the set of prime implicates}
\label{sec:compprimeimpl}

A simple and apparently new proof, that for a clause-set $F$ the computation of all prime implicates is fixed-parameter tractable (fpt) in the number of clauses, is as follows (recall the general discussion in Subsection \ref{sec:prelimresl}). The basic concept here is that of a ``minimal premise set'' as introduced in \cite[Subsection 4.1]{Kullmann2007ClausalFormZII} (further explored in \cite[Subsection 4.1]{GwynneKullmann2013GoodRepresentationsI}), which is a clause-set $F \in \Cls$ such that there is a clause $C \in \Cl$ with $F \models C$, while for all $F' \subset F$ holds $F' \not\models C$. As shown in \cite[Corollary 4.5]{Kullmann2007ClausalFormZII}, for a minimal premise set $F$ there is exactly one minimal such clause $C$, and as shown in \cite[Lemma 4.12]{GwynneKullmann2013GoodRepresentationsI} for boolean clause-sets, we have $C = \purec(F)$ (recall that $\purec(F)$ is the set of pure literals). To determine the prime implicates of $F \in \Cls$, we only need to consider the minimal premise subsets $G \sse F$ and their unique minimal prime implicate $\purec(G)$, and we obtain the following proof:
\begin{lem}\label{lem:primcec}
  For $F \in \Cls$ we have $c(\primec_0(F)) \le 2^{c(F)}-1$, and $\primec_0(F)$ can be computed in time $O(\ell(F)^2 \cdot 2^{2 c(F)})$ and linear output-space.
\end{lem}
\begin{prf}
Let $F_0 := \set{\purec(G) : \top \ne G \sse F \und F \models \purec(G)}$. By definition we have $c(F_0) \le 2^{c(F)}-1$. SAT-decision for a CNF-clause-set $F \in \Cls$ can be done in time $O(\ell(F) \cdot 2^{c(F)})$ and linear space, whence $F_0$ can be computed in time $O(\ell(F)^2 \cdot 2^{2 c(F)})$. We obtain $\primec_0(F) = \rsub(F_0)$ (applying subsumption-elimination), which can be computed (by the trivial algorithm) in time $O(c(F_0)^2 \cdot n(F_0))$ and linear space. \Qed
\end{prf}

Thus computation of the representation $F'' := \primec_0(X_1(F))$ of an XOR-clause-set $F$ with fixed maximal clause-length $p$ is fpt in $c(F)$ (recall Section \ref{sec:transarbxor}):
\begin{lem}\label{lem:fptkc}
  Consider a constant $p \in \NNZ$ and an XOR-clause-set $F \in \Pcls{p}$. The CNF-representation $F' := \primec_0(X_0(F))$ of $F$ is obtained from $F'' := \primec_0(X_1(F))$ via selecting the clauses $C \in F''$ with $\var(C) \sse \var(F)$, where we have $c(F'') \le 16^{p \cdot c(F)}$, and where $F''$ can be computed in time $O(\ell(F) \cdot 4096^{p \cdot c(F)})$.
\end{lem}
\begin{prf}
We have $c(X_1(F)) \le 4 p \cdot c(F)$, and thus by Lemma \ref{lem:primcec} we can compute $F''$ of size $c(F'') \le 2^{4 p \cdot c(F)}$ in time $O(\ell(F) \cdot 2^{12 p \cdot c(F)}$. \Qed
\end{prf}

\end{document}